\documentclass[pre,twocolumn,preprintnumbers,amsmath,amssymb,nofootinbib,floatfix]{revtex4}

\usepackage{graphicx,bm,xcolor, bbold}
\usepackage[normalem]{ulem}
\usepackage{hyperref}

\makeatletter
\def\graphicscale{\twocolumn@sw{0.3}{0.4}}
\def\graphicthreescale{\twocolumn@sw{0.3}{0.4}}

\newcommand{\rev}[1]{\textcolor{black}{#1}}

\begin{document}

\title{Out-of-equilibrium dynamics arising from slow round-trip 
  variations of
  \\ Hamiltonian parameters across quantum and classical critical points}

\author{Francesco Tarantelli}
\affiliation{Dipartimento di Fisica dell'Universit\`a di Pisa
        and INFN, Largo Pontecorvo 3, I-56127 Pisa, Italy}

\author{Ettore Vicari} 
\affiliation{Dipartimento di Fisica dell'Universit\`a di Pisa
        and INFN, Largo Pontecorvo 3, I-56127 Pisa, Italy}

\date{\today}

\begin{abstract}
  We address the out-of-equilibrium dynamics of many-body systems
  subject to  time-dependent round-trip protocols across quantum
  and classical (thermal) phase transitions. They are realized by
  slowly changing one relevant parameter $w$ across its critical point
  $w_c=0$, linearly in time with a large time scale $t_s$, from
  $w_i<0$ to $w_f>0$ and then back to $w_i<0$, thus entailing multiple
  passages through the critical point.  Analogously to the one-way
  Kibble-Zurek protocols across a critical point, round-trip protocols
  develop dynamic scaling behaviors at both classical and quantum
  transitions, put forward within renormalization-group
  frameworks. The scaling scenario is analyzed within some
  paradigmatic models undergoing quantum and classical transitions
  belonging to the two-dimensional Ising universality class, such as
  one-dimensional quantum Ising models and fermionic wires, and
  two-dimensional classical Ising models (supplemented with a purely
  relaxational dynamics).  While the dynamic scaling frameworks are
  similar for classical and quantum systems, substantial differences
  emerge due to the different nature of their dynamics, which is
  purely relaxational for classical systems (implying thermalization
  in the large-time limit at fixed model parameters), and unitary in
  the case of quantum systems. In particular, when the critical point
  separates two gapped (short-ranged) phases and the extreme value
  $w_f>0$ is kept fixed in the large-$t_s$ limit of the round-trip
  protocol, we observe hysteresis-like scenarios in classical systems,
  while quantum systems do not apparently develop a sufficiently
  robust scaling limit along the return way, due to the presence of
  rapidly oscillating relative phases among the relevant quantum
  states.
\end{abstract}

\maketitle

% ========================= BODY =========================

\section{Introduction}
\label{intro}

Many-body systems generally develops out-of-equilibrium phenomena when
they are driven across phase transitions, due to the fact that
large-scale critical modes do not equilibrate, even when the time
scale $t_s$ of the variation of the system parameters is taken very
large.  Out-of-equilibrium dynamic phenomena at phase transitions,
such as hysteresis and coarsening, Kibble-Zurek (KZ) defect
production, aging, etc., have been addressed in a variety of contexts,
both experimentally and theoretically, at classical and quantum phase
transitions (see, e.g., Refs.~\cite{Kibble-76,Kibble-80, Binder-87,
  CDTY-91, BCSS-94, Bray-94, Zurek-96, BBFGP-96, CG-05, BDS-06,
  WNSBD-08, Dziarmaga-10, PSSV-11, Ulm-etal-13, Pyka-etal-13, NGSH-15,
  Biroli-16,Trenkwalder-etal-16,RV-21} and references therein).
Out-of-equilibrium scaling behaviors generally emerge when slowly
crossing a critical point, i.e. in the large-$t_s$ limit. They depend
on the nature of the classical or quantum transition, its universality
class, and the type of critical dynamics in classical systems, see
e.g. Refs.~\cite{Kibble-80,Zurek-85,Zurek-96,Polkovnikov-05, ZDZ-05,
  Dziarmaga-05, PG-08, Dziarmaga-10, DGP-10, GZHF-10, PSSV-11,
  CEGS-12, Dutta-etal-book,PV-16,PRV-18-loc, PRV-18, RV-19-de,
  RV-20,RV-21}.  Therefore, slow (quasi-adiabatic) passages through
critical points allow us to probe the universal features of the
long-range modes emerging at thermal and quantum critical phenomena.

In both classical and quantum contexts, we consider many-body systems
whose Hamiltonian can be written as
\begin{equation} H(t) \equiv H[w(t)] = H_{c} + w(t) \,
  H_{p} \,,
  \label{hlamt}
\end{equation}
where $w(t)$ is a time-dependent Hamiltonian parameter, while $H_c$
and $H_p$ do not depend on time. $H_c$ is supposed to be a critical
Hamiltonian at its transition point, which may be a quantum continuous
transition driven by quantum fluctuations, or a classical continuous
transition driven by thermal fluctuations.  $H_p$ represents a
nontrivial relevant perturbation.  In particular, within quantum
many-body models, one generally assumes that $[H_c,H_p]\neq 0$.  The
tunable parameter $w$ controls the strength of the coupling with the
perturbation $H_p$, and is taken as a relevant parameter driving the
continuous transition.  Therefore $w_c = 0$ corresponds to the
transition point. The scaling properties of the out-of-equilibrium
dynamics across phase transitions can be probed by considering
time-dependent protocols where one of the relevant parameters, such as
$w(t)$, is slowly changed across the transition point $w_c=0$,
linearly in time with a large time scale $t_s$.

Across a phase transition, the growth of an out-of-equilibrium
dynamics is inevitable in the thermodynamic limit, even for very slow
changes of the parameter $w$, because large-scale modes are unable to
equilibrate the long-distance critical correlations emerging at the
transition point, even in the limit of large time scales of the
variations. As a consequence, when starting from equilibrium states at
the initial value $w_i$, the system cannot pass through equilibrium
states associated with the values of $w(t)$ across the transition
point, thus departing from an adiabatic dynamics. Such a departure
from equilibrium develops peculiar out-of-equilibrium dynamic scaling
phenomena in the limit of large time scale $t_s$ of the time variation
of $w(t)$.  \rev{A related issue is the so-called KZ problem, i.e.  the
scaling behavior of the amount of final defects after slow passages
through continuous transitions, from the disorder phase to the order
phase~\cite{Kibble-76, Kibble-80, Zurek-85, Zurek-96, ZDZ-05,
  Polkovnikov-05, Dziarmaga-05, PG-08, Dziarmaga-10, Dutta-etal-book,
  PSSV-11, CEGS-12, RV-21, Damski-05, USF-07, USF-10, NDP-13,
  DZ-14,RDZ-19,RMAKVE-21}.} The general features of the KZ dynamic
scaling, and in particular the KZ predictions for the abundance of
residual defects, have been confirmed by several analytical and
numerical studies, see, e.g., Refs.~\cite{Dziarmaga-10,
  Dutta-etal-book, PSSV-11, CEGS-12, NDP-13, RV-21} and citing
references, and by experiments for various physically interesting
systems, see, e.g., Refs.~\cite{DRGA-99, MMARK-06, SHLVS-06, WNSBD-08,
  CWBD-11, Griffin-etal-12, Ulm-etal-13, Pyka-etal-13, LDSDF-13,
  Braun-etal-15, Chomaz-etal-15, NGSH-15, Cui-etal-16, Gong-etal-16,
  Anquez-etal-16, CFC-16, Keesling-etal-19}.

The out-of-equilibrium scaling behaviors of many-body systems subject
to slow passages across classical and quantum critical points present
notable analogies. They can be discussed within a unified
renormalization-group (RG) framework, like the equilibrium scaling
behaviors that can be related by the quantum to classical mapping, see
e.g. Refs.~\cite{Sachdev-book,RV-21}.  However, we should recall
that, while quantum systems are ruled by the unitary dynamics of
quantum mechanics, the out-of-equilibrium scaling behavior of
classical systems depend also on the particular choice of dynamics,
whether it is purely relaxational or it implies conserved quantities,
which gives generally rise to different dynamic
features~\cite{HH-77,Ma-book,FM-06}.

In this paper we address the effects of slow round-trip variations of
the Hamiltonian parameter $w(t)$ in Eq.~(\ref{hlamt}), entailing
multiple crossings of quantum and thermal transitions.  More
precisely, we consider round-trip protocols where the system starts at
the equilibrium condition (ground state in quantum systems) associated
with the initial value $w_i=w(t_i)<0$, then the out-of-equilibrium
dynamics is driven by linear changes of $w(t)$ up to $w_f>0$, thus
crossing the transition point $w_c=0$, and then by changing it back to
the original value $w_i<0$, again linearly in time, which implies a
further crossing of the transition point. The time scale $t_s$ of the
variations of $w(t)$ is unique, and the slow-crossing regime is
realized in the large-$t_s$ limit.

We address these issues within classical (see
e.g. Ref.~\cite{Cardy-book}) and quantum (see
e.g. Ref.~\cite{Sachdev-book}) continuous transitions, characterized
by emerging long-range correlations. We exploit unified RG
frameworks~\cite{WK-74,Fisher-74,Wegner-76,Ma-book,Parisi-book,
  PV-02,Cardy-book,SGCS-97,Sachdev-book,RV-21}, which allow us to
derive general dynamic scaling behaviors at both classical and quantum
transitions, in the limits of large time scale $t_s$ of the round-trip
KZ protocol and large size $L$ of the model, using standard RG
arguments. For this purpose, we extend the dynamic RG framework
already applied to standard one-way KZ protocols, see
e.g. Refs.~\cite{CEGS-12,RV-21} and references therein.

In this exploratory study of slow round-trip protocols across
continuous transitions, we restrict ourselves to transitions between
gapped phases showing only short-ranged correlations, to avoid the
complications arising from the effects of gapless modes in the ordered
phases.  This is somehow different from the standard KZ protocols
leading to the KZ problem, in which, starting from a disordered phase,
the system is driven to an ordered phases characterized by long-range
correlations, where further important dynamic effects may set in at
large time, such as coarsening phenomena or massless Goldstone
excitations, see e.g. Refs.~\cite{CEGS-12,PV-16}.

In our study we consider some paradigmatic many-body systems
undergoing quantum and classical transitions belonging to the
two-dimensional (2D) Ising universality class:

(i) Quantum one-dimensional (1D) Ising models with an external
time-dependent longitudinal field;

(ii) Quantum Kitaev fermionic wires with a time-dependent chemical
potential;

(iii) Classical 2D lattice Ising models undergoing a
finite-temperature transition, supplemented with a purely relaxational
dynamics driven by an external time-dependent magnetic field.

In all cases we consider time-dependent protocols with round-trip
variations of the Hamiltonian parameter corresponding to $w(t)$ in
Eq.~(\ref{hlamt}), crossing twice the critical point separating
classical or quantum phases with finite correlation lengths, when
$|w(t)|>0$ in Eq.~(\ref{hlamt}).

As we shall see, the analogy of the scaling behaviors emerging from
standard one-way KZ protocols at classical and quantum transitions is
only partially extended to round-trip KZ protocols. Indeed substantial
differences emerge, in particular when the extreme value $w_f>0$ at
the return point (where $w(t)$ stops increasing and starts decreasing)
is kept fixed and finite in the large-$t_s$ dynamic scaling limit of
the round-trip protocol. On the one hand, classical systems show
well-defined scaling phenomena, developing hysteresis-like scenarios;
this is essentially related to the fact that the purely relaxational
stochastic dynamics leads eventually to thermalization in the
large-time limit when keeping the model parameters
fixed~\cite{Parisi-book}.  On the other hand, in quantum systems the
observation of scaling behaviors along the return way turns out to be
more problematic, due to the persistence of rapidly oscillating
relative phases between the relevant quantum states, which make the
return way extremely sensitive to the parameters of the protocol, such
as the extreme value $w_f$ and the size of the system.  This is
essentially related to the quantum unitary nature of the
dynamics. \rev{Indeed we observe some notable similarities with the
behavior of quantum two-level models subject to round-trip protocols,
related to the well-known Landau-Zener-St\"uckelberg
problem~\cite{LZeff,VG-96,SAN-10,ISN-22}.}  Even in this apparently
simple case some features of the behavior along the return way turn
out to be extremely sensitive to the parameters of the round-trip
protocol.

The paper is organized as follows.  In Sec.~\ref{models} we introduce
the above-mentioned quantum and classical models that develop critical
behaviors belonging to the 2D Ising universality class.  In
Sec.~\ref{protocols} we describe the one-way and round-trip KZ
protocols that we consider, across thermal and quantum transitions.
Sec.~\ref{obsdyn} reports the observables that we use to monitor the
dynamic evolution along the KZ protocols in the various models
considered.  Sec.~\ref{fssKZoneway} summarizes the dynamic scaling
theory associated with one-way KZ protocols, within RG frameworks
which apply to both classical and quantum transitions. In
Sec.~\ref{roundtrip} we extend the dynamic scaling theory to
round-trip KZ protocols, emphasizing the possible differences between
classical and quantum behaviors.  Sec.~\ref{numresrotrip} reports the
numerical analyses that support, and further characterize, the
predicted dynamic scaling behaviors, showing also substantial
differences between classical and quantum round-trip KZ
protocols. Finally, in Sec.~\ref{conclu} we summarize and draw our
conclusions.  The App.~\ref{LZlike} analyzes analogous round-trip
protocols within a two-level quantum model with time-dependent
Hamiltonian parameters (similar to that used for the so-called
Landau-Zener-St\"uckelberg problem), which turns out to be useful to
interpret the results obtained for the quantum many-body systems.

\section{The models}
\label{models}

\subsection{Quantum many-body systems}
\label{quamod}

As a paradigmatic quantum many-body system we consider the 1D quantum
Ising models, described by the Hamiltonian
\begin{eqnarray}
  H_{qI}(g,h) = - J \sum_{x=1}^{L} \sigma^{(1)}_{x\phantom{1}}
  \sigma^{(1)}_{x+1} - g \sum_{x=1}^L \sigma^{(3)}_x
  - h \sum_{x=1}^L \sigma^{(1)}_x\,,\quad
  \label{qisingmodel}
\end{eqnarray}
where $L$ is the system size, $\sigma^{(k)}_x$ are the Pauli matrices
on the $x^{\rm th}$ site ($k = 1,2,3$ labels the three spatial
directions). \rev{In the following we consider quantum Ising systems
  with periodic boundary conditions (PBC), obtained by requiring
 $\sigma^{(k)}_{L+1} = \sigma^{(k)}_1$.}

We recall that the quantum Ising model (\ref{qisingmodel}) develops a
quantum critical behavior at $g=g_c=J$ and $h=0$, belonging to the 2D
Ising universality class, see e.g. Ref.~\cite{Sachdev-book}. The model
is always gapped for $h\neq 0$.  The relevant parameters $r\equiv
g-g_c$ and $h$ are respectively associated with even and odd RG
perturbations at the Ising fixed point.  Their RG dimensions are
respectively $y_r=1/\nu=1$ and $y_h=15/8$, so that the length scale
$\xi$ of the critical modes behaves as $\xi\sim |g-g_c|^{-1/y_r}$ for
$h=0$, and $\xi\sim |g-g_c|^{-1/y_h}$ at $g=g_c$.  The dynamic
exponent $z$, controlling the vanishing of the gap $\Delta\sim
\xi^{-z}$ at the transition point, is given by $z=1$. Moreover, we
recall that the RG dimension of the order-parameter field, associated
with the longitudinal operators $\sigma_x^{(1)}$, is given by $y_l=d +
z - y_h=1/8$, while that associated with the transverse operator
$\sigma_x^{(3)}$ is given by $y_t = d + z - y_r = 1$.  In the
following we assume ferromagnetic nearest-neighbour interactions with
$J=1$, thus $g_c=J=1$.

To achieve round-trip protocols between gapped phases, without
degeneration of the lowest quantum states, we consider Ising chains
with PBC at $g=g_c$ driven by a time-dependent longitudinal field
$h(t)$. Therefore,  comparing with Eq.~(\ref{hlamt}), we identify
\begin{eqnarray}
  H_c = H_{qI}(g_c,0)\,,\quad w(t) = h(t)\,, \quad H_p = - \sum_{x}
  \sigma^{(1)}_x\,.
\label{isichoice}
\end{eqnarray}

The quantum Ising Hamiltonian $H_{qI}(g,0)$ for vanishing longitudinal
field $h$ can be mapped into a quadratic model of spinless fermions
through a Jordan-Wigner transformation~\cite{LSM-61, Katsura-62},
obtaining the so-called quantum Kitaev wire:~\cite{Kitaev-01}
\begin{equation}
  H_{K}(\mu) = - \sum_{x} \big( c_x^\dagger c_{x+1} + c_x^\dagger
  c_{x+1}^\dagger + {\rm h.c.}  \big) - \mu \sum_{x} n_x \,,
  \label{kitaev2}
\end{equation}
where $c_x^{(\dagger)}$ is the fermionic annihilation (creation)
operator on site $x$ of the wire, $n_x\equiv c_x^\dagger c_x$ is the
corresponding number operator, and $\mu=-2g$.  The Kitaev model
undergoes a continuous quantum transition at $\mu_c = -2g_c = -2$.  Of
course, it belongs to the 2D Ising universality class as well, so that
$y_\mu= y_r = 1/\nu=1$ (there is no an analogue of the longitudinal
field $h$ of the spin formulation (\ref{qisingmodel}) within the above
fermionic representation).  At the Ising transition the fermionic
operators $c_x$ and the particle density operator $n_x$ acquire the RG
dimensions $y_c=1/2$ and $y_n=1$, respectively.

Although the bulk behaviors of the Ising and Kitaev models in the
infinite-volume limit (and thus their phase diagram) are analogous,
some features of finite-size systems may significantly differ.  As a
matter of fact, the nonlocal Jordan-Wigner transformation of the Ising
chain with PBC does not simply map into the fermionic
model~\eqref{kitaev2} with definte boundary conditions.  Indeed
further considerations apply~\cite{Katsura-62, Pfeuty-70}, leading to
a less straightforward correspondence, which also depends on the
parity of the particle-number eigenvalue.

\rev{The Kitaev quantum wire with antiperiodic boundary
  conditions (ABC), obtained by requiring that $c_{L+1} = -c_1$, turns
  out to be gapped in both phases separated by the quantum transition
  at $\mu_c=-2$.} Indeed, it does not exhibit the lowest-state
degeneracy of the ordered phase of the quantum Ising chain (namely,
the exponential suppression of the gap with increasing $L$).  The
reason for such substantial difference resides in the fact that the
Hilbert space of the former is restricted with respect to that of the
latter, so that it is not possible to restore the competition between
the two vacua belonging to the symmetric/antisymmetric sectors of the
Ising model~\cite{Katsura-62, Kitaev-01, CPV-14, RV-21}.  Therefore, a
continuous quantum transition between gapped phases is also realized
within the Kitaev wire with ABC, by choosing
\begin{eqnarray}
  H_c = H_K(\mu_c)\,,\quad w(t) = \mu(t)-\mu_c\,,
  \quad H_p = - \sum_{x} n_x\,.\;\;
\label{kitchoice}
\end{eqnarray}

\subsection{Classical Ising model}
\label{classmod}

As a classical paradigmatic model undergoing a finite-temperature
continuous transition, we consider the 2D Ising model,
defined on a square lattice by the partition function
\begin{eqnarray}
  &&Z = \sum_{\{s_{\bm x}\}} e^{-\beta H_{cI}}\,,\qquad
  \beta=1/T\,, \label{partfunc}\\
  &&H_{cI}(J,h) = - J \sum_{\langle
    {\bm x} {\bm y} \rangle} s_{\bm x} s_{\bm y} - h
  \sum_{\bm x}
  s_{\bm x}\,,
\label{classisi}  
\end{eqnarray}
where ${\bm x}$ are the sites of the lattice, ${\langle {\bm x} {\bm
    y} \rangle}$ indicates the nearest-neighbour sites of the lattice,
$s_{\bm x}=\pm 1$ are classical spin variables, and $h$ is an external
homogenous magnetic field (we use the same symbol of the external
longitudinal field of the quantum ising model (\ref{qisingmodel}), but
this should not lead to confusion). We consider systems with PBC.  We
again set $J=1$.

The square-lattice Ising model (\ref{classisi})
undergoes a thermal continuous transition at $h=0$ and
$T_c=2/\ln(\sqrt{2}+1)$~\cite{Onsager-44}.  The critical behavior
belongs to the same universality class of the 1D quantum Ising
model. Therefore, it is characterized by the critical exponents
$\nu=1$ and $\eta=1/4$. They are related to the RG dimension $y_t$
associated with the even temperature parameter by $y_t=1/\nu=1$, and
to that associated with the odd external field $h$ by $y_h =
2-\eta/2=15/8$, see e.g. Ref.~\cite{PV-02}.

Since we are going to discuss dynamic behaviors, we must also define
the type of dynamics driving the time evolution of the system.  We
consider a purely relaxational dynamics (also known as model A of
critical dynamics~\cite{HH-77,Ma-book}), which can be realized by
stochastic Langevin equations, or just Metropolis updatings in Monte
Carlo simulations~\cite{Metropolis:1953am}.  The corresponding dynamic
exponent $z$ has been accurately estimated by numerical studies,
obtaining $z\approx 2.167$ with a relative precision that is
apparently better than one per mille. Indeed, some of the most recent
estimates of the dynamic exponent $z$ for purely relaxational dynamics
are $z=2.1667(5)$ from \cite{NB-00}, $z=2.168(5)$ from \cite{WH-97},
$z=2.1665(12)$ from \cite{NB-96}, $z=2.172(6)$ from \cite{G-95},
$z=2.170(6)$ from Ref.~\cite{CV-11}, which have been obtained by
numerical analyses based on Monte Carlo simulations in equilibrium
conditions. In the following we use the estimate $z=2.167(1)$.

One may consider time-dependent KZ protocols also in this classical
context, supplementing the partition function (\ref{partfunc})
defining the classical Ising model with the purely relaxational
dynamics.  Analogously to the quantum case, cf. Eq.~(\ref{isichoice}),
we consider 2D Ising models with PBC at $T_c$ driven by a
time-dependent magnetic field $h(t)$. Therefore, we identify
\begin{eqnarray}
  & H_c = H_{cI}(1,0)\,,\qquad &\beta=\beta_c = {\ln(\sqrt{2}+1)\over
    2} \,, \label{clisichoice}\\ & w(t) = h(t)\,, \qquad &H_p
  = - \sum_{\bm x} s_{\bm x}\,.\nonumber
\end{eqnarray}

\section{One-way and round-trip KZ protocols across transition points}
\label{protocols}

In the following we assume the general Hamiltonian (\ref{hlamt}),
which represents the three models presented in Sec.~\ref{models} with
the identifications in Eqs.~(\ref{isichoice}), (\ref{kitchoice}), and
(\ref{clisichoice}).

\subsection{One-way KZ protocols}
\label{oneway}

KZ-like protocols have been largely employed to investigate the
dynamics of critical systems, at quantum transitions when the
many-body system is subject to unitary time evolutions, and at
classical (thermal) transitions considering, for example, a purely
relaxational dynamics that can be implemented by standard Langevin
equations~\cite{HH-77}.

\subsubsection{Quantum KZ protocols}
\label{quoneway}

In the case of quantum many-body systems, quasi-adiabatic passages
through the continuous quantum transition are obtained by slowly
varying $w$ across $w_c = 0$, following, e.g., the standard KZ
procedure:

(i) One starts from the ground state of the many-body system at
  $w_i < 0$, that is $|\Psi(t=0)\rangle \equiv |\Psi_0(w_i)\rangle$.
  
(ii) Then the out-of-equilibrium unitary dynamics, ruled by the
Schr\"odinger equation
  \begin{equation}
    {{\rm d} \, |\Psi(t)\rangle \over {\rm d} t} =
    - i \, \hat H[w(t)] \, |\Psi(t)\rangle \,,
    \label{unitdyn}
  \end{equation}
  arises from a linear time dependence of the Hamiltonian parameter
  $w(t)$, such as
  \begin{equation}
    w(t) = t/t_s \,,
    \label{wtkz}
  \end{equation}
  up to a final value $w_f>0$. Therefore the KZ protocol starts at
  time $t_i = t_s \, w_i<0$ and stops at $t_f= t_s \, w_f>0$.  The
  parameter $t_s$ denotes the time scale of the slow variations of the
  Hamiltonian parameter $w$.

Across a continuous transition, the growth of an out-of-equilibrium
dynamics is inevitable in the thermodynamic limit, even for very slow
changes of the parameter $w$, because large-scale modes are unable to
equilibrate as the system changes phase. Indeed, when starting from
the ground state associated with the initial value $w_i$, the system
cannot pass adiabatically through the ground states associated with
$w(t)$ across the transition point (in the infinite volume limit),
thus departing from an adiabatic dynamics.  Note that, in the quantum
cases that we consider, cf. Eqs.~(\ref{isichoice}) and
(\ref{kitchoice}), the slow variation of the longitudinal field $w$
brings the system from a gapped condition at $w_i<0$ to another gapped
condition for $w_f>0$. This somehow differs from the standard
situation of the KZ problem related to the defect production going
from disorder to order phases, see e.g. Refs.~\cite{Kibble-76,
  Kibble-80, Zurek-85, Zurek-96, ZDZ-05, Polkovnikov-05, Dziarmaga-05,
  PG-08, Dziarmaga-10, Dutta-etal-book, CEGS-12, PSSV-11, Damski-05,
  USF-07, USF-10, NDP-13, DZ-14}.

\subsubsection{Classical KZ protocols}
\label{cloneway}

In the case of many-body systems at classical transitions, one can
again assume that slow passages through the continuous transition are
obtained by slowly varying $w$ across $w_c = 0$, following the
classical KZ procedure:

(i) One starts from an equilibrium thermalized configuration at $w_i <
0$.
  
(ii) Then the out-equilibrium classical dynamics, ruled by the
relaxational Langevin equation~\cite{HH-77}, or a standard Metropolis
upgrading~\cite{Metropolis:1953am} of lattice configurations, arises
from linear changing of the parameter $w(t)$, as $w(t) = t/t_s$, up to
a final value $w_f>0$.  In the case of Metropolis-like dynamics, this
can be achieved by incrementing the time by one unity after one global
sweep of the lattice variables (Metropolis upgrading of all lattice
spin variables).  Again the KZ protocol starts at time $t_i = t_s \,
w_i<0$ and stops at $t_f= t_s \, w_f>0$.

Since the above protocol involves a stochastic relaxational
process, results are obtained by averaging over an ensemble of
trajectories (starting from an ensenble of thermalized configurations
at $w_i$), obtained following the above protocol.

We remark again that the classical out-of-equilibrium phenomena
associated with the above protocol occurs between two phases, for
$w<0$ and $w>0$, with short-ranged correlations.  This is again
different from standard classical protocols associated with the KZ
problem, in which one passes from a disordered to an ordered phase
characterized by long-range correlations, where further important
dynamic effects may set in, in particular when the global symmetry is
preserved by the KZ protocol and its initial state, such as coarsening
phenomena or massless Goldstone excitations, see
e.g. Ref.~\cite{CEGS-12}.

\subsection{Round-trip KZ protocols}
\label{rtpro}

We now consider round-trip protocols in which the Hamiltonian
parameter $w(t)$ varies linearly from $w_i<0$ to $w_f>0$, which is
analogous to the one-way KZ protocol, and then it returns back to the
original value, crossing twice the transition point. In the case of
quantum systems the round-trip KZ protocol follows the steps:

(i) One starts at $t=t_i$ from the ground state of the many-body
system at $w_i < 0$, given by $|\Psi(t_i)\rangle \equiv
|\Psi_0(w_i)\rangle$.
  
(ii) The out-equilibrium unitary dynamics, ruled by the Schr\"odinger
equation (\ref{unitdyn}), is driven by linearly increasing $w(t)$: as
$w(t) = t/t_s$ from $w_i<0$ (at time $t_i=w_i t_s<0$) to $w_f>0$ (at
time $t_f = w_f t_s>0$).

(iii) Then, for $t>t_f$ the dynamics is ruled by the Schr\"odinger
equation (\ref{unitdyn}) with an external field $w(t)$ that decreases
linearly with the same time scale $t_s$, from $w_f>0$ to the original
value $w_i<0$, closing the cycle.

To simplify the protocol, reducing its number of parameters, we
consider a {\em symmetric} round-trip KZ protocol (an extension of the
later results to the most general case is straightforward) in which we
fix
\begin{equation}
  w_\star = w_f = - w_i \,,
  \label{wstardef}
  \end{equation}
and write the time dependence of $w(t)$ as
  \begin{eqnarray}
  w(t) = {{\cal T}(t)\over t_s}\quad {\rm for}\;\; t_i=-t_\star \le t
  \le 3t_\star\,,
\label{wtrtripdef}
\end{eqnarray}
where
\begin{equation}
  {\cal T}(t) = t_\star - |t-t_\star|
\label{triafunc}
\end{equation}
is the {\em triangular} function going linearly from ${\cal
  T}(-t_\star)=-t_\star$ to ${\cal T}(t_\star)=t_\star$, and then back
to ${\cal T}(3t_\star)=-t_\star$. The parameter $t_s$ represents the
time scale of the variation. The parameter $t_\star>0$ controls the
extension, i.e.  the starting and final times, of the protocols, from
$t_i=-t_\star$ to $t_f = 3 t_\star$, and also the interval of
variation of $w(t)$, from $w(t_i)=-t_\star/t_s$ to $w(t_\star) =
t_\star/t_s$.

Analogously to the quantum case, we extend the one-way KZ protocol for
classical systems to {\em symmetric} round-trip KZ protocols, by
taking the time-dependent parameter $w(t)$ as in
Eq.~(\ref{wtrtripdef}), with the same definitions.

\rev{We finally mention that similar cyclic protocols have been also
  considered in various contexts and phase transitions, see e.g.
  Refs.~\cite{PV-16,HZFKXJD-11,GAQRJ-16,ONL-18,PYLNL-20,BA-20}, in
  particular at first-order phase transitions to show the emergence of
  hysteresis phenomena~\cite{Binder-87}. As we shall see, round-trip
  KZ protocols of classical systems will also lead to the emergence of
  a scaling hysteresis-like scenarios, however their nature and
  scaling propeties are substantially different from that arising at
  first-order transitions. In this paper we will not pursue hysteresis
  issues at first-order classical and quantum transitions; however,
  they may be worth further investigation, as we will mention in the
  conclusive section.}

\section{Observables to monitor the out-of-equilibrium dynamics}
\label{obsdyn}

\subsection{Quantum case}
\label{quobs}

The resulting out-of-equilibrium evolution of quantum many-body
systems can be investigated by monitoring observables and correlations
at fixed time.  To characterize the departure from adiabaticity along
the slow dynamic across the continuous transition, we monitor the
adiabaticity function
\begin{eqnarray}
  A(t) = |\langle \, \Psi_0[w(t)] \, | \, \Psi(t) \, \rangle|\,,
  \label{adtfunc}
\end{eqnarray}
where $|\,\Psi_0[w(t)]\,\rangle$ is the ground state of the
Hamiltonian $H[w(t)]$, i.e. at instantaneous values of $w(t)$,
while $|\,\Psi(t)\,\rangle$ is the actual time-dependent state
evolving according to the Schr\"odinger equation (\ref{unitdyn}).

The adiabaticity function measures the overlap of the time-dependent
state with the corresponding ground state of the Hamiltonian at the
same $w(t)$. Of course, the adiabaticity function for an adiabatic
evolution takes the value $A(t)=1$ at any time.  Since the KZ protocol
starts from the ground state associated with $w_i=w(t_i)$, we have
$A(t_i) = 1$ initially.  In general protocols crossing transition
points, $A(t)$ is expected to depart from the initial value
$A(t_i)=1$, due to the impossibility of the system to adiabatically
follow the changes of the function $w(t)$ across its critical value
$w=0$.  Note however that this is strictly true in the infinite-volume
limit.  In systems of finite size $L$, there is always a sufficiently
large time scale $t_s$, so that the system can evolve adiabatically,
essentially because finite-size systems are strictly gapped, although
the gap $\Delta$ at the continuous quantum transition gets suppressed
as $\Delta \sim L^{-z}$. The interplay between the size $L$ and the
time scale $t_s$ gives rise to nontrivial out-of-equilibrium scaling
behaviors, which can be studied within finite-size scaling (FSS)
frameworks~\cite{RV-21,RV-20}.

Another general observable is related to the surplus energy of the
system with respect to its instantaneous ground state at the given
$w(t)$, i.e.
\begin{equation}
  E_s(t) = \langle \Psi(t) | \, H \, | \Psi(t) \rangle - \langle
  \Psi_0[w(t)] | \, H \,| \Psi_0[w(t)] \rangle \,.
  \label{etdiff}
  \end{equation}
Since the protocols considered start from a ground state at $t_i$, the
surplus energy $E_s(t)$ vanishes along adiabatic evolutions, while
nonzero values $E_s(t)>0$ are related to the degree of
out-of-equilibrium of the dynamics across the transition.

To monitor the out-of-equilibrium dynamics in the case of Ising models
in the presence of a time-dependent longitudinal field $w(t)$, one may
consider the evolution of the local and global average magnetization
\begin{equation}
  m_x(t) \equiv \langle \Psi(t) | \, \sigma_x^{(1)} \, | \Psi(t)\rangle\,,
  \;\;\; M(t) \equiv {1\over L} \sum_x m_x(t)\,,
  \label{magnt}
\end{equation}
as well as the fixed-time correlation function of the order-parameter
operator and its space integral,
\begin{equation}
  G(t,x,y) \equiv \langle \Psi(t) | \,  \sigma_{x}^{(1)} \, 
  \sigma_{y}^{(1)}\,| \Psi(t)\rangle\,.
  \label{twopointt}
\end{equation}
Taking into account the translation invariance due to the absence of
boundaries (such as the cases with PBC or ABC), we trivially have
$m_x(t) = M(t)$ and $G(t,x,y) \equiv G(t,x-y)$.
We also consider the transverse magnetization 
\begin{equation}
N(t) \equiv {1\over L} \sum_x \langle \Psi(t) | \, \sigma_x^{(3)} \,
| \Psi(t)\rangle\,,
  \label{magntt}
\end{equation}
and the  related subtracted quantity
\begin{eqnarray}
  N_s(t) = N(t) - N_c\,,\label{subdef}
\end{eqnarray}
where $N_c$ is the ground-state transverse magnetization at the
critical point, i.e.~\cite{Pfeuty-70}
\begin{eqnarray}  
  N_c = \lim_{L\to\infty}
  \langle \Psi_0, w=0 | \sigma_x^{(3)} | \Psi_0, w=0\rangle =
  {2\over \pi}\,.\label{mzc}
 \end{eqnarray}

In the case of the Kitaev model with ABC subject to a time-dependent
chemical potential, one may consider the particle density, and in
particular the subtracted definition
\begin{equation}
  \rho_s(t) \equiv \langle \Psi(t) | \,  n_x \, |
  \Psi(t)\rangle - \rho_c\,,
  \label{rhot}
\end{equation}
which is independent of $x$ due to translation invariance, and, for
convenience, we have subtracted its known critical ground-state value
in the infinite volume limit, which is given by~\cite{Pfeuty-70}
$\rho_c = (\pi-2)/(2\pi)=0.18169011...$. One may also consider 
fermionic correlation functions, such as
\begin{eqnarray}
%  P(x,t) &\equiv& \langle \Psi(t) | \, c_j^\dagger 
%  c_{j+x}^\dagger + c_{j+x} c_{j} \, | \Psi(t)\rangle\, ,
%  \nonumber \\
  C(x,t) & \equiv  &\langle \Psi(t) |
  \, c_j^\dagger c_{j+x} 
  + c_{j+x}^\dagger c_{j} \, | \Psi(t) \rangle \, ,
  \label{eq:corr}
%  \\
%  D(x,t)& \equiv &
%  \langle \Psi(t) | \, n_j n_{j+x} \, |\Psi(t) \rangle_c \,, 
%  \nonumber
\end{eqnarray}
where $j,x \in [1,L/2]$, we have taken into account the translation
invariance of systems with ABC.

\subsection{Classical case}
\label{clobs}

In the case of the classical 2D Ising systems we consider
the magnetization
  \begin{equation}
    m_{\bm x}(t) \equiv \langle s_{\bm x} \rangle_t\,,
      \;\;\; M(t) \equiv {1\over L^2} \sum_{\bm x} m_{\bm x}(t)\,,
  \label{magnt2d}
\end{equation}
as well as the fixed-time correlation function of the order-parameter
operator and its space integral,
\begin{equation}
  G(t,{\bm x},{\bm y}) \equiv \langle s_{\bm x}\, s_{\bm y}\,
  \rangle_t \,.
  \label{twopointtcl}
\end{equation}
The symbol $\langle \; \rangle_t$ indicates the average over
trajectories at time $t$.  Taking into account the translation
invariance due to the absence of boundaries (such as the cases with
PBC), we trivially have $m_{\bm x}(t) = M(t)$ and $G(t,{\bm x},{\bm
  y}) = G(t,{\bm x}-{\bm y})$.

\section{Dynamic scaling along the one-way KZ protocol}
\label{fssKZoneway}

In this section we outline the main features of the dynamic scaling
behavior that is expected to emerge at the one-way KZ protocol of the
models introduced in the previous sections, driven by the time
dependent $w(t)=t/t_s$, starting from equilibrium conditions at
$w_i=w(t_i)<0$.

\subsection{Dynamic FSS for quantum KZ protocols}
\label{qfssoneway}

\rev{
We first present an overview of the dynamic scaling behavior emerging
at quantum one-way KZ protocols. We discuss it within a dynamic RG
framework.  The RG arguments leading to the dynamic scaling framework
of KZ protocols at quantum transitions have been reviewed in
Ref.~\cite{RV-21} (see in particular its chapter 9).
Dynamic scaling laws are expected to develop in the limit of
  large time scale $t_s$ of the driven parameter $w(t)$, and large
  size $L$ of the system. They must describe the interplay of the
  various dimensionful scales of the problem, such as the time $t$ and
  time scale $t_s$ of the KZ protocol, the size $L$ of the system, and
  the energy scale $\Delta \sim L^{-z}$ of the system at the critical
  point.}

\rev{ Let us consider observables constructed from a local operator
  $O({\bm x})$ with RG dimension $y_o$.  The dynamic FSS of its
  expectation value $O_s$ and its two-point correlation function $G_O$
  are expected to obey homogeneous scaling laws, such as~\cite{RV-21}
  \begin{eqnarray}
&&    O_s(t,t_s,w_i,L) \equiv\langle \Psi(t)| O({\bm x})
  |\Psi(t)\rangle
    \nonumber\\
&& \quad \approx b^{-y_o} {\cal O}(b^{-z} t, b^{y_w} w(t), b^{y_w} w_i,
    b^{-1} L)\,,
    \label{Oscadynwt}  \\ 
    &&    G_O({\bm x},t,t_s,w_i,L) \equiv
\langle \Psi(t)| O({\bm x}_1)\,O({\bm x}_2)
  |\Psi(t)\rangle
    \nonumber\\
&& \quad \approx
    b^{-2y_o} \, {\cal G}(b^{-1} {\bm x}, b^{-z} t,
    b^{y_w} w(t), b^{y_w} w_i, b^{-1} L)\,, \qquad \label{g12scadynwt}
  \end{eqnarray}
  where $b$ is an arbitrary (large) length scale, and we assumed
  translation invariance, i.e., systems without boundaries such as PBC
  or ABC, so that $O_s$ does not depend on ${\bm x}$, and the
  two-point function depends on the difference ${\bm x}\equiv
  {\bm x}_1 - {\bm x}_2$ only.  The scaling functions ${\cal O}$ and ${\cal
    G}_O$ are expected to be universal, i.e.  largely independent of
  the microscopic details of the models and the KZ protocols. Their
  arguments take into account the RG dimensions of the various
  relevant parameters $t$, $w(t)$, $w_i$ at the equilibrium quantum
  transition~\cite{RV-21}.  }

\rev{To derive a dynamic scaling theory, it is possible to exploit the
  arbitrariness of the scale parameter $b$, by fixing it as $b=L$ (see
  e.g. Ref.~\cite{RV-21} for the optimal choice to derive the dynamic
  scaling laws in the infinite-volume {\rm thermodynamic} limit).
  Then, the asymptotic dynamic FSS behavior is obtained by taking
  $t_s\to\infty$ and $L\to\infty$, while appropriate scaling variables
  are kept fixed, such as~\cite{RV-21}}
\begin{eqnarray}
  &K = w(t) L^{y_w}\,, \qquad &\Upsilon = t_s/L^{\zeta}\,,  
  \label{KZscavar}\\
  &\Theta_i
  = w_i\, t_s^{1-\kappa} \,,\qquad &\Theta = w(t) \,
  t_s^{1-\kappa} = t / t_s^{\kappa} \,,\nonumber
\end{eqnarray}
where
\begin{eqnarray}
\zeta = y_w + z\,,\qquad 
\kappa = {z/\zeta} \,,\qquad
1-\kappa = {y_w/\zeta}\,.\label{KZexps}
\end{eqnarray}
Note that $\Theta\ge \Theta_i$, $K= \Upsilon^{\kappa-1}\Theta$, and
that the exponents $\kappa$ and $1-\kappa$ are both positive and
smaller than one. \rev{Note that the most natural time scaling
  variable $t \,\Delta$, where $\Delta \sim L^{-z}$ is the critical
  gap of the system, can be straightforwardly related to $\Theta$ and
  $\Upsilon$ by $t \, \Delta \sim \Theta \Upsilon$.}

Then the dynamic FSS of the generic observables introduced in
Eqs.~(\ref{Oscadynwt}) and (\ref{g12scadynwt}) is given
by~\cite{RV-21}
\begin{eqnarray}
  && O_s(t,t_s,w_i,L) \approx L^{-y_o} {\cal
    O}(\Upsilon,\Theta,\Theta_i)\,,
  \label{magscao}\\
&&    G_O(x,t,t_s,w_i,L) \approx L^{-2y_o}\,
  {\cal G}_O(X,\Upsilon,\Theta,\Theta_i)\,,\qquad
  \label{G2scao}
\end{eqnarray}
where $X\equiv x/L$. The above scaling behaviors are expected to
describe the dynamics within the interval $t_i\le t \le t_f$,
corresponding to the interval $w_i\le w(t) \le w_f$, therefore the
scaling variable $\Theta$ takes values within the interval
\begin{equation}
  \Theta_i \le \Theta \le \Theta_f = w_f t_s^{1-\kappa}>0\,.
  \label{intomega}
  \end{equation}
Since the dynamic FSS limit at fixed $\Theta<\Theta_f$ does not depend
on $\Theta_f$, but only on $\Upsilon$ and $\Theta_i$, in the following
of this section dedicated to one-way KZ protocols, we omit the
dependence on $\Theta_f$. Of course, if we keep $w_f$ fixed in the
large-$t_s$ limit, i.e. if we do not scale $w_f$ to zero to keep
$\Theta_f$ fixed, then $\Theta_f\to\infty$.

We also mention that the scaling functions may have a nontrivial
large-$\Theta$ behavior.  But we postpone this discussion when we will
consider round-trip protocols, where the impact of the extreme value
$w_f$, and therefore $\Theta_f$, will be important for the return
trajectories in quantum models.

Using the above general dynamic scaling ansatz, we can derive the
dynamic FSS of the longitudinal magnetization $M$, the correlation
function $G$, and the transverse magnetization $N_s$ of the quantum
Ising systems, cf. Eq.~(\ref{magnt}), (\ref{twopointt}),
(\ref{magntt}), and (\ref{subdef}),
\begin{eqnarray}
&&  M(t,t_s,w_i,L) \approx L^{-y_l} {\cal
    M}(\Upsilon,\Theta,\Theta_i)\,,
  \label{magsca}\\
&&    G(x,t,t_s,w_i,L) \approx L^{-2y_l}\,
  {\cal G}(X,\Upsilon,\Theta,\Theta_i)\,,
  \label{G2sca}\\
    &&  N_s(t,t_s,w_i,L) \approx L^{-y_t} {\cal
    N}(\Upsilon,\Theta,\Theta_i)\,,
  \label{magscaz}
\end{eqnarray}
where ${\cal M}$, ${\cal G}$ and ${\cal N}$ are appropriate scale
  functions, and we recall that $y_w=y_h=15/8$, $y_l=1/8$, and
  $y_t=1$, for the 2D Ising universality class.

An analogous scaling behavior is put forward for the adiabaticity
function in quantum systems, cf. Eq.~(\ref{adtfunc}),
\begin{eqnarray}
  A(t,t_s,w_i,L) \approx {\cal A}(\Upsilon,\Theta,\Theta_i)
  = \widetilde{\cal A}(\Upsilon,K,\Theta_i)\,.
\label{wsca2}
\end{eqnarray}
Due to the initial condition of the KZ protocol, we must have
${\cal A}(\Upsilon,\Theta_i,\Theta_i)=1$.
Moreover, since $\Upsilon\to\infty$
keeping $K$ fixed corresponds to the adiabiatic limit within the FSS
framework, we must also have that
\begin{equation}
\widetilde{\cal A}(\Upsilon\to\infty,K,\Theta_i)=1\,.
\label{adlim}
\end{equation}
Using standard RG arguments, we may also derive an ansatz for the
dynamic scaling behavior of the surplus energy defined in
Eq.~(\ref{etdiff}), which turns out to be
\begin{eqnarray}
E_s(t,t_s,w_i,L) \approx L^{-z} {\cal E}(\Upsilon,\Theta,\Theta_i)\,,
\label{essca}
\end{eqnarray}
where $z=1$ is the RG exponent associated with the energy differences
of the lowest states of the spectrum.  Note that the leading analytic
background contributions~\cite{CPV-14,RV-21}, generally arising at the
critical point, get cancelled by the difference of the two terms in
the definition of $E_s$, cf. Eq.~(\ref{etdiff}).

\rev{We now note that, with increasing $L$, the dynamic FSS occurs
  within a smaller and smaller interval $\delta w$ of values of $|w|$
  around $w=0$: since the time interval of the out-of-equilibrium
  process described by the scaling laws scales as $t_{\rm KZ}\sim
  t_s^{\kappa}$, the relevant interval $\delta w$ of values of $|w|$,
  where a nontrivial out-of-equilibrium scaling behavior is observed,
  shrinks as
\begin{equation}
  \delta w \sim {t_{\rm KZ}/t_s}\sim L^{-y_w}\,,
  \label{deltaw}
\end{equation}
when keeping $\Upsilon$ fixed.} Therefore, assuming that the KZ protocol
starts from a gapped phase, such as the case of Ising rings with any
$|w|>0$, and that the initial $w_i<0$ is kept fixed in the dynamic
scaling limit (corresponding to $\Theta_i\to -\infty$), the same
dynamic FSS limit is expected to hold, irrespective of the value of
$w_i$.  Thus, the dynamic FSS behavior at fixed $w_i<0$ in
Eqs.~\eqref{magsca} and (\ref{G2sca}) simplify to
\rev{\begin{eqnarray}
  &&M(t,t_s,w_i,L) \approx L^{-y_l} {\cal M}_i(\Upsilon,\Theta)\,,
  \label{magsca2}\\
    &&G(x,t,t_s,w_i,L) \approx L^{-2y_l}\, {\cal
      G}_i(X,\Upsilon,\Theta)\,,
  \label{G2sca2}\\
      &&  N_s(t,t_s,w_i,L) \approx L^{-y_t} {\cal
    N}_i(\Upsilon,\Theta)\,,
  \label{magscaz}
\end{eqnarray}
They are expected to match the $\Theta_i\to -\infty$ limit, for
example
\begin{equation}
{\cal
  M}_i(\Upsilon,\Theta) = {\cal M}(\Upsilon,\Theta,\Theta_i\to
-\infty)\,,
\label{thetaiinf}
\end{equation}
and analogously for the scaling functions ${\cal G}_i$ and ${\cal
  N}_i$.  Analogous limiting scaling functions ${\cal A}_i$ and ${\cal
  E}_i$ can be defined for the adiabaticity
function and the surplus energy, respectively.}

Note that, in the limit $\Upsilon\to\infty$, the evolution as a
function of $w(t)=t/t_s$ corresponds to an adiabatic dynamics.
Indeed, since the finite size $L$ guarantees the presence of a gap
between the lowest states, one may adiabatically cross the critical
point if $\Upsilon\to\infty$, passing through the ground states of the
finite-size system for $w(t)$. The adiabatic evolution across the
transition point is prevented only when $L\to\infty$ (before the limit
$t_s\to\infty$), i.e., when the time scale $t_{r}$ of the critical
correlations diverges, as $t_r\sim \Delta^{-1}\sim L^z$.  Within the
FSS framework, the adiabatic limit is achieved by taking the
$\Upsilon\to\infty$ limit keeping $K$ fixed, cf. Eq.~(\ref{KZscavar}).

The scaling behavior in the infinite size {\em thermodynamic} limit
can be straightforwardly obtained by taking the $L\to\infty$ limit of
the FSS equations, therefore in the limit $\Upsilon\to 0$ keeping
$\Theta$ fixed.  Thus, taking the large-$t_s$ limit keeping the
initial value $w_i$ fixed, we expect the asymptotic dynamic scaling
behavior
\rev{\begin{eqnarray}
  M(t,t_s,w_i,L\to\infty) &\approx&
  \lambda^{-y_l} {\cal M}_\infty(\Theta)\,,
  \label{magsca3}\\
    G(x,t,t_s,w_i,L\to\infty) &\approx& \lambda^{-2y_l}\, {\cal
      G}_\infty(x/\lambda,\Theta)\,,
  \label{G2sca3}
\end{eqnarray}
}where
\begin{equation}
  \lambda = t_s^{1/\zeta} \,
  \label{xit}
\end{equation}
is the KZ length scale arising from the linear time-dependence of the
Hamiltonian parameter across the transition.  Note that
\begin{eqnarray}
  {\cal M}_\infty(\Theta) = \lim_{\Upsilon\to 0}
  \Upsilon^{y_l/\zeta} {\cal M}_i(\Upsilon,\Theta)\,.
  \label{minftyrel}
\end{eqnarray}
An analogous relation can be derived for the two-point function.
Moreover for the adiabaticity function we obtain
\begin{eqnarray}
  A(t,t_s,w_i,L\to\infty) \approx {\cal A}_\infty(\Theta) =
  {\cal A}_i(\Upsilon\to 0,\Theta)\,.
\label{wsca3}
\end{eqnarray}

\rev{ The above dynamic scaling behaviors are expected to apply in the
  large-$t_s$ and large-$L$ limits. These asymptotic behaviors are
  expected to be approached with power-law suppressed corrections.}
  Scaling corrections to the asymptotic dynamic scaling limit arises
  for finite time scales $t_s$, in particular for moderately large
  $t_s$.  They are expected to be generally controlled by the leading
  irrelevant perturbations at the 2D Ising fixed point, which get
  suppressed as $\xi^{-\omega}$ (where $\xi$ is diverging correlation
  length, or the KZ length scale $\lambda$) with the universal
  exponent $\omega=2$~\cite{CHPV-02,CCCPV-00,CH-00,CPRV-98,CPV-14},
  and also from analytical contribution which dominates the
  corrections arising from the leading irrelevant
  perturbation~\cite{PV-02,RV-21}. However, typically the leading
  corrections in out-of-equilibrium dynamic phenomena arising from KZ
  protocols are suppressed as $\lambda^{-1}$, cf. Eq.~(\ref{xit}), or
  equivalently as $1/L$ in the dynamic FSS~\cite{RV-21}.

Analogous dynamic scaling behaviors are expected for the protocol
within the Kitaev model, essentially replacing $w(t) = \mu(t)-\mu_c$,
and $y_w = y_r = 1$, and using $y_c=1/2$ and $y_n=1$ (instead of
$y_l$) for the scaling prefactor of the two-point functions defined in
Eqs.~(\ref{eq:corr}).

\subsection{Dynamic FSS for classical KZ protocols}
\label{cfssoneway}

\rev{The dynamic FSS framework at classical thermal continuous
  transitions is essentially analogous, so we do not outline its
  derivation, which can be found in Ref.~\cite{PV-16}.}  We introduce
  the same scaling variables (\ref{KZscavar}) with the corresponding
  critical exponents, see Sec.~\ref{classmod}. In particular, the
  dynamic exponent associated with the purely relaxational dynamics is
  given by $z=2.167(1)$. Then the dynamic FSS of the observables
  introduced in Sec.~\ref{clobs} is the same as that reported in
  Eqs.~(\ref{magsca}) and (\ref{G2sca}). Analogous considerations
  concern protocols at finite fixed $w_i$, whose scaling behavior must
  match that obtained in the limit $\Theta_i\to -\infty$, thus leading
  to the scaling ansatzes reported in Eqs.~(\ref{magsca2}) and
  (\ref{G2sca2}), and also Eqs.~(\ref{magsca3}), (\ref{G2sca3}), and
  (\ref{xit}) in the infinite-volume limit, formally obtained in the
  $\Upsilon\to 0$ limit. The adiabatic limit is analogously obtained
  by taking the limit $\Upsilon\to\infty$.

In one-way KZ protocols between phases with short-range correlations,
the purely relaxational dynamics leads to thermalization for
sufficiently long times~\cite{Parisi-book}, after the
out-of-equilibrium regime across the transition.  The limit
$\Theta\to\infty$ of ${\cal M}(\Upsilon,\Theta,\Theta_i)$ at fixed
$\Upsilon$ is expected to lead to the infinite-volume equilibrium
value of the magnetization.  To infer it, note first that, in a finite
volume, the slowest time scale scales as $\tau_r \sim L^{z}$ where $z$
is the dynamic exponent. A necessary condition to obtain equilibrium
results is therefore that $t_s \gg \tau_r$, i.e., $t_s L^{-z} \to
\infty$. At fixed $\Upsilon$, since $\Upsilon=t_s/L^{\zeta}$ and
$\zeta=y_w+z$, we have $t_s L^{-z} = \Upsilon L^{y_w}$ and hence the
condition is satisfied for $L\to \infty$. Since we take the limit
$\Theta\to\infty$, we are considering the system at times $t$ much
larger than the time scale at which the out-of-equilibrium behavior
occurs, so that the system is in equilibrium. Therefore, the scaling
function ${\cal M}$ should match its equilibrium counterpart ${\cal
  M}_e(K)$. Finally, since $K = w L^{y_w} =
\Upsilon^{1-\kappa}\Theta$, in the limit $\Theta\to\infty$ at fixed
$\Upsilon$ we have $K \to \infty$, i.e., we are considering the
behavior in the infinite-volume limit.

The above considerations, arising from the eventual thermalization
under relaxational dynamics, turn out to be the key point
distinguishing round-trip protocols within classical and quantum
contexts, see below.

\section{Dynamic scaling for the round-trip KZ protocol}
\label{roundtrip}

We now address the out-of-equilibrium dynamics at the round-trip
protocols outlined in Sec.~\ref{rtpro}.  The scaling arguments of the
one-way protocol can be extended to the case of round trip. For
round-trip protocols we expect a further nontrivial dependence on the
upper extreme value $w_f$ of $w$, through the scaling variable
$\Theta_f=w_f t_s^{1-\kappa}$.  In the following we consider the {\em
  symmetric} round-trip protocol with $w_f = - w_i = w_\star$, thus
$\Theta_i = - \Theta_f$.

\subsection{Quantum dynamic FSS}
\label{qfssKZroundtrip}

Analogously to the one-way KZ protocol, we define the scaling
variables
\begin{eqnarray}
  \Upsilon = t_s/L^{\zeta}\,, \quad
  \Theta = w(t) \, t_s^{1-\kappa}\,,\quad
  \Theta_\star =  w_\star\, t_s^{1-\kappa} \,,
 \label{scalvar2}
  \end{eqnarray}
where $|\Theta|\le\Theta_\star$, and the exponents $\zeta$ and
$\kappa$ are reported in Eq.~(\ref{KZexps}).  We also define $K=w(t)
L^{y_w}$ and $X = x/L$. Note that now $\Theta$ is nonmonotonic, like 
$w(t)$, cf. Eq.~(\ref{wtrtripdef}), i.e. it takes the same
value twice. For this reason, we divide the time evolution into two parts:
the first {\em outward} time evolution ($a$), from
$\Theta=-\Theta_\star<0$ to $\Theta=\Theta_\star>0$ (corresponding to
$-t_\star\le t \le t_\star$), and then the second {\em return}
evolution ($b$), from $\Theta=\Theta_\star$ to $\Theta=-\Theta_\star$
(corresponding to $t_\star\le t \le 3t_\star$).

Again, the dynamic FSS behavior is expected to be obtained by taking
$t_s\to\infty$ and $L\to\infty$, while keeping the scaling variables
$\Upsilon$, $\Theta$, $\Theta_\star$, $K$ and $X$ fixed.  Then, the
expectation value $O_s$ and correlation function $G_O(x-y)$ of a
generic local observable $O(x)$ are expected to behave as
\begin{eqnarray}
  &&  O_s^{(a/b)}(t,t_s,w_\star,L) \approx 
  L^{-y_o} {\cal O}^{(a/b)}(\Upsilon,\Theta,\Theta_\star)\,,
  \label{genOscart}\\
&&    G_O^{(a/b)}(x,t,t_s,w_\star,L) \approx L^{-2y_o}\,
  {\cal G}_O^{(a/b)}(X,\Upsilon,\Theta,\Theta_\star)\,,\qquad
  \nonumber
\end{eqnarray}
\rev{
  where the superscripts $(a)$ and $(b)$ indicate the outward and
  return trajectories.}
Note that the values of the observables after the full cycle do not
generally equal those at the beginning, i.e. for finite $\Upsilon$
\begin{equation}
  {\cal O}_s^{(b)}(\Upsilon,-\Theta_{\star},\Theta_\star)
  \neq {\cal O}_s^{(a)}(\Upsilon,-\Theta_{\star},\Theta_\star)\,,
  \label{disacy}
  \end{equation}
unless we consider the adiabatic limit $\Upsilon\to\infty$.  The above
scaling ansatz apply to any observable introduced in Sec.~\ref{obsdyn}
for the quantum models considered.  In particular, the adiabaticity
function and the surplus energy are expected to behave as
\begin{eqnarray}
  &&  A^{(a/b)}(t,t_s,w_\star,L) \approx
  {\cal A}^{(a/b)}(\Upsilon,\Theta,\Theta_\star)\,,
  \label{asca3}\\
  &&  E_s^{(a/b)}(t,t_s,w_\star,L) \approx L^{-z}
        {\cal E}_s^{(a/b)}(\Upsilon,\Theta,\Theta_\star)\,.
  \label{esca3}  
\end{eqnarray}
Concerning the approach to the above asymptotic scaling behaviors, we
expect scaling corrections analogous to those mentioned in the case of
the one-way KZ protocol, at least when $\Theta_\star$ is kept finite.

The above scaling behaviors appear quite similar to those already
emerging at the one-way KZ protocols. However, a nontrivial issue
concerns the existence of the large-$\Theta_\star$ limit, and the
existence of a scaling limit of the return trajectories when
$w_\star>0$ is kept fixed in the round-trip protocol. As we shall see,
classical and quantum systems turn out to behave differently. On the
one hand, the relaxational dynamics of classical system lead to a well
defined dynamic scaling when keeping $w_\star>0$ fixed, developing a
hysteresis-like scenario. On the other hand, for quantum systems, thus
unitary dynamics, such a limit turns out to be problematic, due to
rapid oscillations which make the return somehow chaotic, and
extremely sensitive to the protocol parameters, such as $w_\star$,
$L$, etc...

\subsection{Classical dynamic FSS}
\label{cfssKZroundtrip}

The RG framework allows us to describe also the dynamic FSS arising
from round-trip KZ protocols in classical systems. Indeed, analogous
scaling relations apply. We introduce the same scaling variables as in
Eq.~(\ref{scalvar2}), and the scaling Eqs.~(\ref{genOscart}) for
generic observables, such as those defined in Eqs.~(\ref{magnt2d}) and
(\ref{twopointtcl}).  An important difference between quantum and
classical systems is related to the large-$\Theta_\star$ limit of
these scaling equations.

The large-$\Theta_\star$ limit is expected to be well defined for
classical systems driven across transitions between phases with
short-ranged correlations. This is essentially related to the fact
that the purely relaxational dynamics is able to thermalize the system
at finite $w=O(1)$, i.e. outside the critical region, for sufficiently
large $t_s$. When $w(t)>0$ thermalization is achieved after a
sufficiently large time $t_{\rm th}$. Therefore an equilibrium
behavior is realized for any $t>t_{\rm th}$, thus depending on the
actual value $w(t)$ only, independently of the versus of the time
changes of $w(t)$.  At the turning point the system is thermalized and
ready to follow an equivalent trajectory toward $w=-w_\star$, starting
from an equilibrium condition as the initial one.  Of course, due to
the inevitable out-of-equilibrium when crossing the transition, the
return trajectory with decreasing $w(t)$ differs from the one with
increasing $w(t)$, and the size of the area within the two curves
somehow quantifies the degree of out-of-equilibrium.  Therefore, for
classical systems we expect that the limits $\Theta_\star\to\infty$ of
the scaling functions exist, i.e.
\begin{eqnarray}
  \lim_{\Theta_\star\to\infty}
      {\cal M}^{(a/b)}(\Upsilon,\Theta,\Theta_\star) \equiv
      {\cal M}_i^{(a/b)}(\Upsilon,\Theta)\,,
      \label{limthetastar}
\end{eqnarray}
and analogously for the correlation
functions. Moreover, such limit is expected to be realized by
round-trip protocols with finite $w_\star>0$.  Moreover, the symmetry
under ${\mathbb Z}_2$ reflection implies that
\begin{eqnarray}
  {\cal M}_i^{(b)}(\Upsilon,\Theta) = - {\cal
    M}^{(a)}_i(\Upsilon,-\Theta)\,.
  \label{partityimp}
  \end{eqnarray}
Since the outward ($a$) and return ($b$) trajectories give rise to a
close area, to achieve a quantitative indication of how far the system
is out of equilibrium in the large-$t_s$ limit, we may
define~\cite{PV-16}
\begin{equation}
  I_A(t_s,w_\star,L) = - t_s^{-\kappa} \oint dt\, M(t,t_s,w_\star,L)\,,
\label{aharea}
\end{equation}
where the integration is over the time from the beginning to the end
of the round-trip protocol.  Assuming that the $\Theta_\star\to\infty$
is well defined, and the system develops a critical hysteresis, i.e. a
closed area, during the whole round-trip protocol, the scaling
behavior of $I_A$ must be independent of the actual finite value of
$w_\star>0$. Using the dynamic FSS framework outlined above, we obtain
the scaling prediction
\begin{eqnarray}
&&  I_A(t_s,w_\star,L) \approx L^{-y_l} {\cal I}_A(\Upsilon)   
\label{IAsca}\\
&& = - L^{-y_l} 
  \int_{-\infty}^\infty 
  d\theta \left[ {\cal M}_i^{(a)}(\Upsilon,\theta) -
  {\cal M}_i^{(b)}(\Upsilon,\theta)\right]
  \nonumber\\
  && = - L^{-y_l} 
  \int_{-\infty}^\infty 
  d\theta \left[{\cal M}_i^{(a)}(\Upsilon,\theta) +
  {\cal M}_i^{(a)}(\Upsilon,-\theta)\right]\,.
  \nonumber
\end{eqnarray}
As we shall see, the numerical result will confirm that the scaling
function ${\cal I}_A(\Upsilon)$ is well defined and finite.  Note also
that such scaling hysteresis area is expected to shrink in the
adiabatic limit, i.e. for $\Upsilon\to\infty$.

\section{Numerical results}
\label{numresrotrip}

In this section we report numerical analyses for the various quantum
and classical models introduced in Sec.~\ref{models}, subject to the
one-way and round-trip KZ protocols outlined in Sec.~\ref{protocols}.

\subsection{Along the quantum one-way KZ protocol}
\label{qoneway}

\begin{figure}[!htb]
    \includegraphics[width=0.95\columnwidth]{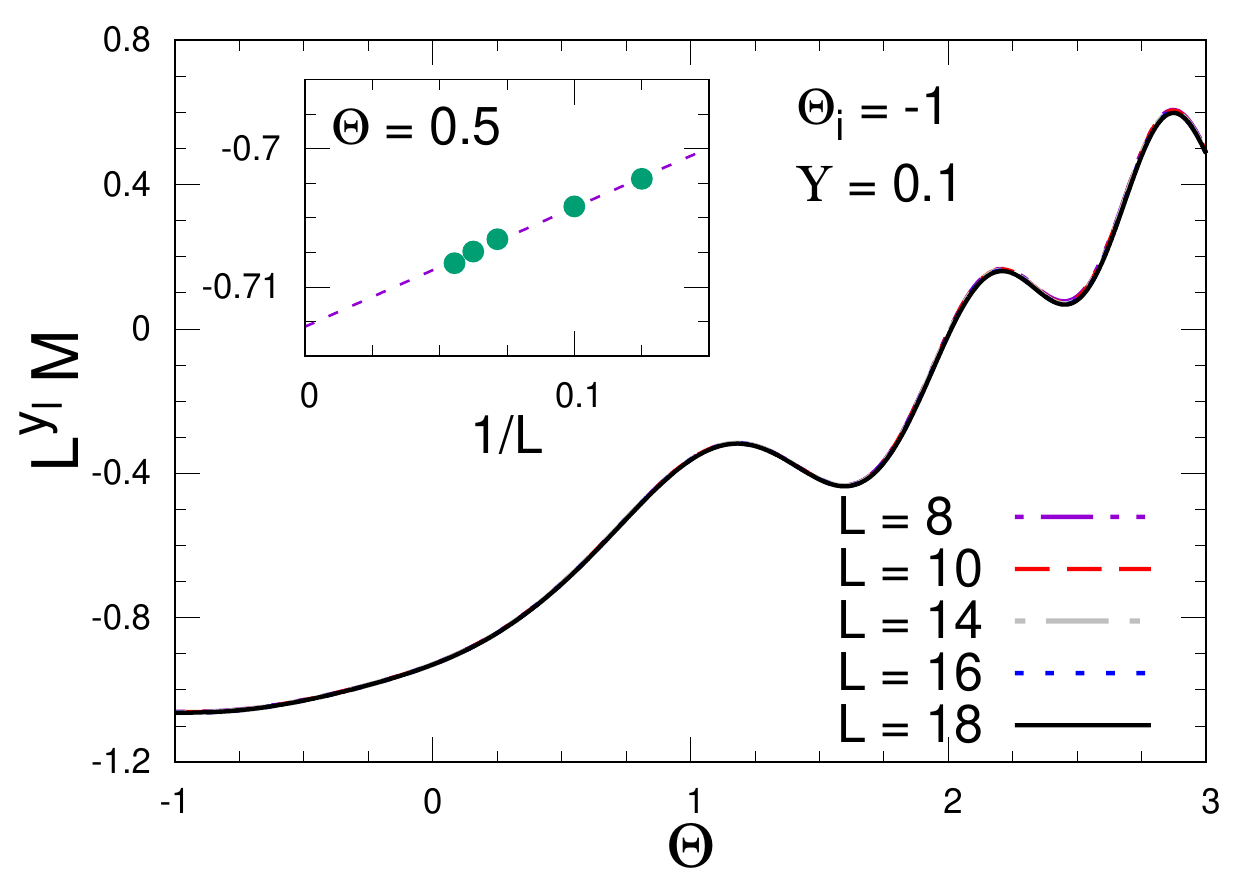}
  \includegraphics[width=0.95\columnwidth]{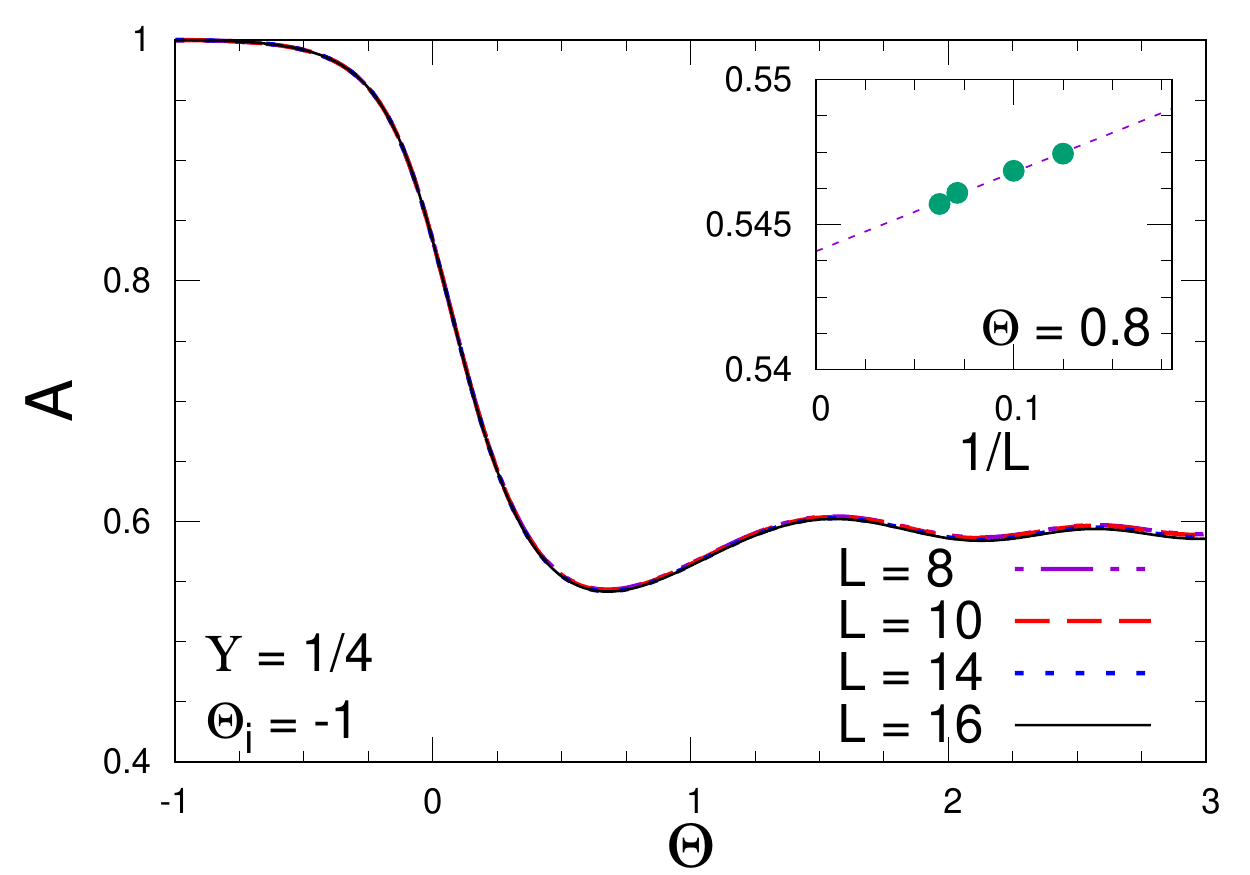}
  \caption{Dynamic FSS of the quantum Ising chain along the one-way KZ
    protocol at fixed $\Theta_i\equiv w_i L^{y_w}$.  We show results
    for the adiabaticity function $A(t,t_s,w_i,L)$ at fixed
    $\Upsilon=t_s/L^\zeta=1/4$ and $\Theta_i=-1$ up to $L=16$ (bottom)
    and the longitudinal magnetization $M(t,t_s,w_i,L)$ at fixed
    $\Upsilon=0.1$ and $\Theta_i=-1$ up to $L=18$ (top), versus
    $\Theta = t/t_s^\kappa$. The exponents $y_w$, $\zeta$, and
    $\kappa$ are reported in Eq.~(\ref{qisiexps}).  The approach to
    the large-$t_s$ asymptotic behavior is globally characterized by
    $O(1/L)$ corrections (apart from small superimposed wiggles), as
    shown by the insets (where the line is drawn to
    guide the eyes).  }
  \label{dfssthetai}
\end{figure}

\begin{figure}[!htb]
  \includegraphics[width=0.95\columnwidth]{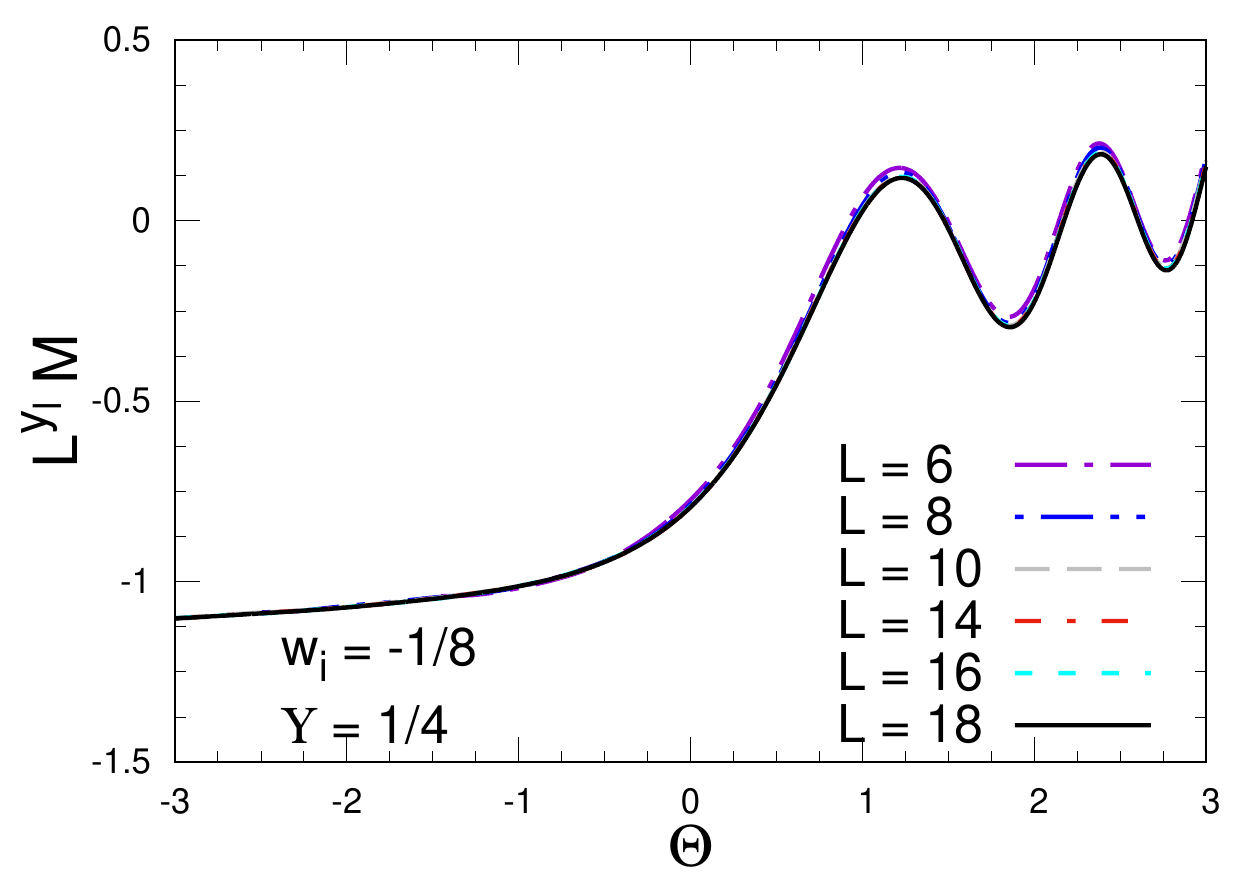}
  \includegraphics[width=0.95\columnwidth]{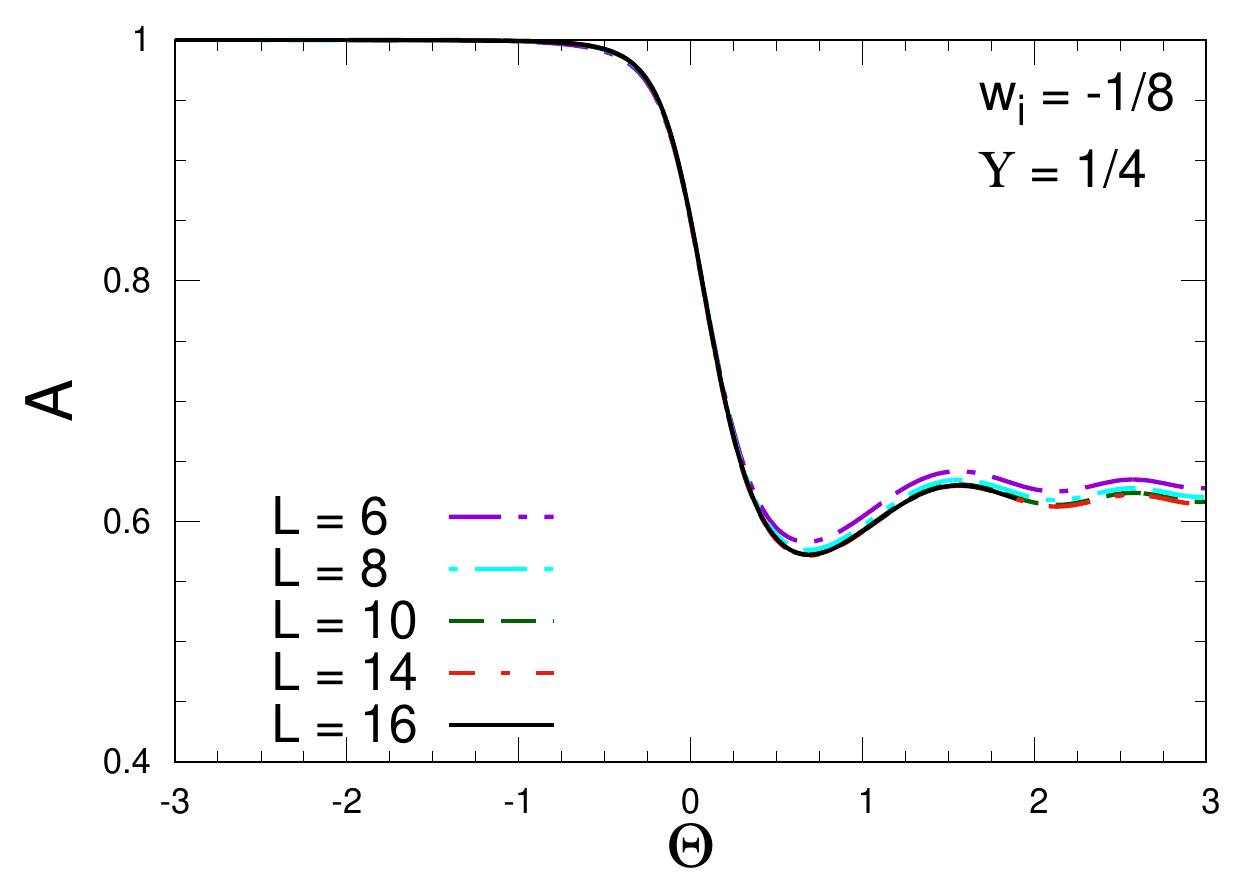}
  \caption{ Dynamic FSS of the quantum Ising chain along the one-way
    KZ protocol at fixed $w_i<0$.  We show the adiabaticity function
    $A(t,t_s,w_i,L)$ up to $L=16$ (bottom) and the longitudinal
    magnetization $M(t,t_s,w_i,L)$ up to $L=18$ (top), at fixed
    $\Upsilon=1/4$ and $w_i=-1/8$, versus $\Theta$. As explained in
    the text, the scaling behavior emerging at fixed $w_i<0$ matches
    that obtained in the $\Theta_i\to -\infty$ limit.
    }
  \label{dfsswi}
\end{figure}

The numerical analyses of quantum Ising chains (\ref{qisingmodel})
with a time-dependent longitudinal field is based on exact
diagonalization. The corresponding Schr\"odinger equation is solved
using a $4^{\rm th}$ order Runge-Kutta method. This approach allows us
to compute the out-of-equilibrium dynamics for lattice size $L\lesssim
20$, which, as we shall see, turns out to be sufficient to achieve a
robust evidence of the dynamic FSS outlined in the previous sections,
and their problematic aspects.

We want to check the dynamic FSS put forward in Sec.~\ref{qfssoneway}.
In the case of the quantum 1D Ising model (\ref{isichoice}), the
exponents entering the definitions of the scaling variables
(\ref{KZscavar}) are 
\begin{equation}
y_w = 15/8\,,\qquad \zeta = 23/8\,,
\qquad \kappa = 8/23\,.
\label{qisiexps}
\end{equation}
Some results for the one-way protocol are reported in
Figs.~\ref{dfssthetai} and \ref{dfsswi}, for the adiabaticity
function, defined in Eq.~(\ref{adtfunc}), and the longitudinal
magnetization, defined in Eq.~(\ref{magnt}), at fixed $\Theta_i$
(Fig.~\ref{dfssthetai}) and fixed $w_i$ (Fig.~\ref{dfsswi}), for
lattice sizes up to $L=16$ and $L=18$ respectively (this difference is
due to the fact that the computation of the adiabaticity function is
heavier).  Although the system sizes of the available results are only
moderately large, we clearly observe a collapse toward asymptotic
scaling curves, thus a robust evidence of the dynamic FSS outlined in
Sec.~\ref{qfssoneway}.  In particular, the dynamic FSS emerging from
the data at fixed $w_i<0$ turns out to be independent of the actual
fixed value $w_i<0$, as predicted by the scaling arguments reported in
Sec.~\ref{qfssoneway} (in Fig.~\ref{dfsswi} we only show results for
$w_i=-1/8$, but we have explicitly checked the independence of $w_i<0$
of the scaling curves). We note that, as expected, the adiabaticity
function significantly drops crossing the quantum transition at finite
values of $\Upsilon$, while it remains close to one, i.e. the value
corresponding to adiabatic evolutions, for large values of $\Upsilon$.
We also note that the data show that the convergence to the asymptotic
dynamic FSS is globally consistent with $O(1/L)$ corrections (apart
from superimposed wiggles), see the insets of Fig.~\ref{dfssthetai}.
\rev{Analogous corrections are observed for other values of the
  parameters, in particular when keeping the starting point $w_i$
  fixed as in Fig.~\ref{dfsswi}.}

\rev{ We remark that the boundary conditions are not particularly
  relevant for the dynamic scaling behavior of quantum Ising systems
  when the KZ protocol is driven by the longitudinal field.  Analogous
  scaling behaviors are expected for systems with boundaries, such as
  open boundary conditions. Note however that, while the power laws
  are not changed, the dynamic FSS functions depend on the boundary
  conditions, moreover the presence of boundaries gives rise to further
  $O(1/L)$ scaling corrections~\cite{CPV-14}.}

\rev{Analogous results are obtained for the quantum Kitaev wire, with
  driving chemical potential. We recall that in this case the choice
  of the boundary conditions, such as ABC, is essential to guarantee
  that the KZ protocol connects two gapped phases~\cite{RV-21}.} The
corresponding exponents, cf. Eq.~(\ref{KZexps}), entering the
definitions of the scaling variables (\ref{KZscavar}), are
\begin{equation}
y_w = 1, \qquad \zeta = 2\,,\qquad \kappa =
1/2\,.
\label{kexps}
\end{equation}
The simpler {\em integrable} nature of the quantum Kitaev wire
(\ref{kitaev2}) allows us to easily consider much larger systems, up
to $L\approx 10^3$, using standard procedures after Fourier
transforming to the momentum space. Again the resulting data (not
shown) for the adiabaticity function, energy surplus, particle
density, and the two-point functions, nicely
support the dynamic FSS outlined in Sec.~\ref{qfssoneway}, see also
Ref.~\cite{RV-21}.

We finally mention that other results for one-way KZ protocols within
quantum 1D Ising systems can be found in the literature, see
e.g. Refs.~\cite{Dziarmaga-10, Dutta-etal-book,RV-21} and references
therein.

\subsection{Along the classical round-trip KZ protocol}
\label{clroundtrip}

\begin{figure}[!htb]
  \includegraphics[width=0.95\columnwidth]{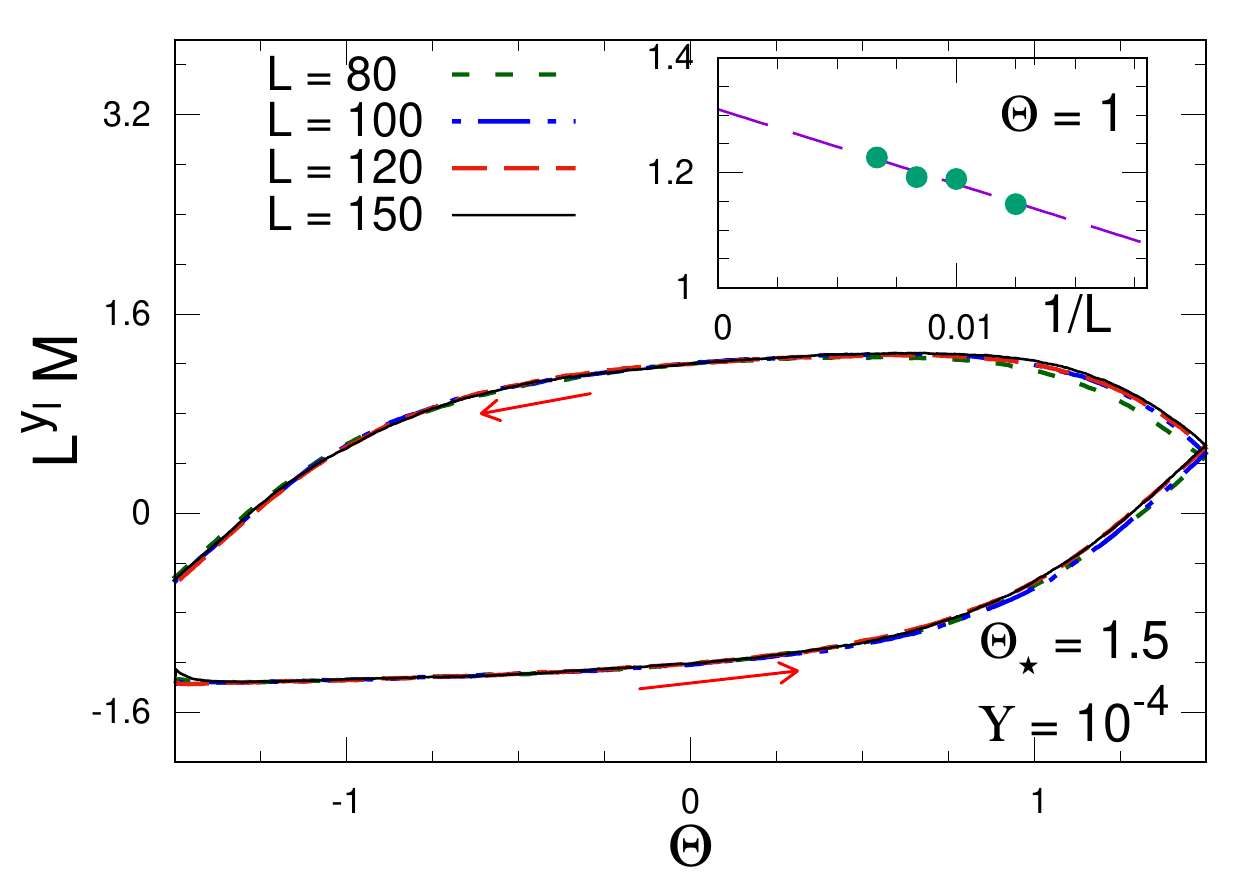}
  \includegraphics[width=0.95\columnwidth]{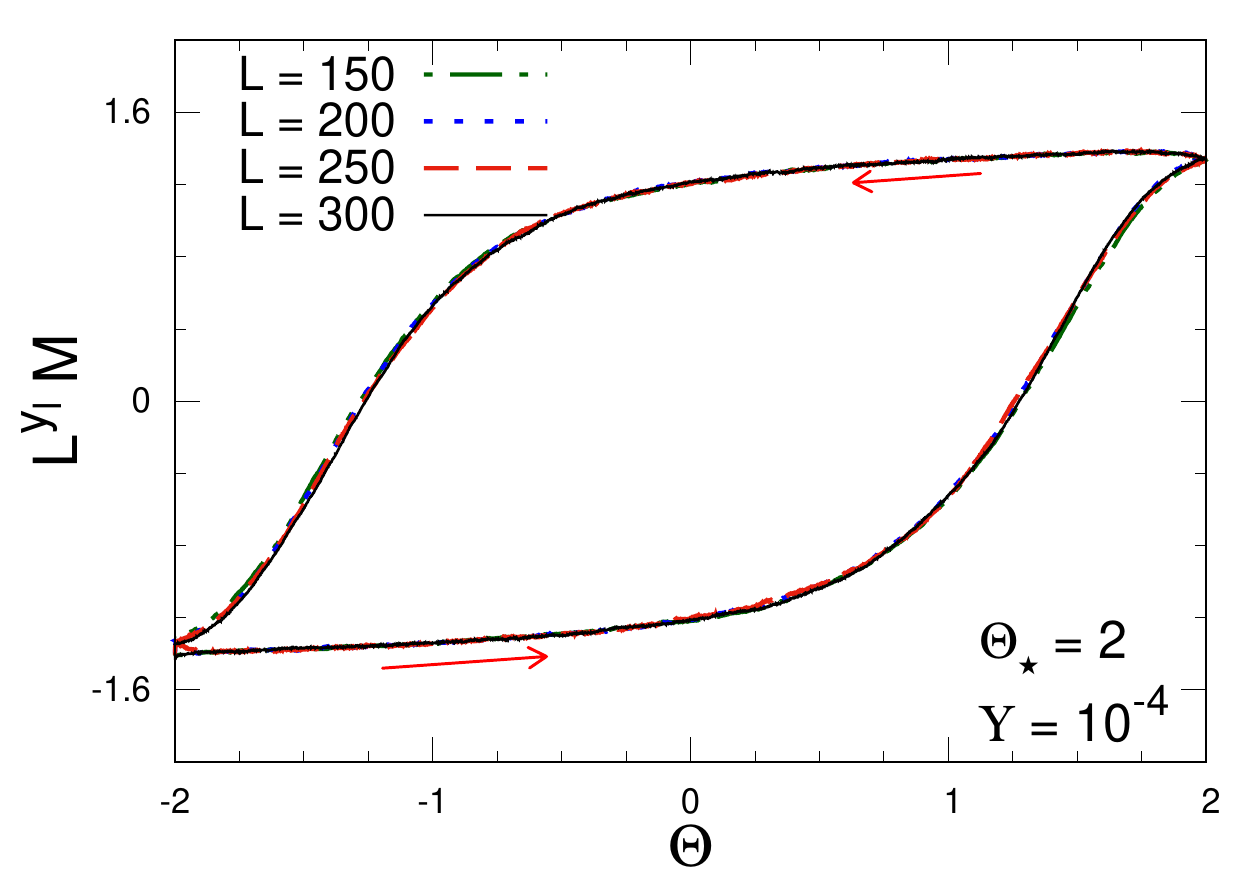}
  \caption{ Dynamic FSS behavior of $M(t,t_s,w_\star,L)$ for the
    classical 2D Ising model along the round-trip KZ protocol. Data
    are obtained at fixed $\Upsilon=10^{-4}$, fixed $\Theta_\star =
    1.5$ (top) and $\Theta_\star = 2$ (bottom), and are plotted versus
    $\Theta=w(t) t_s^{1-\kappa}$. \rev{The arrows indicate the
      direction of the protocol along the outward and return trip.}
    The values of the exponents $y_w$, $\zeta$, and $\kappa$ are
    reported in Eq.~(\ref{clisiexps}). Statistical errors are
    typically smaller than the thickness of the lines.  The
    convergence to the asymptotic scaling behavior is globally
    consistent with an $1/L$ approach, see for example the inset of
    the top figure.  Notice that the return trip goes from right to
    left, because increasing time corresponds to decreasing $\Theta$.
    We note that the magnetization at the end of the protocol differs
    from that at the beginning, i.e.  for $\Theta=-\Theta_\star$ along
    the outward and backward trip, see Eq.~(\ref{disacy2}). Of course,
    the values at $\Theta=\Theta_\star$ coincide for the two
    trajectories.  }
  \label{roundtripM}
\end{figure}

\begin{figure}[!htb]
    \includegraphics[width=0.95\columnwidth]{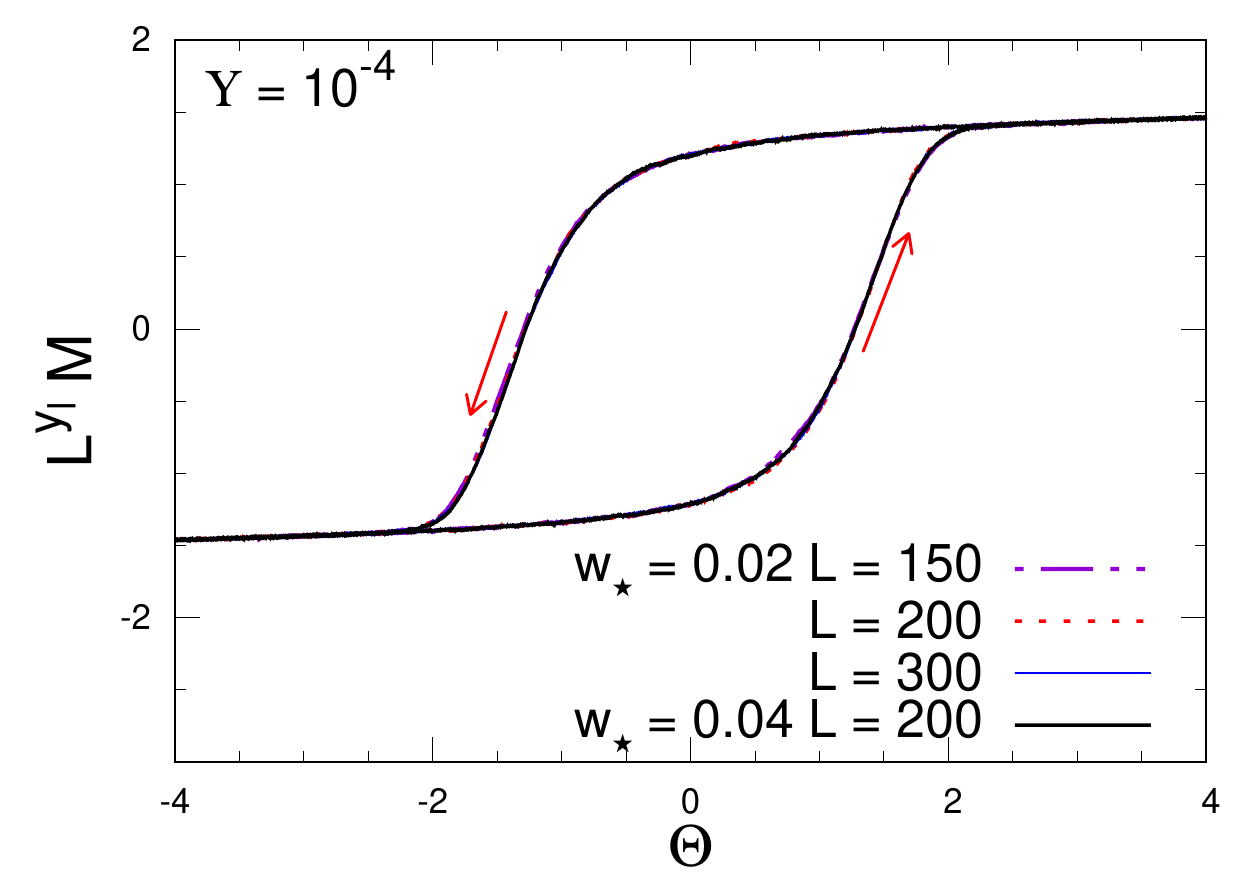}
  \caption{ Dynamic FSS behavior of $M(t,t_s,w_\star,L)$ for the
    classical 2D Ising model along the round-trip KZ protocol for
    fixed $\Upsilon=10^{-4}$, and fixed $w_\star = 0.02$ and $w_\star
    = 0.04$. Statistical errors are typically smaller than the
    thickness of the lines. \rev{The arrows indicate the direction of
      the protocol along the outward and return trip.}  These results
    clearly support the predicted scaling behaviors, see
    Sec.~\ref{cfssKZroundtrip}, and their independence of the finite
    value of $w_\star>0$.}
  \label{roundtripMW}
\end{figure}

\begin{figure}[!htb]
    \includegraphics[width=0.95\columnwidth]{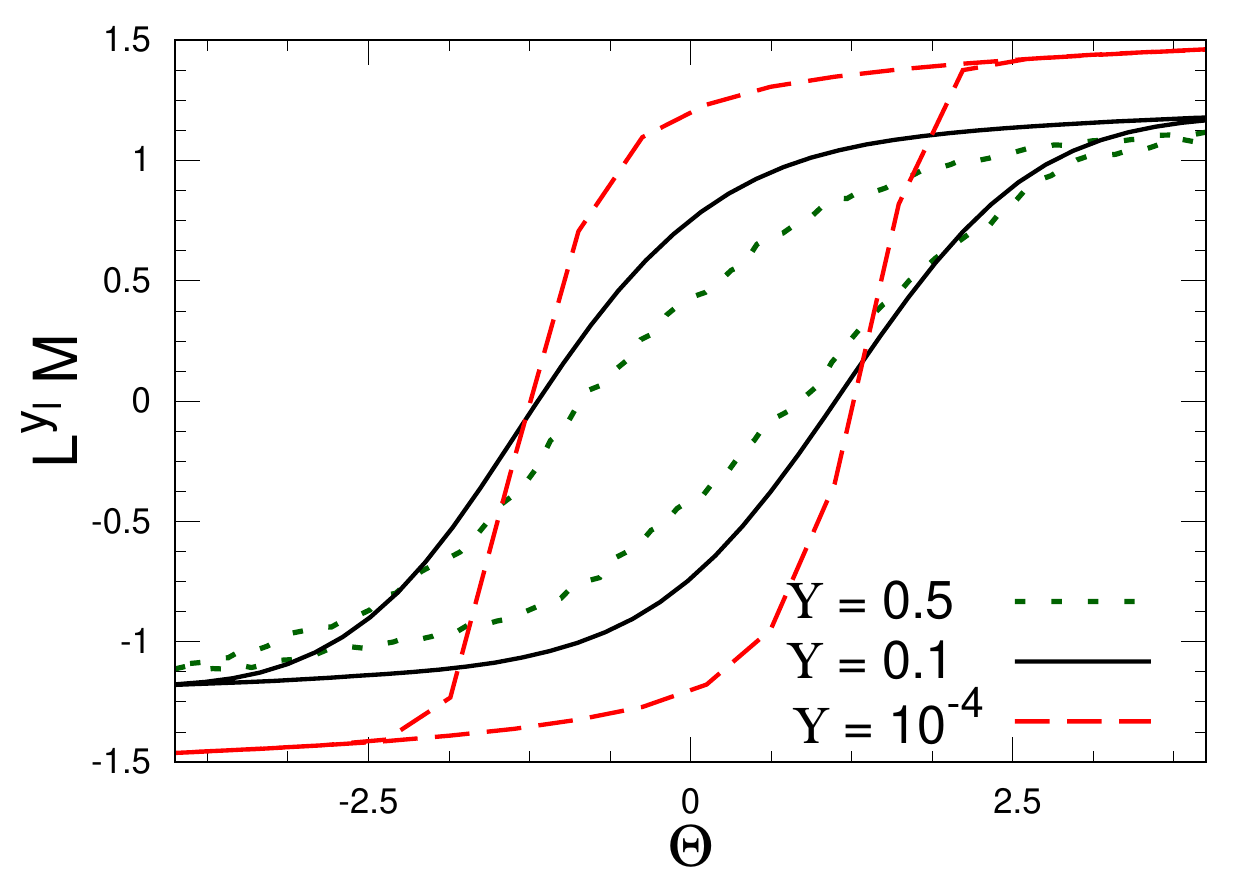}
    \caption{Histeresis curves of the magnetization
      $M(t,t_s,w_\star,L)$ for the classical 2D Ising model along the
      round-trip KZ protocol for various values of fixed $\Upsilon$.
      They confirm that the hysteresis area decreases as $\Upsilon$
      increases. The curve for $\Upsilon=10^{-4}$ is taken from the
      data shown in Fig.~\ref{roundtripMW}, those for $\Upsilon=0.1$
      and $\Upsilon=0.5$ are obtained from simulations for $L=50$,
      whose size is already sufficient to provide a good approximation
      of the asymptotic large-$L$ scaling curves (note that Monte
      Carlo simulations becomes more demanding with increasing
      $\Upsilon$). }
  \label{roundtripMW2}
\end{figure}

The numerical analysis of the classical Ising model is based on
standard Monte Carlo simulations based on local Metropolis upgrading
procedures~\cite{Metropolis:1953am}, which provide a purely
relaxational dynamics without conserved quantities, that is model A
according to the standard classification reported in
Ref.~\cite{HH-77}.  The time unit of this dynamics is represented by a
global sweep of upgradings of all $L\times L$ spin variables. We
perform the single-site update sequentially, moving from one site to
one of its neighbours in a typewriter fashion.  The results along the
time-dependent protocols are obtained by averaging over a sample of
trajectories (tipically of order $10^3$), starting from an ensemble of
thermalized configurations at the initial parameter values. Also in
this case relatively large systems can be simulated, typically for
$L\gtrsim 10^2$.

The dynamic scaling arising from the one-way protocol is quite
analogous to that observed at quantum transitions, with corresponding
scaling behaviors, characterized by the static Ising critical exponents
supplemented by the purely relaxational dynamic exponent $z = 2.167(1)$.
The corresponding relevant exponents, cf. Eq.~(\ref{KZexps}), entering
the definitions of the scaling variables (\ref{KZscavar}), are
\begin{equation}
  y_w = 15/8\,,\quad \zeta = 4.0420(1)\,,\quad
  \kappa = 0.5361(1)\,.
  \label{clisiexps}
  \end{equation}
In the following we only report results for the symmetric round-trip
KZ protocols, taking also into account that its first part is
equivalent to the one-way KZ protocol.

The dynamic scaling behavior of the magnetization,
cf. Eq.~(\ref{genOscart}), is fully supported by the data reported in
Fig.~\ref{roundtripM}, for a fixed $\Upsilon=10^{-4}$ and two
different values of $\Theta_\star$.  Analogous results are obtained
for other values of $\Upsilon$.  As expected the round-trip cycle does
not close the curves for finite values of $\Upsilon$ and
$\Theta_\star$, see Eq.~(\ref{disacy}), leaving a finite gap between
the initial and final values of the cycle, i.e.
\begin{equation}
  {\cal M}^{(b)}(\Upsilon,-\Theta_{\star},\Theta_\star) -
  {\cal M}^{(a)}(\Upsilon,-\Theta_{\star},\Theta_\star)>0\,,
  \label{disacy2}
\end{equation}
which becomes smaller and smaller with increasing $\Theta_\star$.

As argued in Sec.~\ref{cfssKZroundtrip}, the outward and return
trajectories close in the large-$\Theta_\star$ limit, and therefore
for finite $w_\star>0$, giving rise to a critical hysteresis
phenomenon.  This is clearly demonstrated by the results shown in
Fig.~\ref{roundtripMW} for two different finite values of $w_\star>0$,
whose scaling curves coincide. The outward and return curves for large
$|\Theta|$ tend to coincide, differing only within an interval around
$\Theta=0$, which becomes smaller and smaller with increasing
$\Upsilon$, and vanishes in the adiabatic limit
$\Upsilon\to\infty$. Such a dependence on $\Upsilon$ is demonstrated
by the curves reported in Fig.~\ref{roundtripMW2}, showing the
magnetization hysteresis for various values of $\Upsilon$. They
confirm the scaling law (\ref{IAsca}) of the hysteresis
area. Moreover, we mention that the data at small values of $\Upsilon$
(not shown) hint at a convergence of the scaling hysteresis area
${\cal I}_A(\Upsilon)$ to a constant for $\Upsilon\to 0$.

As we shall see, these peculiar behaviors of round-trip protocols
developing scaling hysteresis do not have a quantum counterpart, being
strictly connected with the fact that the classical purely
relaxational dynamics leads eventually to thermalization in the
large-time limit when keeping the model parameters fixed.

We also stress that the above hysteresis scenario arises from the
round-trip protocols between phases with short-ranged
correlations. More complicated situations are expected to occur when
round-trip protocols involve ordered phases, where coarsening
phenomena may drastically change the picture, in particular
along the return trip, in the large-$\Theta_\star$ limit.

\rev{ We finally remark that the boundary conditions do not play a
  relevant role, indeed analogous scenarios are expected to emerge in
  classical Ising systems with boundaries, such as open boundary
  conditions.}

\subsection{Along the quantum round-trip KZ protocol}
\label{qroundtrip}

\subsubsection{Scaling for finite $\Theta_\star$}
  \label{scafinthetastar}

\begin{figure}[!htb]
\includegraphics[width=0.95\columnwidth]{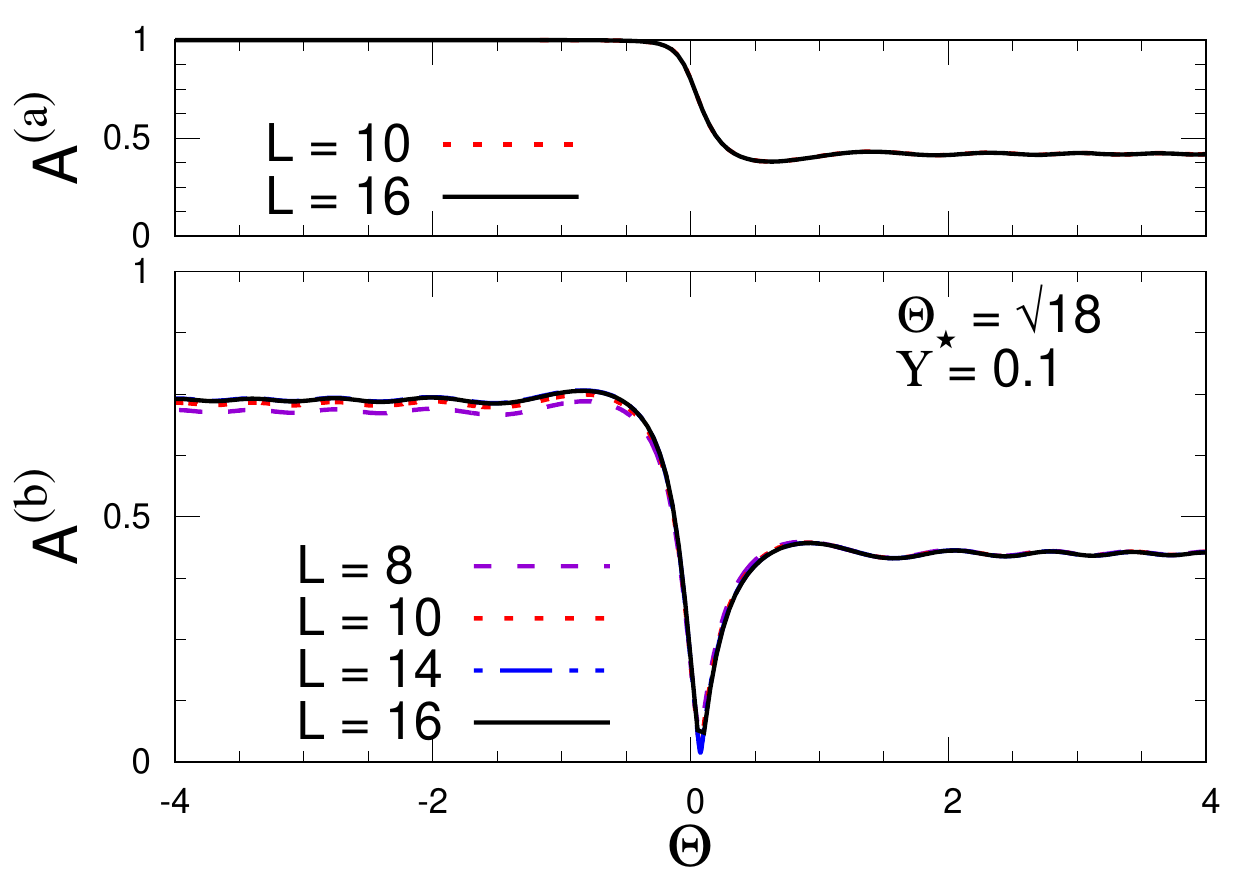}
\caption{ Round-trip dynamic FSS of the quantum Ising chain,
  cf. Eq.~(\ref{isichoice}, for a finite $\Theta_\star$.  We show
  results for the adiabaticity function $A(t,t_s,w_\star,L)$ at fixed
  $\Upsilon = t_s/L^\zeta= 0.1$ and $\Theta_\star = w_\star
  L^{1-\kappa}=\sqrt{18}$, for the outward (top) and return (bottom)
  branches of the round-trip KZ protocol, versus
  $\Theta=w(t)L^{1-\kappa}$, for various size $L$ up to $L=16$.  The
  values of the exponents $y_w$, $\zeta$, and $\kappa$ are reported in
  Eq.~(\ref{qisiexps}).  Notice that the return trip goes from right
  to left, because increasing time corresponds to decreasing
  $\Theta$. The collapse of the curves along both outward and
  return trips clearly support the dynamic scaling behavior given in
  Eq.~(\ref{asca3}).  }
    \label{roundtripA}
\end{figure}

\begin{figure}[!htb]
  \includegraphics[width=0.95\columnwidth]{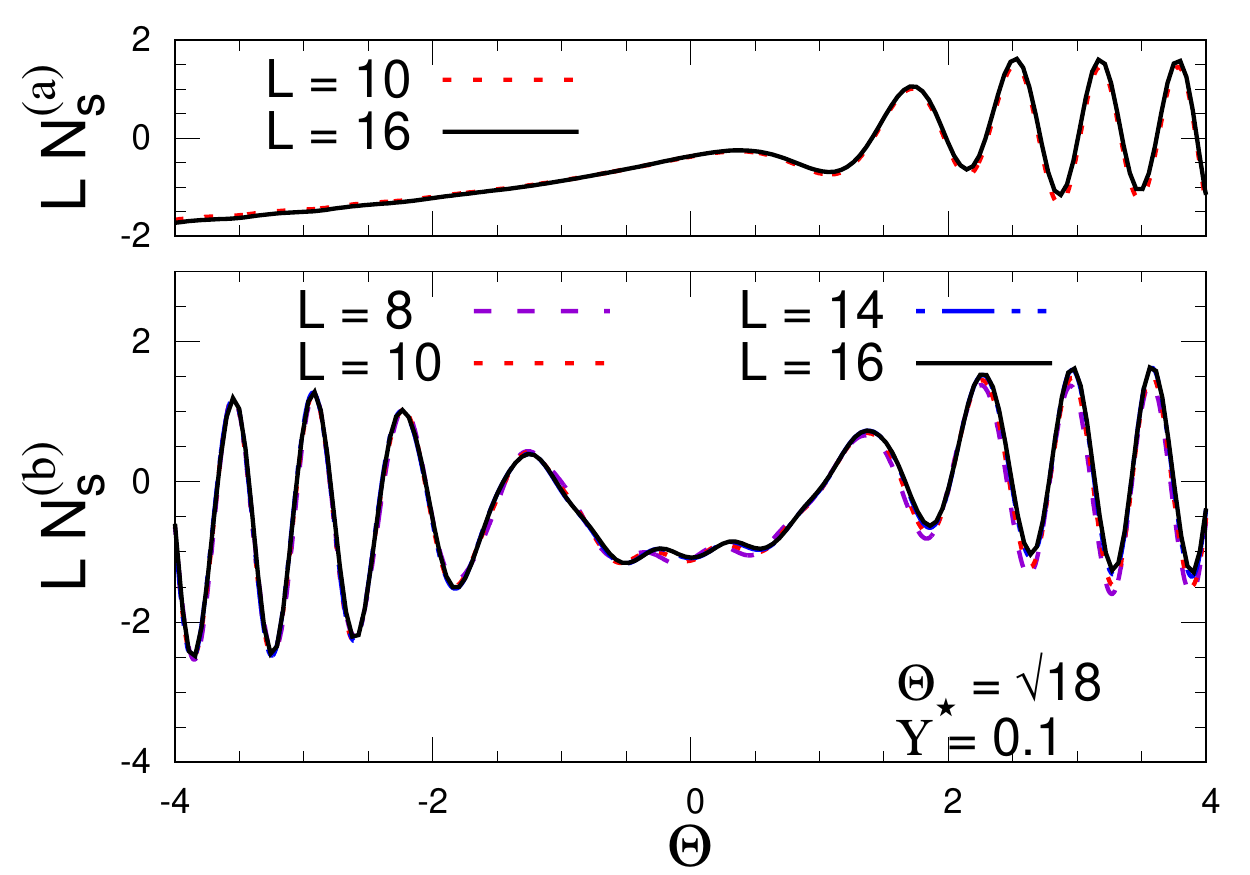}
  \includegraphics[width=0.95\columnwidth]{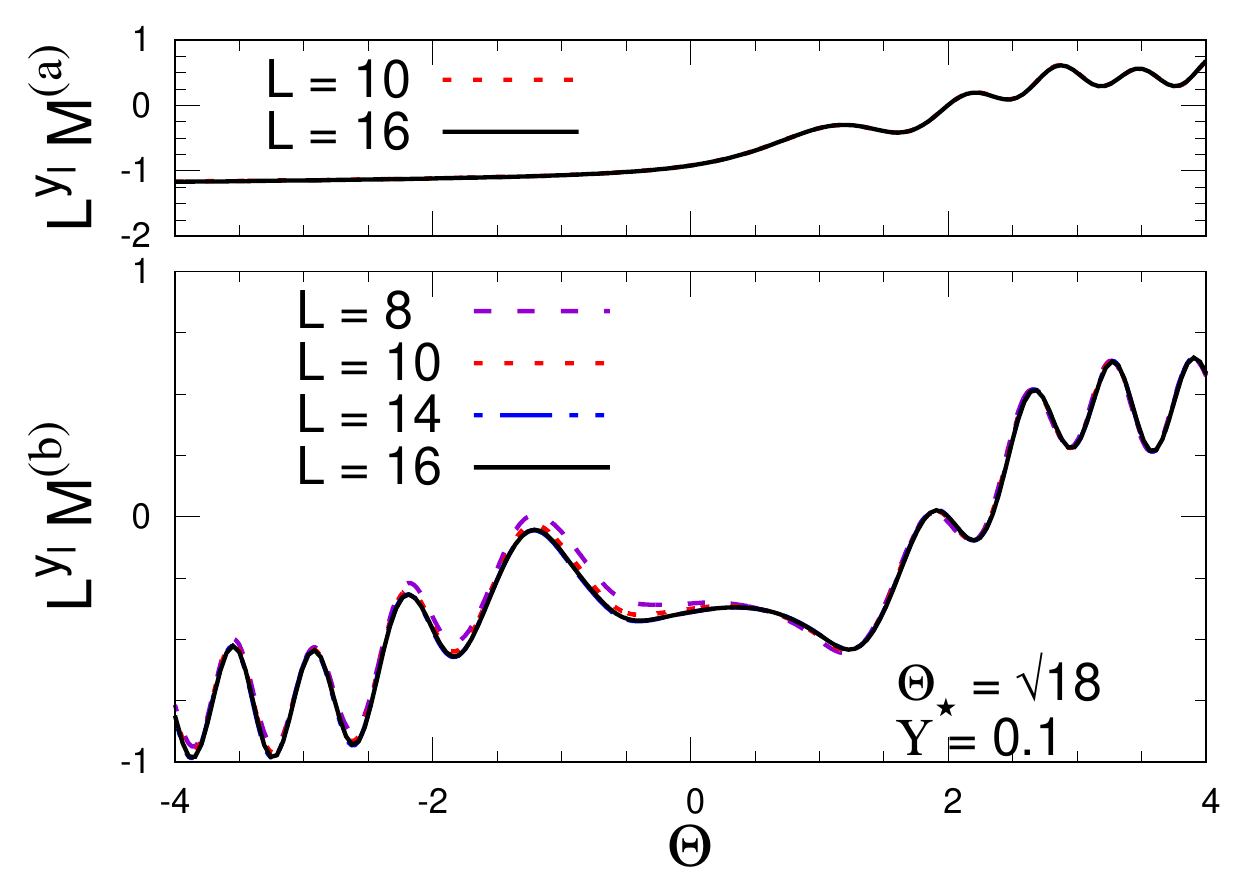}
  \caption{ Round-trip dynamic FSS of the longitudinal magnetization
    $M(t,t_s,w_\star,L)$ (bottom figures) and subtracted transverse
    magnetization $N_s(t,t_s,w_\star,L)$ (top figures),
    cf. Eq.~(\ref{subdef}), in the quantum Ising chain at fixed
    $\Upsilon=0.1$ and $\Theta_\star = \sqrt{18}$, for the outward
    (top) and return (bottom) branches of the round-trip KZ protocol,
    versus $\Theta$, for various size $L$ up to $L=16$.  The results
    clearly support the dynamic scaling behavior given in
    Eq.~(\ref{genOscart}).}
  \label{roundtripMN}
\end{figure}

\begin{figure}[!htb]
  \includegraphics[width=0.95\columnwidth]{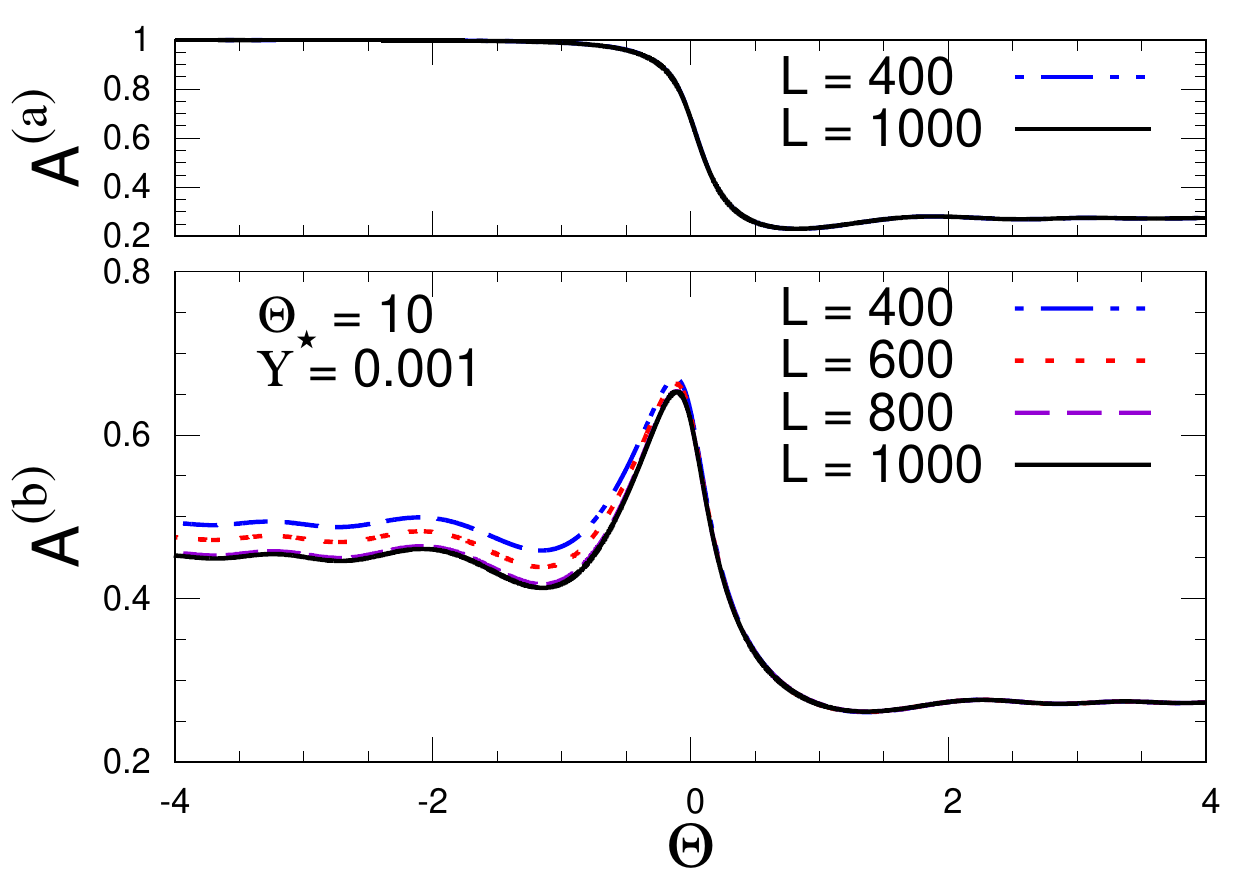}
  \caption{ Round-trip dynamic FSS within the quantum Kitaev wire for
    a finite $\Theta_\star=10$.  We show results for the adiabaticity
    function $A(t,t_s,w_\star,L)$ at fixed $\Upsilon =t_s/L^\zeta =
    0.001$ and $\Theta_\star = w_\star L^{1-\kappa}=10$, for the
    outward (top) and return (bottom) branches of the round-trip KZ
    protocol, versus $\Theta=w(t)L^{1-\kappa}$, for various size $L$
    up to $L=1000$.  The values of the exponents $y_w$, $\zeta$, and
    $\kappa$ are reported in Eq.~(\ref{kexps}).  The numerical results
    clearly support the dynamic scaling behavior given in
    Eq.~(\ref{asca3}).}
  \label{roundtripdfssA}
\end{figure}

\begin{figure}[!htb]
  \includegraphics[width=0.95\columnwidth]{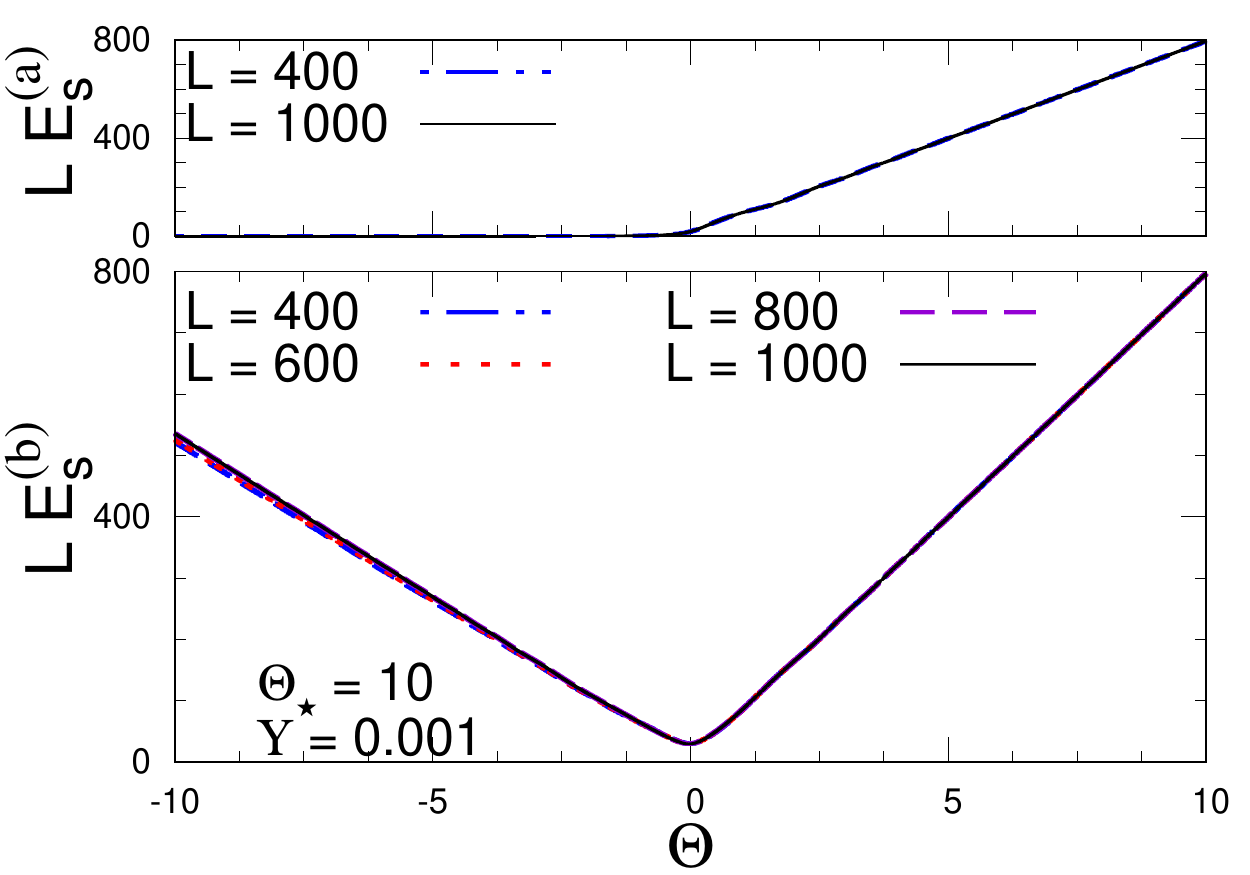}
  \caption{ Round-trip dynamic FSS within the quantum Kitaev wire for
    a finite $\Theta_\star=10$.  We show results for the surplus energy
    $E_s(t,t_s,w_\star,L)$ defined in Eq.~(\ref{etdiff}), at $\Upsilon
    =t_s/L^\zeta = 0.001$, and $\Theta_\star = w_\star
    L^{1-\kappa}=10$, for the outward (top) and return (bottom)
    branches of the round-trip KZ protocol, versus
    $\Theta=w(t)L^{1-\kappa}$, for various size $L$ up to $L=1000$.
    The results clearly support the dynamic scaling behavior given in
    Eq.~(\ref{esca3}).  }
  \label{roundtripdfssE}
\end{figure}

\begin{figure}[!htb]
  \includegraphics[width=0.95\columnwidth]{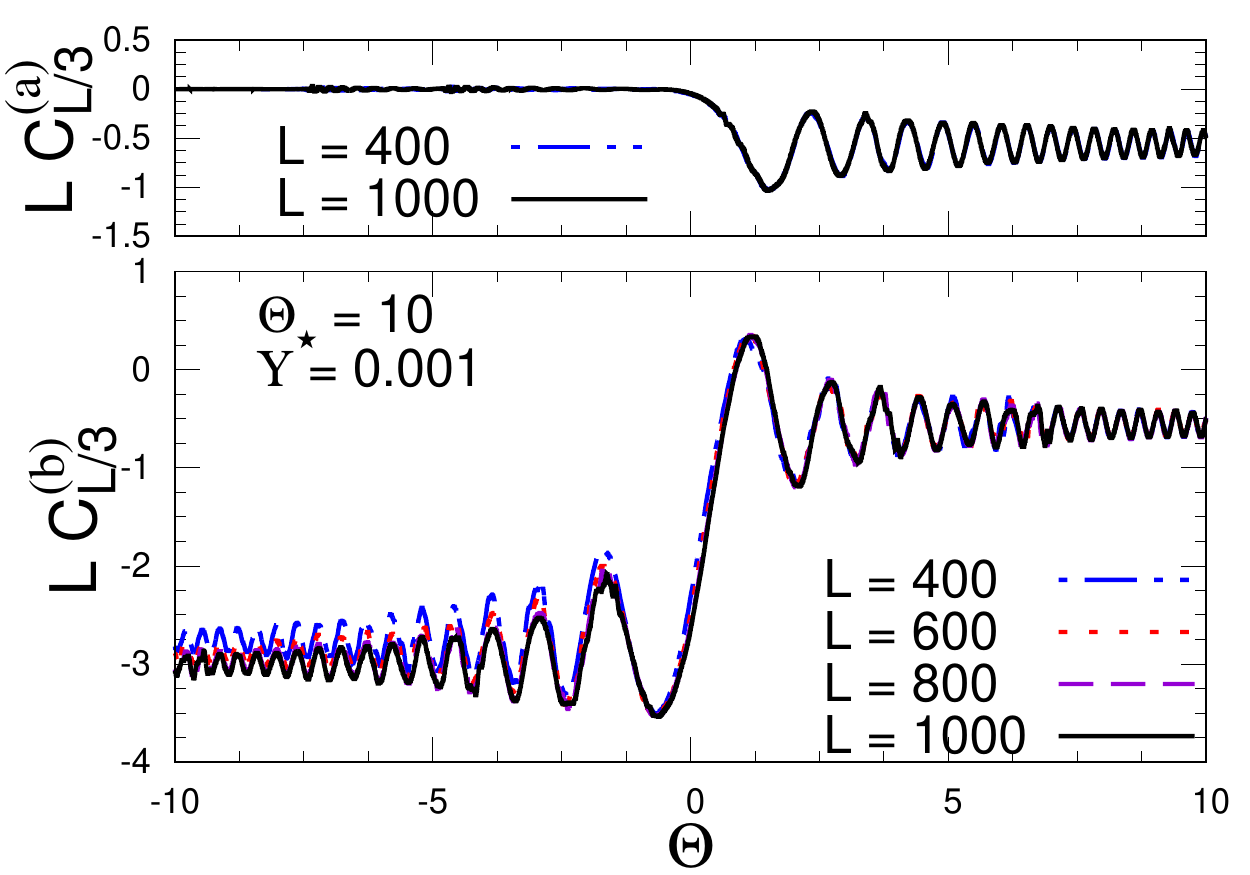}
  \caption{ Round-trip dynamic FSS within the quantum Kitaev wire for
    a finite $\Theta_\star$.  We show results for the two-point
    function $C(x,t,t_s,w_\star,L)$, cf. Eq.~(\ref{eq:corr}), at fixed
  $X=x/L=1/3$, $\Upsilon =t_s/L^\zeta = 0.001$, and $\Theta_\star =
  w_\star L^{1-\kappa}=10$, for the outward (top) and return (bottom)
  branches of the round-trip KZ protocol, versus
  $\Theta=w(t)L^{1-\kappa}$, for various size $L$ up to $L=1000$.
  }
  \label{roundtripdfssC}
\end{figure}

To begin with, we show results for round-trip KZ protocols for the
quantum Ising chain, cf. Eq.~(\ref{isichoice}), when keeping
$\Theta_\star$ finite, see Figs.~\ref{roundtripA} and
\ref{roundtripMN}, respectively for the adiabaticity function and the
longitudinal and transverse magnetizations.  Analogous results are
obtained for other values of $\Upsilon$ and $\Theta_\star$.  Analogous
results are also obtained for the quantum Kitaev wire,
cf. Eq.~(\ref{kitchoice}), see for example the results shown in
Figs.~\ref{roundtripdfssA}, \ref{roundtripdfssE}, and
\ref{roundtripdfssC}, respectively for the adiabaticity function, the
surplus energy $E_s$ defined in Eq.~(\ref{etdiff}), and the two point
function defined in Eq.~(\ref{eq:corr}). These results fully support
the dynamic FSS put forward in Sec.~\ref{qfssKZroundtrip} when keeping
$\Theta_\star$ finite.

\subsubsection{The limit $\Theta_\star\to\infty$}
  \label{scafinthetastarinf}

\begin{figure}[!htb]
  \includegraphics[width=0.95\columnwidth]{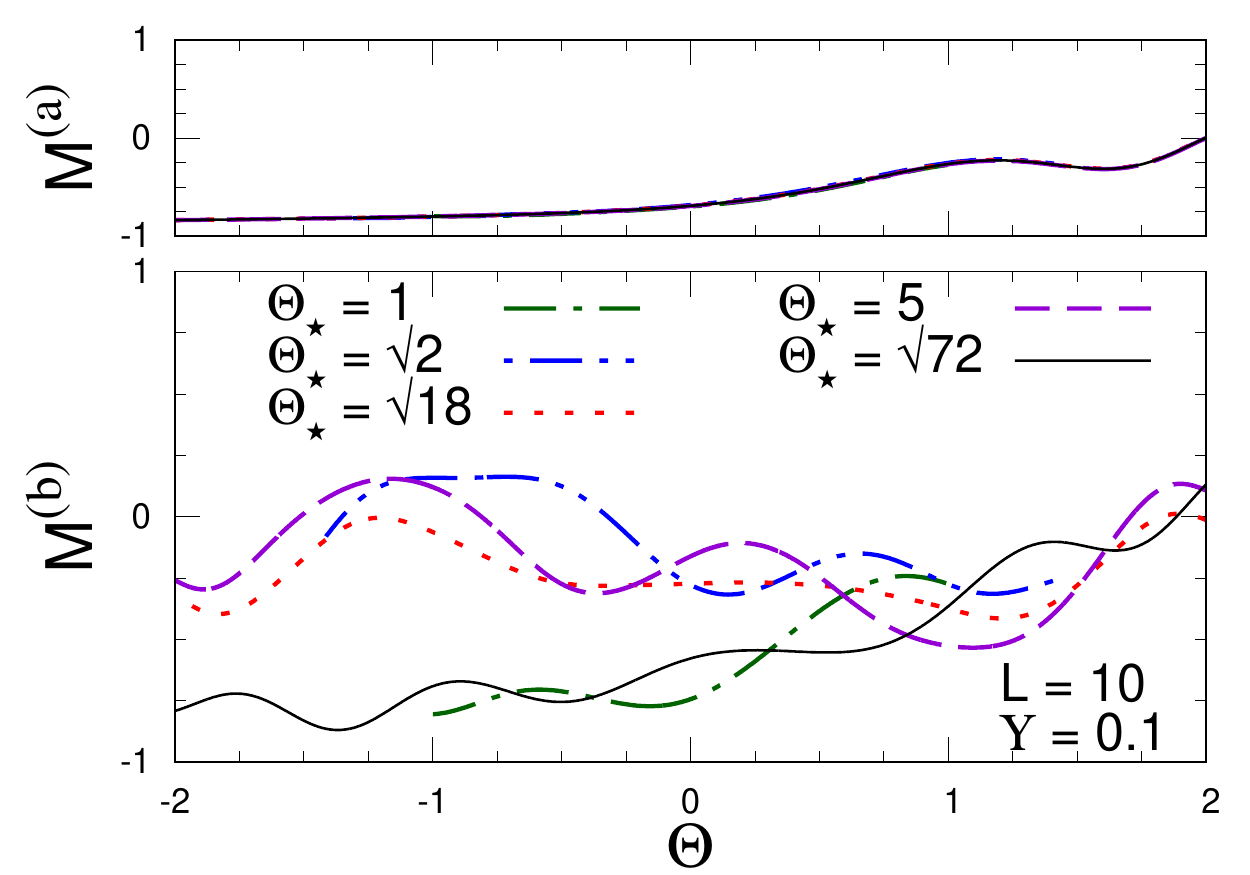}
 \caption{ Behavior of $M(t,t_s,w_\star,L)$ for fixed $L = 10$,
   $\Upsilon = 0.1$ for the one way trip (top) and for the return trip
   (bottom), versus $\Theta$, for various $\Theta_\star$ up to
   $\Theta_\star = 6\sqrt{2}$.  We note that along the outward path
   the large-$\Theta_\star$ convergence of the curves is rapid (it is
   essentially related to the convergence with respect to
   $\Theta_i=-\Theta_\star$ of the one-way protocol); on the other
   hand the curves do not appear to approach a large-$\Theta_\star$
   limit along the return path.}
  \label{diffThetaStar}
\end{figure}

We now discuss the large-$\Theta_\star$ limit, and also the related
case in which we keep $w_\star>0$ fixed in the round-trip protocols.
This limit turns out to be quite problematic in quantum round-trip KZ
protocols.

Some hints at the absence of a well defined large-$\Theta_\star$ limit
of the dynamic scaling behavior are shown by the plots of
Fig.~\ref{diffThetaStar} reporting the longitudinal magnetization of a
quantum Ising system of size $L=10$ for various $\Theta_\star$.  When
increasing $\Theta_\star$, the curves along the outward way show a
good convergence, while no apparent convergence is observed along the
return paths.

\begin{figure}[!htb]
  \includegraphics[width=0.95\columnwidth]{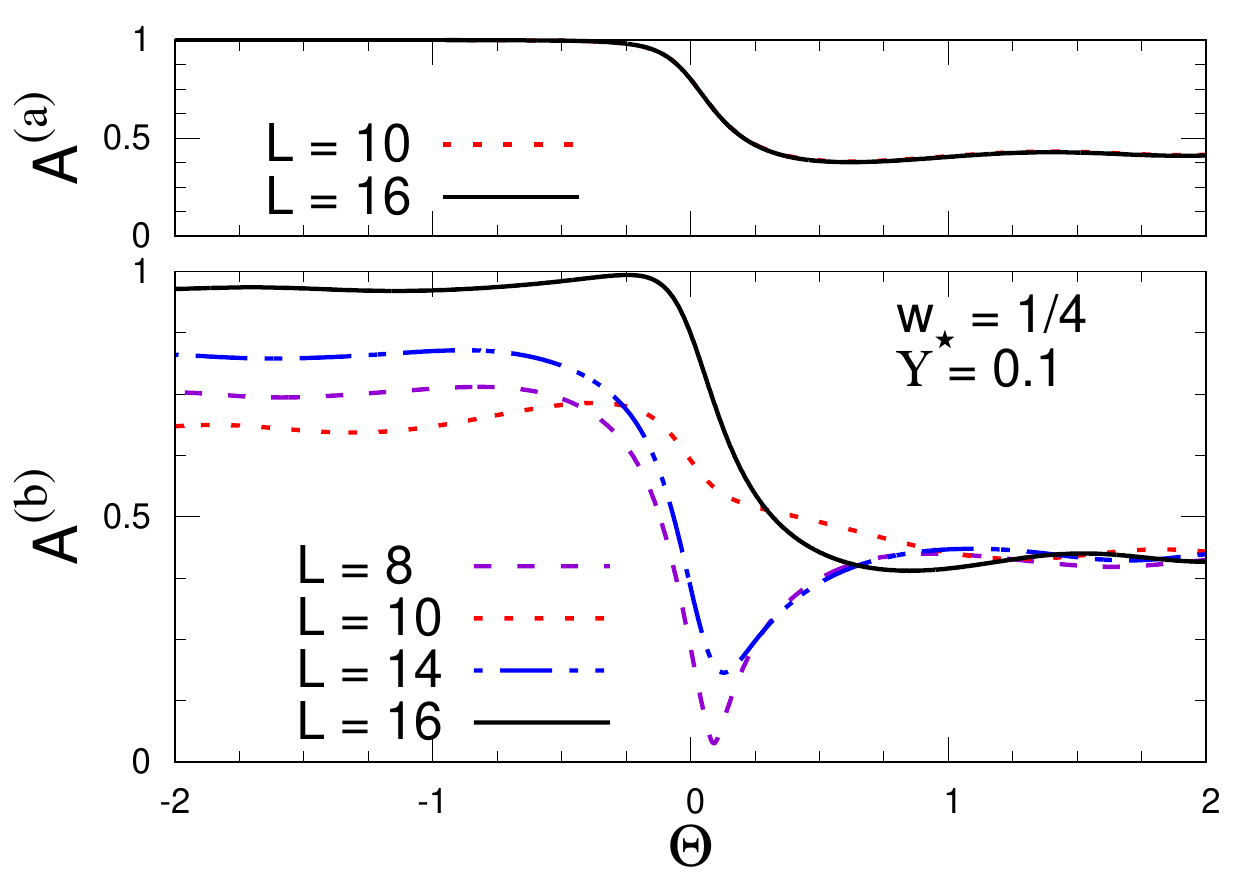}
  \caption{ The adiabaticity function $A(t,t_s,w_\star,L)$ of quantum
    Ising chains along round-trip protocols, for fixed $\Upsilon =
    0.1$ and $w_\star = 1/4$, for the outward (top) and return
    (bottom) branches of the round-trip protocol, versus $\Theta$, for
    various size $L$ up to $L=16$.  We note that along the outward
    path the large-$t_s$ convergence is rapid, unlike the return way
    where no evidence of convergence is observed.}
  \label{roundtripAW}
\end{figure}

\begin{figure}[!htb]
  \includegraphics[width=0.95\columnwidth]{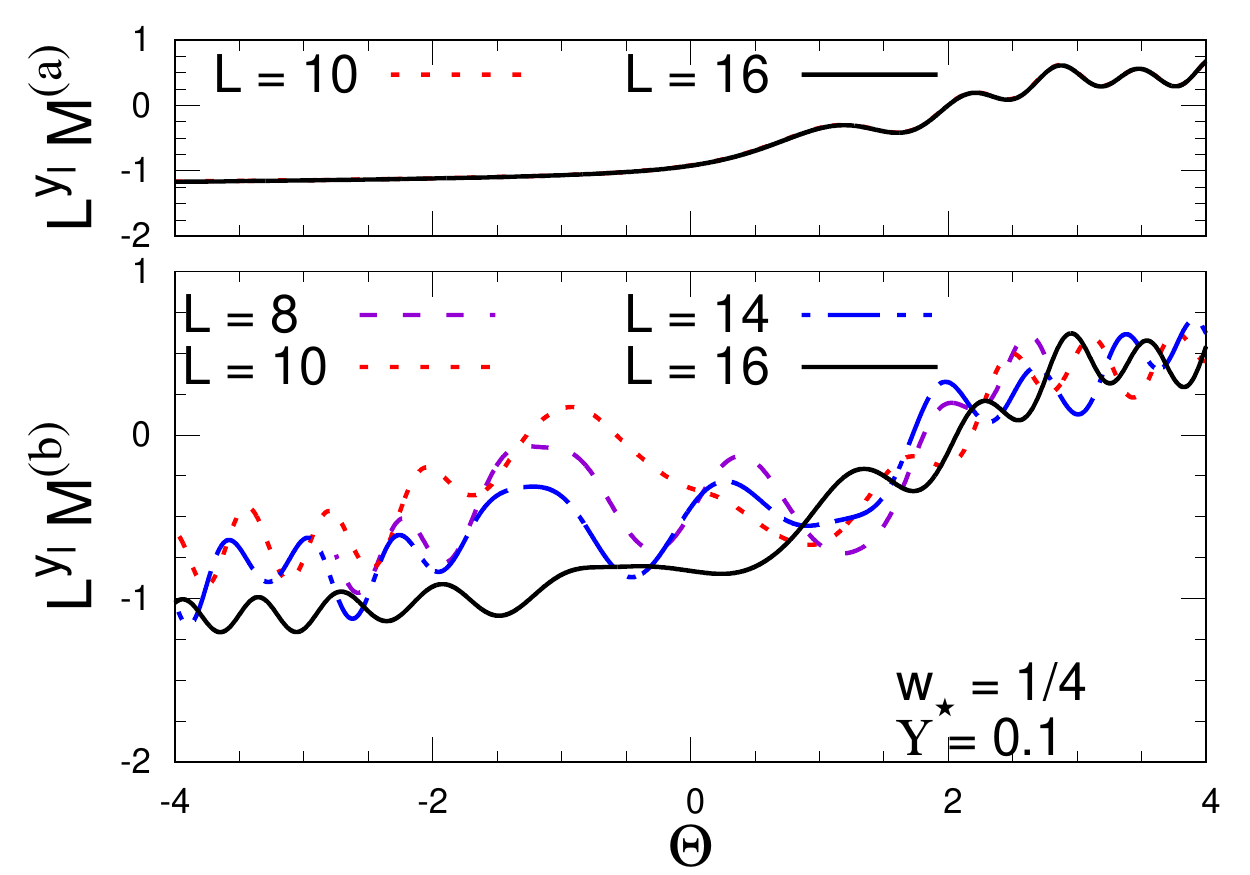}
 \caption{ The longitudinal magnetization $M(t,t_s,w_\star,L)$ along
   the round-trip protocol, for fixed $w_\star = 1/4$, $\Upsilon =
   0.1$ for the outward (top) and return (bottom) branches of the
   round-trip protocol, versus $\Theta$, for various size $L$ up to
   $L=16$.  }
  \label{roundtripMxW}
\end{figure}

\begin{figure}[!htb]
  \includegraphics[width=0.95\columnwidth]{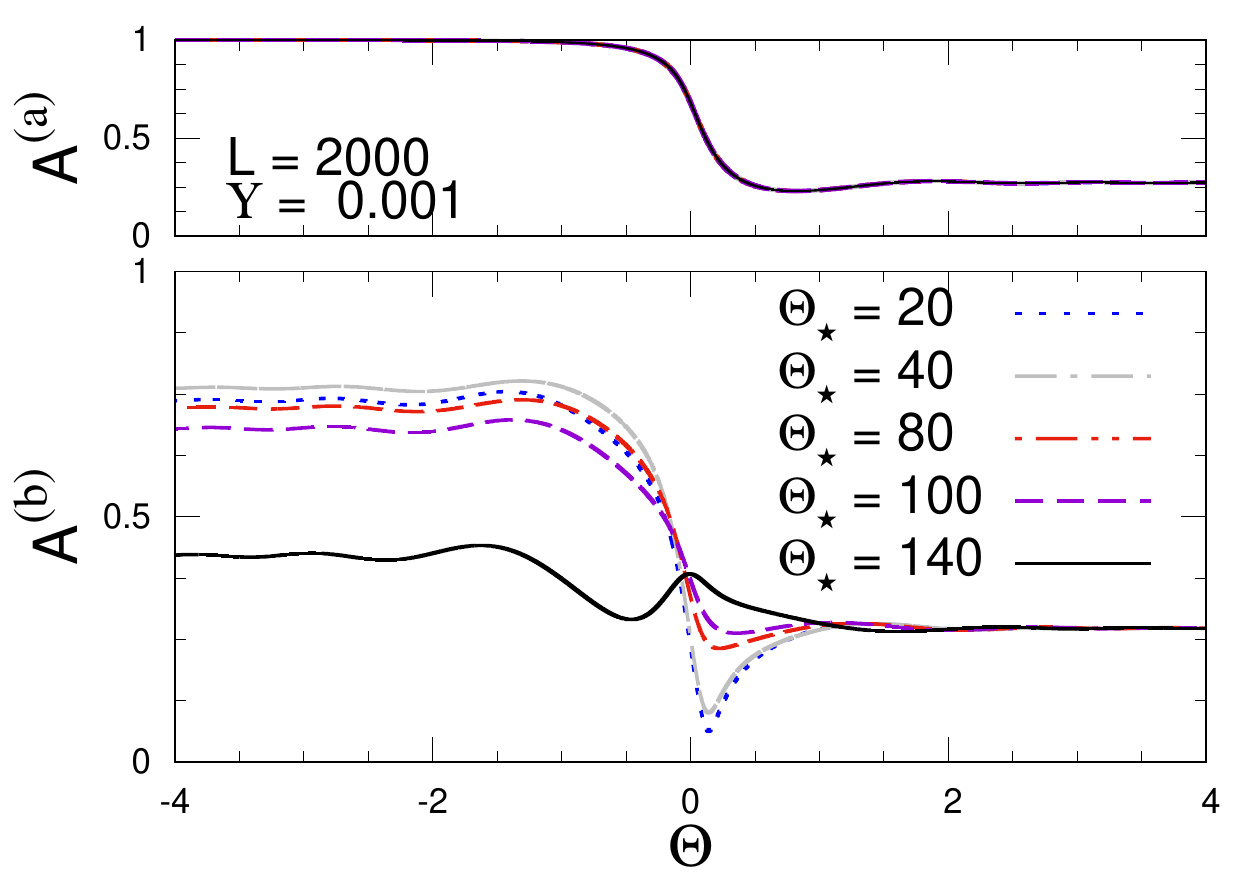}
 \caption{The adiabaticity function $A(t,t_s,w_\star,L)$ for the
   quantum Kitaev wire at $L=2000$ and $\Upsilon = 0.001$ for the
   outward (top) and return (bottom) branches of the round-trip KZ
   protocol, versus $\Theta$, for various $\Theta_\star$ up to
   $\Theta_\star = 140$.  We note that along the outward path the
   convergence is large-$\Theta_\star$ convergence is rapid (it is
   essentially related to the convergence with respect to
   $\Theta_i=-\Theta_\star$ in the one-way KZ protocol), along the
   return path the curves do not appear to approach a
   large-$\Theta_\star$ limit.}
  \label{diffThetaStarA}
\end{figure}

When we keep $w_\star$ fixed and finite, our computations do not show
evidence of convergence along the return trajectories in the
large-$t_s$ and large-$L$ dynamic scaling limit.  This is shown by the
curves of the adiabaticity function along the return branch of the
round-trip protocol, see Fig.\ref{roundtripAW}, for $w_\star=1/4$ and
$\Upsilon=0.1$.  While convergence is clearly observed along the
outward path, as expected because the one-way KZ protocol showed a
well defined limit in the large-$|\Theta_i|$ limit, the return path
does not show a stable convergence pattern. The same behavior is also
shown by the longitudinal and transverse magnetizations $M$ and $N$,
see for example Fig.~\ref{roundtripMxW}.  Analogous results are also
obtained for the quantum Kitaev wire, see Fig.~\ref{diffThetaStarA},
where we report results for the adiabaticity function at
$\Upsilon=0.001$ and various large values of $\Theta_\star$, for a
large lattice size $L=2000$.

To interpret, and understand, the above instability emerging in
quantum systems subject round-trip KZ protocols, it is useful to make
a comparison with the dynamic behavior of two-level models subject to
analogous round-trip protocols, discussed in
App.~\ref{LZlike}. Analogously to the Landau-Zener-St\"uckelberg
problem~\cite{LZeff,SAN-10}, we consider a time-dependent two-level
Hamiltonian
\begin{equation}
  H_{2\ell}(t) = - \beta(t) \sigma^{(3)}
  + {\Delta\over 2} \sigma^{(1)}\,,
\label{hrdef2}
\end{equation}
where $\Delta$ is a constant, 
\begin{eqnarray}
  \beta(t) = {{\cal T}(t)\over t_s}\quad
       {\rm for}\;\; t_i=-t_\star \le t \le 3t_\star\,,
\label{betadef}
\end{eqnarray}
and $ {\cal T}(t) = t_\star - |t-t_\star|$ is the triangular
function. The quantities $\tau={\cal T}(t)/\sqrt{t_s}$ and
$\tau_\star=t_\star/\sqrt{t_s}$ play the same role of the scaling
variables $\Theta$ and $\Theta_\star$ describing the round-trip KZ
protocols in quantum many-body systems. The corresponding
Schr\"odinger equation can be analytically solved in terms 
of parabolic cylinder functions $D_\nu(x)$~\cite{VG-96}, see
App.~\ref{LZlike}.

The resulting behavior of the expectation values of $\sigma^{(3)}$ and
the adiabatic function show that the large-$\tau_\star$ limit is
problematic, being characterized by large $O(1)$ oscillations with
frequencies increasing proportionally to $\tau_\star$, roughly. See
App.~\ref{LZlike} for details.  They turn out to be related to the
rapid changes of the relative phase between the relevant states of the
two-level system at the extreme values $\tau=\tau_\star$ when
$\tau_\star$ becomes large, increasing as $\tau_\star^2$. Since the
quantum evolution along the return trajectory turns out to be very
dependent on such phase, it becomes extremely sensitive to the value
of $\tau_\star$, showing analogous oscillations. As a consequence, the
value of all observables along the return trajectory, from
$\tau=\tau_\star$ down to the return point $\tau = - \tau_\star$, do
not show a well defined limit for $\tau_\star\to\infty$.  The size of
the oscillations depend on the value of the scaling variable $\upsilon
= t_s \Delta^2$, which plays the same role of $\Upsilon$ in the quantum
many-body systems, and tend to be suppressed in the adiabatic limit
$\upsilon\to\infty$.

\begin{figure}[!htb]
  \includegraphics[width=0.95\columnwidth]{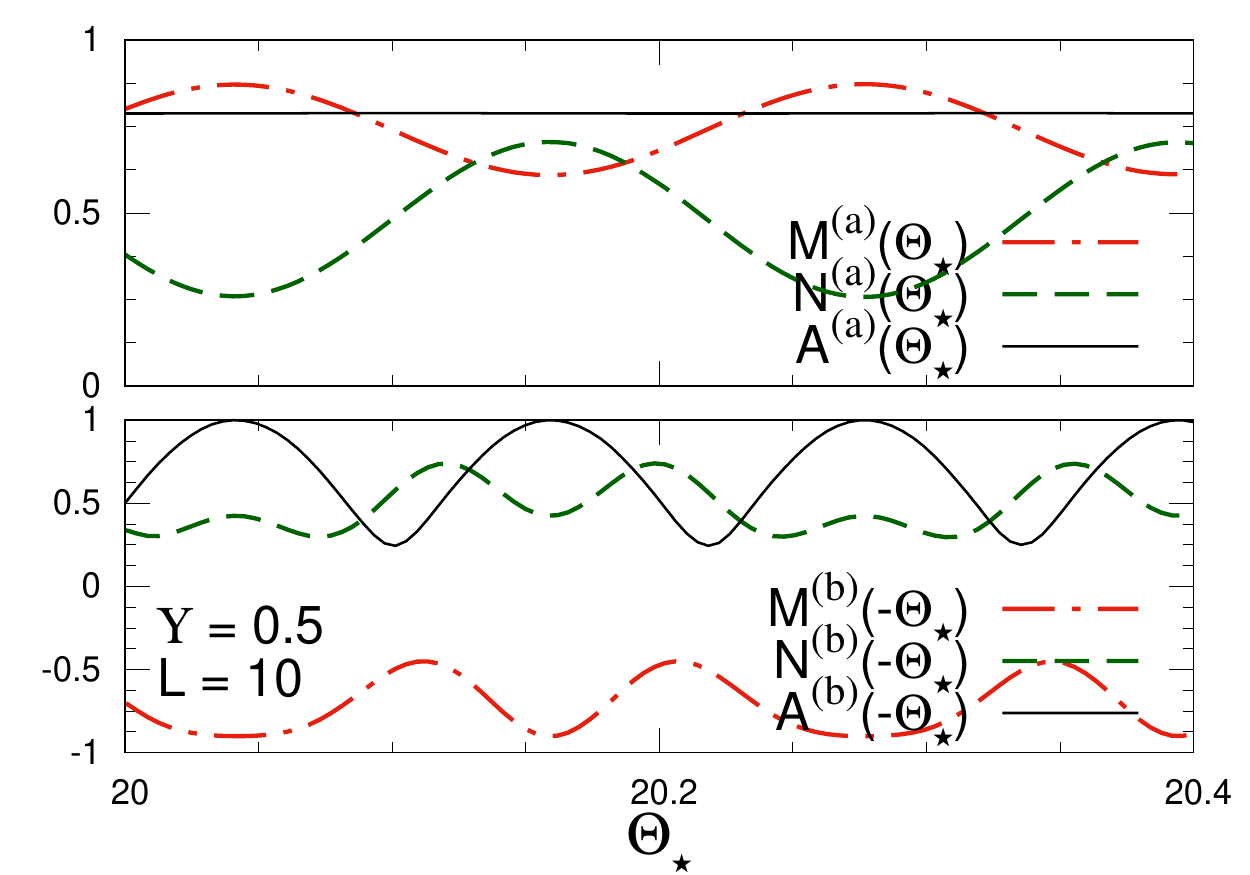}
  \includegraphics[width=0.95\columnwidth]{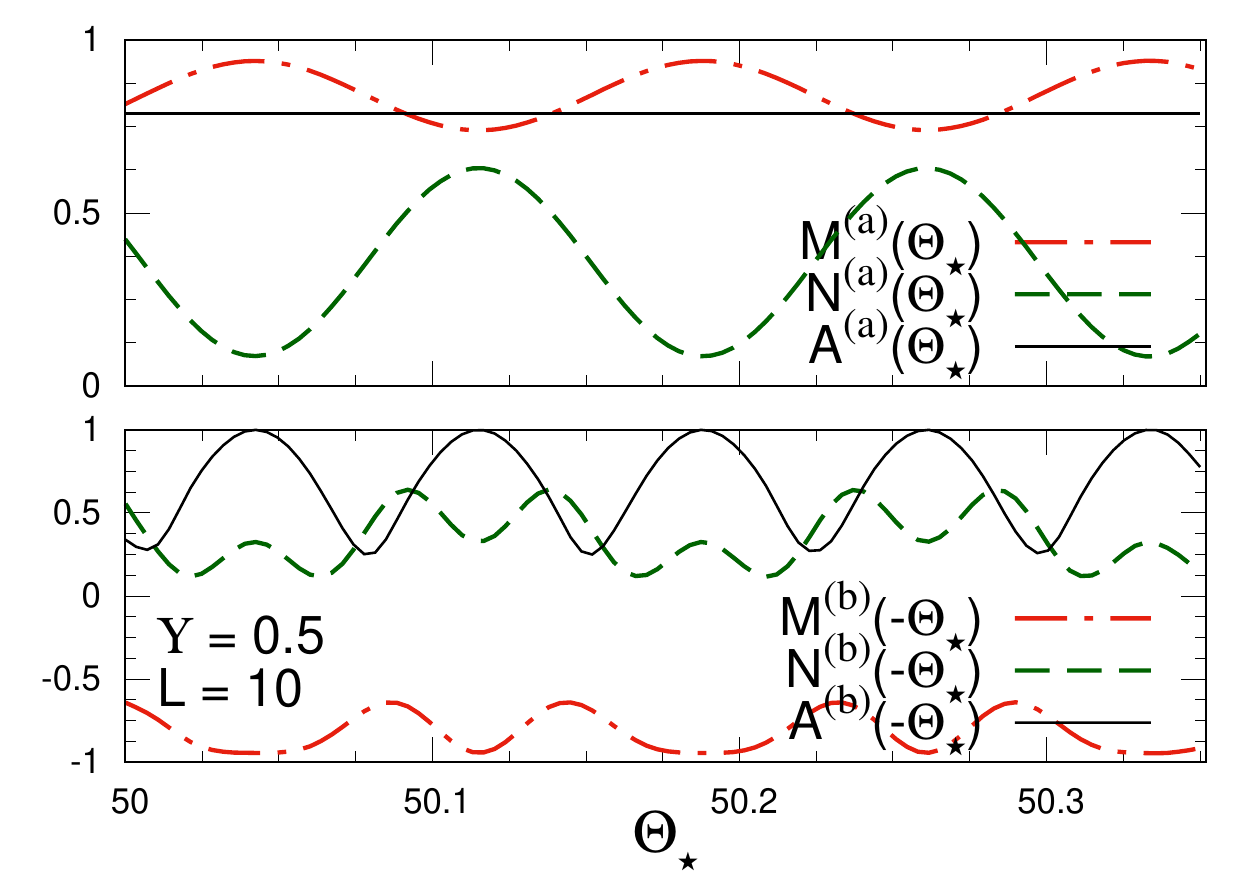}
 \caption{ Results for $M,\, N$ and $A$ for fixed $L = 10$, $\Upsilon
   = 0.5$ versus $\Theta_\star$, close to $\Theta_\star = 20$ (top
   figure) and $\Theta_\star = 50$ (bottom figure). In each figure,
   the top plot the values of $M^{(a)}$, $N^{(a)}$ and $A^{(a)}$ at
   the end of the outward branch, corresponding to
   $\Theta=\Theta_\star$, while the bottom plot shows the values of
   $M^{(b)}$, $N^{(b)}$ and $A^{(b)}$ at the end of the return branch,
   corresponding to $\Theta=-\Theta_\star$. The comparison of the top
   and bottom figures show that the oscillations tend to become more
   frequent with increasing $\Theta_\star$ (note that the interval of
   the abscissa is different).  Analogous results are obtained for
   other values of $\Upsilon$.}
  \label{roundtripDiffTheta}
\end{figure}

\begin{figure}[!htb]
   \includegraphics[width=0.95\columnwidth]{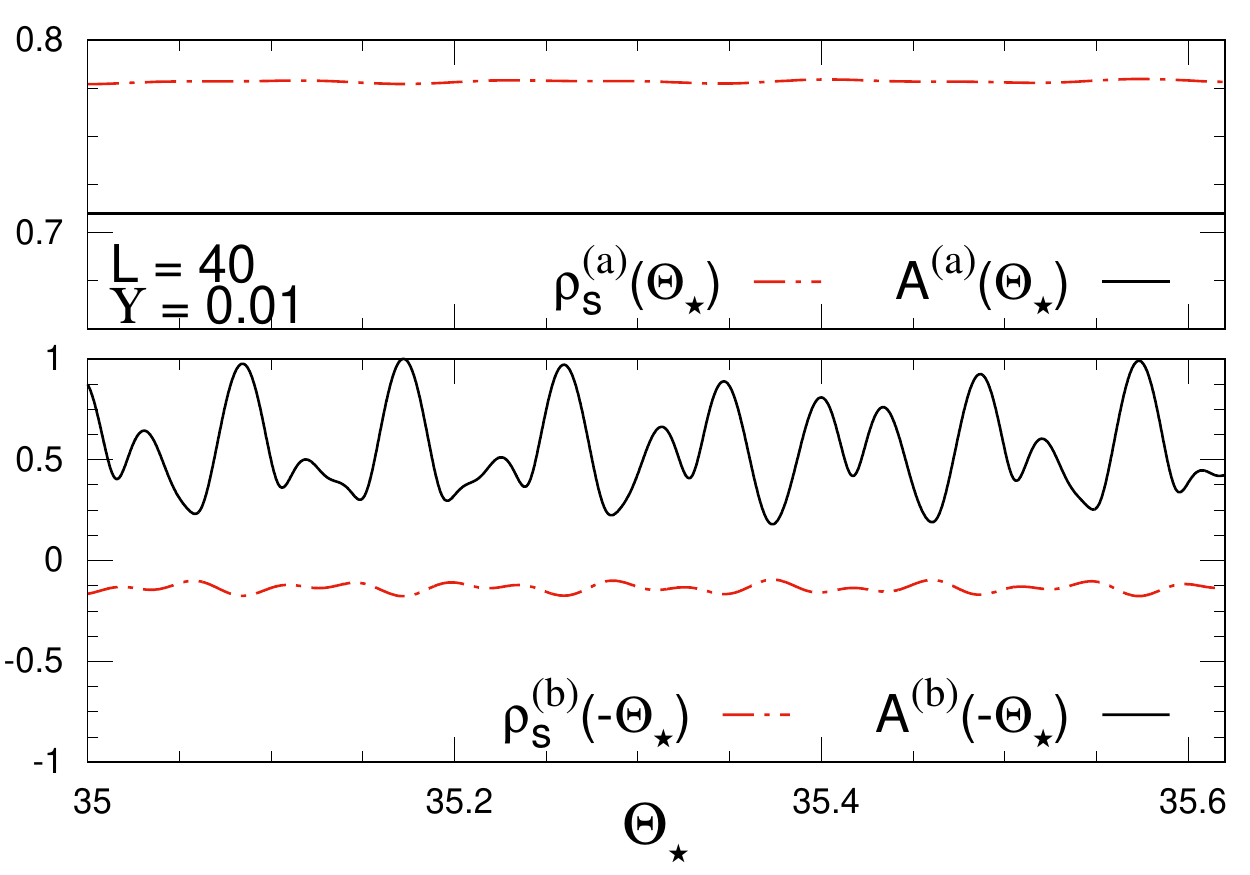}
  \includegraphics[width=0.95\columnwidth]{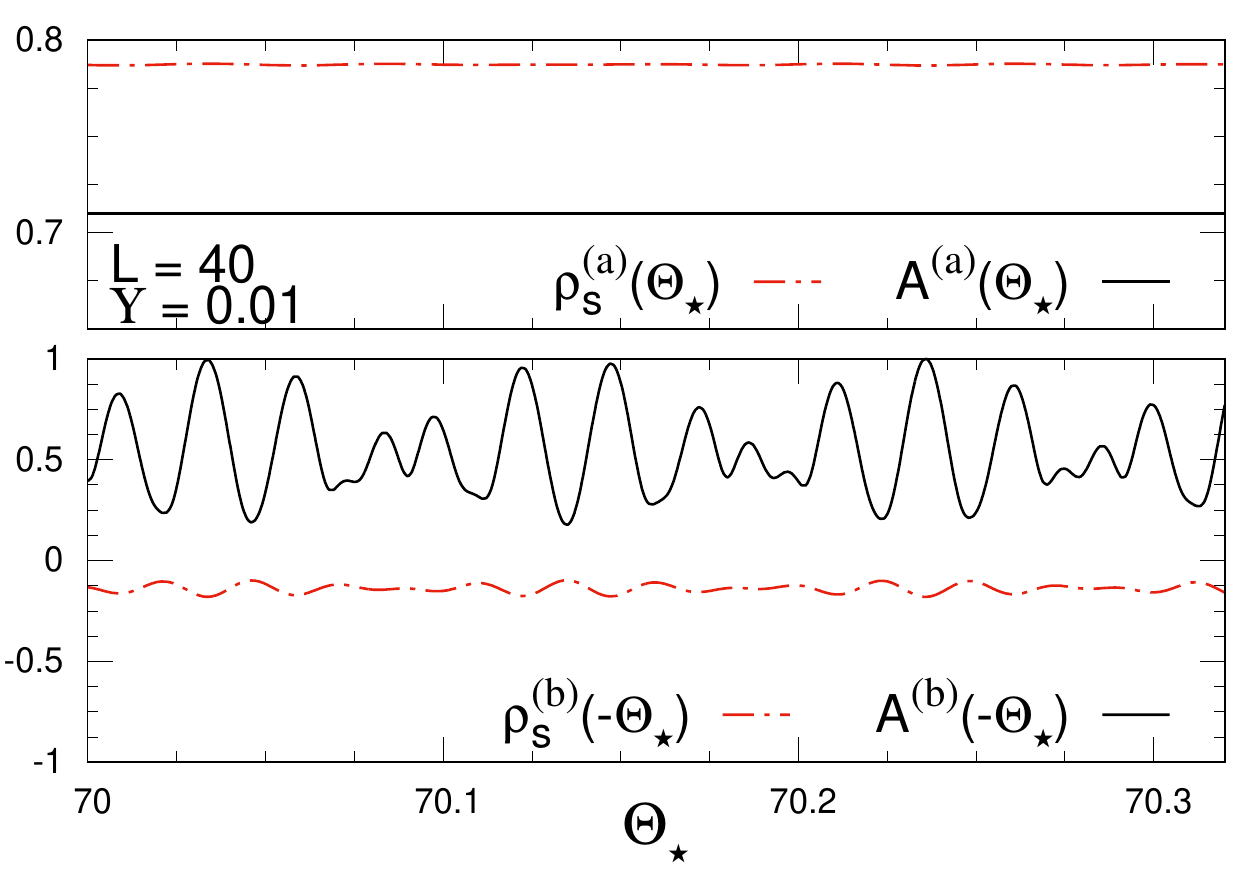}
  \caption{ Behavior of the subtracted particle density $\rho_s$,
    cf. Eq.~(\ref{rhot}), and the adiabaticity function $A$ for the
    Kitaev wire, for fixed $L = 40$, $\Upsilon = 0.01$ versus
    $\Theta_\star$, close to $\Theta_\star = 70$ (bottom figure) and
    $\Theta_\star=35$ (top figure).  In each figure, the top plot the
    values of $\rho_s^{(a)}$ and $A^{(a)}$ at the end of the outward
    branch, corresponding to $\Theta=\Theta_\star$, while the bottom
    plot shows the values of $\rho_s^{(b)}$ and $A^{(b)}$ at the end
    of the return branch, corresponding to $\Theta=-\Theta_\star$.
    Again, the comparison of the top and bottom figures show that the
    oscillations tend to become more frequent with increasing
    $\Theta_\star$.  }
  \label{roundtripDiffThetaAR}
\end{figure}

We observe a similar behavior in the quantum many-body systems.  This
scenario is demonstrated by the results shown in
Fig.~\ref{roundtripDiffTheta}, where we report the values of
$A^{(a)}$, $M^{(a)}$, and $N^{(a)}$ at end of the outward branch and
$A^{(b)}$, $M^{(b)}$, and $N^{(b)}$ at the end of the return branch,
for KZ protocols with different $\Theta_\star$, to check their
large-$\Theta_\star$ convergence, for some interval of values of
$\Theta_\star$ around large values of $\Theta_\star$ and fixed $L=10$.
Similarly to the results obtained for two-level model, the observables
at the end of the outward branch oscillate, with a frequency that
becomes larger and larger with increasing $\Theta_\star$, and the
oscillations observed after the whole cycle are strongly correlated to
those at the end of the first branch, doubling the frequency.
Analogous results are obtained for other values of $\Upsilon$.  We
also note that the oscillations tend to be suppressed in the adiabatic
$\Upsilon\to\infty$ limit. We believe that this extreme sensitivity to
$\Theta_\star$ makes also problematic the large-$L$ limit after the
limit $\Theta_\star\to\infty$ shown by the numerical data.  Similar
results are also obtained for the quantum Kitaev wire, see
Fig.~\ref{roundtripDiffThetaAR} where we show results for the
adiabaticity function and the particle density.  In this case the
values at the end of the outward way appear quite stable, but the
return way is again characterized by large (less regular) oscillations
with larger and larger frequencies with increasing $\Theta_\star$.

The above results for both the quantum Ising rings and Kitaev wires
strongly suggest that in quantum many-body systems the
large-$\Theta_\star$ limit of the dynamic KZ scaling does not exist
along the return trajectories, and, as a consequence, no dynamic
scaling is observed along the return trip when $w_f>0$ is kept fixed
and finite in the round-trip KZ protocols.  In this respect, there are
notable similarities with the behavior of two-level model
(\ref{hrdef2}) subject to round-trip protocols. We believe that this
issue deserves further investigation, for example addressing the
possibility of obtaining well defined scaling behavior after some
average procedures over the oscillations induced by large values of
$\Theta_\star$, to obtain a well defined large-$\Theta_\star$ limit.

However, we stress that the dynamic scaling behavior is nicely
observed when keeping $\Theta_\star$ fixed, even along the return
trajectory. This may be related to fact that, when keeping
$\Theta_\star$ fixed, the time scaling variable $\Theta$ remains
finite, therefore the time variable is always rescaled consistently
with the time scale of the equilibrium quantum transition, provided by
the inverse gap at the transition, i.e. $\Delta \sim L^{-z}$ at the
critical point, or $\Delta \sim \lambda^{-z}$ in the thermodynamic
limit, where $\lambda$ is the KZ length scale (\ref{xit}).  \rev{As a
  consequence, the interval of values of $w(t)$ remains limited within
  a small interval around the transition, which becomes smaller and
  smaller in the large-size limit, as $|w|\lesssim L^{-y_w}$, and the
  relative quantum phases behave consistently with the scaling laws.}

\section{Conclusions}
\label{conclu}

We have studied the out-of-equilibrium behavior of many-body systems
when their time-dependent Hamiltonian parameters slowly cross phase
transition points, where systems at equilibrium develop critical modes
with long-range correlations. Earlier studies have already shown the
emergence of several interesting out-of-equilibrium phenomena, such as
hysteresis, coarsening, KZ defect production, aging, etc..  In this
paper we present an exploratory study of out-of-equilibrium behaviors
arising from round-trip protocols across classical and quantum phase
transitions.

We consider classical and quantum many-body systems described by the
general Hamiltonian (\ref{hlamt}), and study the out-of-equilibrium
evolution arising from cyclic variations of the parameter $w$ driving
the equilibrium transition, entailing multiple crossings of the
transition point $w_c=0$.  More precisely, we consider round-trip
protocols where the many-body system starts from equilibrium
conditions at a given value $w_i<0$, and the out-of-equilibrium
dynamics is driven by changing the parameter $w(t)$ in
Eq.~(\ref{hlamt}) linearly in time up to $w_f>0$, thus crossing the
critical point $w_c=0$, and then by changing it back to the original
value $w_i<0$, again linearly in time, which implies a further
crossing of the transition point.  The round-trip protocol is
characterized by a unique large time scale $t_s$, see
Sec.~\ref{rtpro}. We limit our study to the cases where the transition
point separates phases with short-range correlations.  The more
complicated situations of classical and quantum transitions between
disordered and ordered phases with ungapped excitations is left to
future works.

We address these issues within many-body models undergoing classical
and quantum transitions, exploiting a unified RG framework, where
general dynamic scaling laws are derived in the large-$t_s$ and
large-$L$ limits, see Secs.~\ref{fssKZoneway} and \ref{roundtrip}. In
particular, we extend the RG framework already developed for standard
one-way KZ protocols, see e.g. Refs.~\cite{CEGS-12,RV-21}.

As paradigmatic models, we consider classical and quantum systems that
undergo classical and quantum transitions belonging to the 2D Ising
universality class: (i) Classical 2D Ising models undergoing a
finite-temperature transition, supplemented with a purely relaxational
dynamics driven by an external time-dependent magnetic field; (ii)
Quantum 1D Ising models with an external time-dependent longitudinal
field; (iii) Quantum 1D Kitaev fermionic wires with a time-dependent
chemical potential. In all cases we analyze the out-of-equilibrium
behavior arising from round-trip linear variations of the Hamiltonian
parameters, crossing twice the transition point.  We report various
numerical analyses of one-way and round-trip KZ protocols within the
above models, see Sec.~\ref{numresrotrip}.  They generally support the
dynamic FSS behaviors in the large time-scale ($t_s$) limit, put
forward within the RG frameworks.

However, while the general dynamic scaling picture may appear similar,
there are also important differences between classical and quantum
systems.  Indeed, the analogy of the scaling behaviors for one-way KZ
protocols at classical and quantum transitions is only partially
extended to round-trip KZ protocols. Substantial differences emerge,
in particular when the extreme value $w_f>0$ of the outward variation
of $w(t)$ is kept fixed and finite in the large-$t_s$ limit. On the
one hand, classical systems show a well-defined dynamic scaling limit,
developing scaling hysteresis-like scenarios, essentially because the
purely relaxational stochastic dynamics leads eventually to
thermalization at fixed model parameters.  On the other hand, in
quantum systems the observation of scaling behavior along the return
way turns out to be more problematic, due to the persistence of
rapidly oscillating relative phases between the relevant quantum
states. They make the return way extremely sensitive to the parameters
of the protocol, such as the extreme value $w_f$ and the size $L$ of
the system.  This is essentially related to the quantum nature of the
dynamics. Indeed there are some notable similarities with the behavior
of quantum two-level models subject to round-trip protocols, analogous
the well-known Landau-Zener-St\"uckelberg
problem~\cite{LZeff,SAN-10,ISN-22}, see App.~\ref{LZlike}.  Even in
the simple two-level quantum model some features of the behavior along
the return way turn out not to be smooth. Indeed, they develop ample
oscillations with larger and larger frequencies when increasing the
interval of the round-trip variation of the parameters, showing
chaotic-like behaviors due to the extreme sensitivity to the protocol
parameters. We believe that this issue calls for further
investigation, to achieve a better understanding of these phenomena.

The emerging dynamic scaling scenario put forward for round-trip KZ
protocols across critical points is expected to hold for generic
classical and quantum transitions separating phases with short-range
correlations, in any spatial dimension. Further investigations are
called for round-trip protocols between disordered and ordered phases,
when the ordered phase has gapless excitations.  Round-trip KZ
protocols in these systems may show further interesting features.

In this paper we have focused on continuous transitions.  Analogous
issues may be investigated at first-order classical and quantum
transitions, where dynamic scaling behaviors emerge as well, although
they turn out to significantly depend on the nature of the boundary
conditions (see e.g.
Refs~\cite{Binder-87,RV-21,PRV-18,CNPV-14,PV-17,PV-17-a,PRV-20,DOV-09}
for studies at classical and quantum transitions). Further interesting
issues may concern the effects of dissipation due to the interaction
with an environment, which are inevitable in realistic quantum
devises, and can induce some further relevant effects in the dynamics
of systems subject to round-trip KZ protocols, see
e.g. Refs.~\cite{RV-21,FFO-07,PSAFS-08,PASFS-09,YMZ-14,NVC-15,
  KMSFR-17,SVPKD-17,ABRS-18,NRV-19-dis,RV-19-dis,RV-20}.

We remark that round-trip protocols at classical and quantum
transitions should also be of experimental relevance. Indeed they
represent a straightforward extension of the one-way KZ protocols,
which have been already investigated experimentally at both thermal
and quantum transitions, as already mentioned in the introduction.

Our results may turn out to be particularly relevant for quantum
simulations and quantum computing, where important experimental
advances have been achieved recently, see
e.g. Refs.~\cite{CZ-12,BDN-12,BR-12,AW-12,HTK-12,GAN-14}.  In
particular, our results imply some limitations to the observation of a
round-trip dynamics across quantum transitions in many-body models.
We also note that the dynamic scaling behavior put forward in this
work have been observed in numerical simulations of systems of
moderately large size.  This suggests the possibility that the dynamic
scaling scenario may be accessed by experiments with quantum
simulators in laboratories, e.g., by means of trapped
ions~\cite{Islam-etal-11, Debnath-etal-16}, ultracold
atoms~\cite{Simon-etal-11, Labuhn-etal-16}, or superconducting
qubits~\cite{Salathe-etal-15, Cervera-18}.

\acknowledgements

We thank Alessio Franchi, Davide Rossini, and Stefano Scopa for useful
discussions on issues related to this paper.

\appendix

\section{Round-trip Landau-Zener protocols in two-level models}
\label{LZlike}

In this section we study time-dependent round-trip protocols within a
paradigmatic two-level model, described by the Hamiltonian
(\ref{hrdef2}).  Their quantum evolution is ruled by the Schr\"odinger
equation
\begin{eqnarray}
&&i \, \partial_t \Psi(t) = H_{2\ell}(t) \Psi(t)\,,
  \label{hrdef}\\
 &&   H_{2\ell}(t) = - \beta(t) \sigma^{(3)}
  + {\Delta\over 2} \sigma^{(1)}\,.
\nonumber
\end{eqnarray}
The parameter $\Delta$ corresponds to the energy difference of the
Hamiltonian eigenstates at $\beta(t)=0$.  To describe the states
$\Psi(t)$ of the system, we consider the {\em diabatic} basis provided
by the eigenvectors $|+ \rangle$ and $|-\rangle$ of $\sigma^{(3)}$,
\rev{with eigenvalues $1$ and $-1$, respectively}.
Therefore, we may write
\begin{equation}
  \Psi(t) = \phi_1(t)|+\rangle +  \phi_2(t)|-\rangle  \,,
  \label{psitbas}
\end{equation}
and define $\Psi(t)\equiv [\phi_1(t),\phi_2(t)]$.
It is convenient to define
\begin{equation}
  \eta(t) = {2\beta(t)\over \Delta}\,,
  \label{etadef}
\end{equation}
so that
\begin{equation}
 H_{2\ell}(t) = {\Delta\over 2} \, \widetilde{H}_{2\ell}(t)\,,
  \quad
\widetilde{H}_{2\ell}(t) = - \eta(t) \sigma^{(3)} + \sigma^{(1)}\,. 
\label{redefH}
\end{equation}
Adiabatic time evolutions, i.e. for sufficiently slow changes of the
Hamiltonian parameter $\eta(t)$, pass through the stationary
eigenstates of $H_{2\ell}$ at fixed $\eta(t)=\eta$, which are given by
\begin{eqnarray}
 &&|\Psi_0,\eta \rangle = {\cal N}_0(\eta) \left[ (-\eta -
    \sqrt{1+\eta^2}) |+ \rangle + |-\rangle
    \right]\,,\nonumber\\ &&E_0 = - {\Delta\over 2} \sqrt{1 +
    \eta^2}\,,\label{gslz}\\ &&|\Psi_1,\eta\rangle = {\cal N}_1(\eta)
  \left[ (-\eta + \sqrt{1+\eta^2}) |+ \rangle + |-\rangle
    \right]\,,\nonumber\\ &&E_1 = {\Delta\over 2} \sqrt{1 +
    \eta^2}\,, \label{exlz}
\end{eqnarray}
where ${\cal N}_i(\eta)$ are appropriate normalizations so that
$\langle 0| 0 \rangle=\langle 1| 1\rangle=1$.

In the following we consider a linear time dependence of the
Hamiltonian parameter $\beta(t)$, and round-trip linear protocols.  We
start at $t_i=-t_\star$ from the ground state $|\Psi_0,\eta_i\rangle
\equiv [\phi_1^{(0)},\phi_2^{(0)}]$ of the system for
$\beta(t_i)$. Then the system evolves according to the Schr\"odinger
equation (\ref{hrdef}) with $\beta(t)$ given by the
Eq.~(\ref{betadef}), i.e. $\beta(t) = {{\cal T}(t)/t_s}$ for
$t_i=-t_\star \le t \le 3t_\star$, where $ {\cal T}(t) = t_\star -
|t-t_\star|$ is the {\em triangular} function going linearly from
${\cal T}(-t_\star)=-t_\star$ to ${\cal T}(t_\star)=t_\star$, and then
back to ${\cal T}(3t_\star)=-t_\star$. The parameter $t_s$ represents
the time scale of the variation. The parameter $t_\star>0$ controls
the extension (i.e.  the starting and final times) of the protocols,
from $t_i=-t_\star$ to $t_f = 3 t_\star$, and also the interval of
variation of $\beta(t)$, from $\beta(t_i)=-t_\star/t_s$ to
$\beta(t_\star) = t_\star/t_s$.  \rev{An analogous
  cyclic time dependence is considered in the so-called
  Landau-Zener-St\"uckelberg problem, see
  e.g. Refs.~\cite{SAN-10,ISN-22} and references therein.}

To solve this problem, it is convenient to introduce the variables
\begin{eqnarray}
&&\tau = {{\cal T}(t)\over\sqrt{t_s}}\,, \qquad \tau_\star =
       {t_\star\over \sqrt{t_s}}\,,\qquad
       \upsilon = t_s \Delta^2\,, \qquad \label{scalingvar}\\
       && \kappa = {2 \tau\over \sqrt{\upsilon}} = {2\beta(t)\over \Delta}
       \,,\qquad
       \kappa_\star = {2 \tau_\star\over \sqrt{\upsilon}}\,.
       \label{scaloingvar2}
\end{eqnarray}
Then the time evolution can be straightforwardly determined using the
results of Ref.~\cite{VG-96}, in terms of parabolic cylinder functions
$D_\nu(x)$~\cite{Abrafunc}.  Along the first branch from $-t^\star$
  to $t^\star$, we write
\begin{equation}
  \phi_i^{(1)}(\tau) = U_{ij}(\tau,\tau_i) \phi_j^{(0)}\,,
  \label{fbev}
\end{equation}
where $\tau=t/\sqrt{t_s}$ with $-t_\star\le t\le t_\star$, $\tau_i
=-\tau_\star$, and the evolution matrix elements are~\cite{VG-96}
\begin{eqnarray}
&& U_{11}(\tau,\tau_i) = {\Gamma(1-i\upsilon/8)\over \sqrt{2\pi}}
  \times \label{uij}\\ && \quad \Big[
    D_{i\upsilon/8}(\sqrt{2}e^{-i\pi/4}\tau)\,
    D_{-1+i\upsilon/8}(\sqrt{2}e^{i3\pi/4}\tau_i) + \nonumber\\ &&\quad
    \;\;\; D_{i\upsilon/8}(\sqrt{2}e^{i3\pi/4}\tau)\,
    D_{-1+i\upsilon/8}(\sqrt{2}e^{-i\pi/4}\tau_i) \Big]\,, \nonumber
  \\ && U_{12}(\tau,\tau_i) = {2\Gamma(1-i\upsilon/8)e^{i\pi/4}\over
    \sqrt{\pi\upsilon}} \times \nonumber\\ && \quad \Big[ 
    -D_{i\upsilon/8}(\sqrt{2}e^{-i\pi/4}\tau)\,
    D_{i\upsilon/8}(\sqrt{2}e^{i3\pi/4}\tau_i) + \nonumber\\ && \quad 
    \;\;\; D_{i\upsilon/8}(\sqrt{2}e^{i3\pi/4}\tau)\,
      D_{i\upsilon/8}(\sqrt{2}e^{-i\pi/4}\tau_i)
      \Big]\,,\nonumber\\
    &&U_{21} = - U_{12}^*\,, \qquad U_{22} = U_{11}^*\,.
    \nonumber
\end{eqnarray}
Using the properties of the evolution matrix $U$ under the
transformation $\eta(t)\to -\eta(t)$~\cite{VG-96}, we can write the
evolution for $t>t^\star$ as
\begin{equation}
  \phi_i^{(2)}(\tau) = V_{ij}(\tau_b,\tau_i)\phi_j^{(1)}(\tau_\star)\,,
  \label{bev}
  \end{equation}
where $\tau$ is defined as in Eq.~(\ref{scalingvar}), thus it is
decreasing from $\tau_\star$ to $-\tau_\star$, again
$\tau_i=-t_\star/\sqrt{t_s}$, $\tau_b=t_b/\sqrt{t_s}$ with
$t_b=t-2t_\star$, and the functions $V_{ij}$ are closely related to
$U_{ij}$:~\cite{VG-96}
\begin{eqnarray}
  &V_{11} = U_{11}^*\,,\qquad
  &V_{12}= - U_{12}^*\,, \label{velmm}\\
&V_{22}= U_{22}^*\,,\qquad  &V_{21}=
  -U_{21}^*\,. \quad 
  \nonumber
  \end{eqnarray}
\rev{Note that these expressions are consistent with those used for
  the Landau-Zener-St\"uckelberg problem in the presence of
  Hamiltonian parameters with cyclic time dependence as in
  Eq.~(\ref{hrdef}), see e.g.  Refs.~\cite{SAN-10,ISN-22}.}

\begin{figure}[!htb]
  \includegraphics[width=0.95\columnwidth]{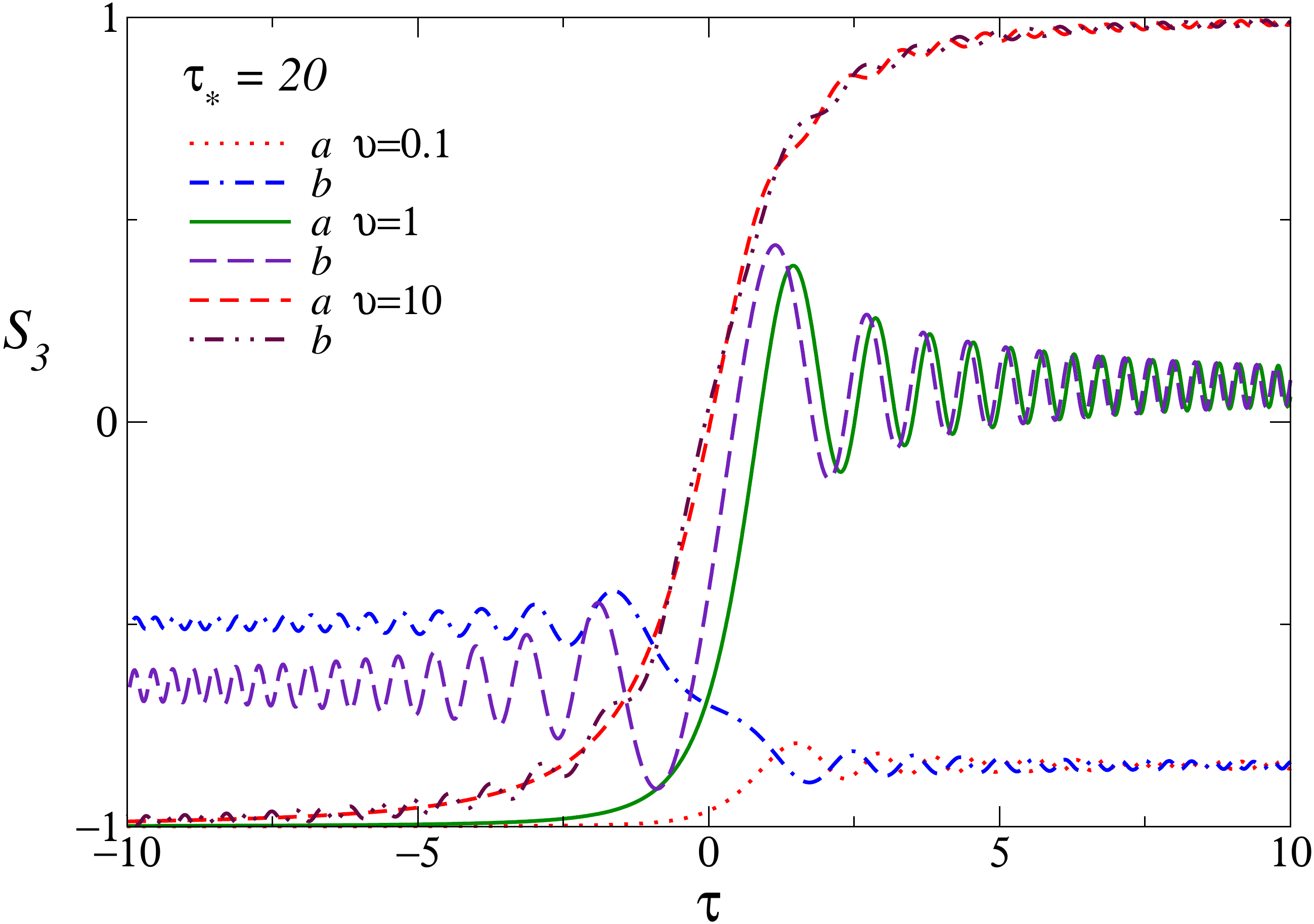}
      \includegraphics[width=0.95\columnwidth]{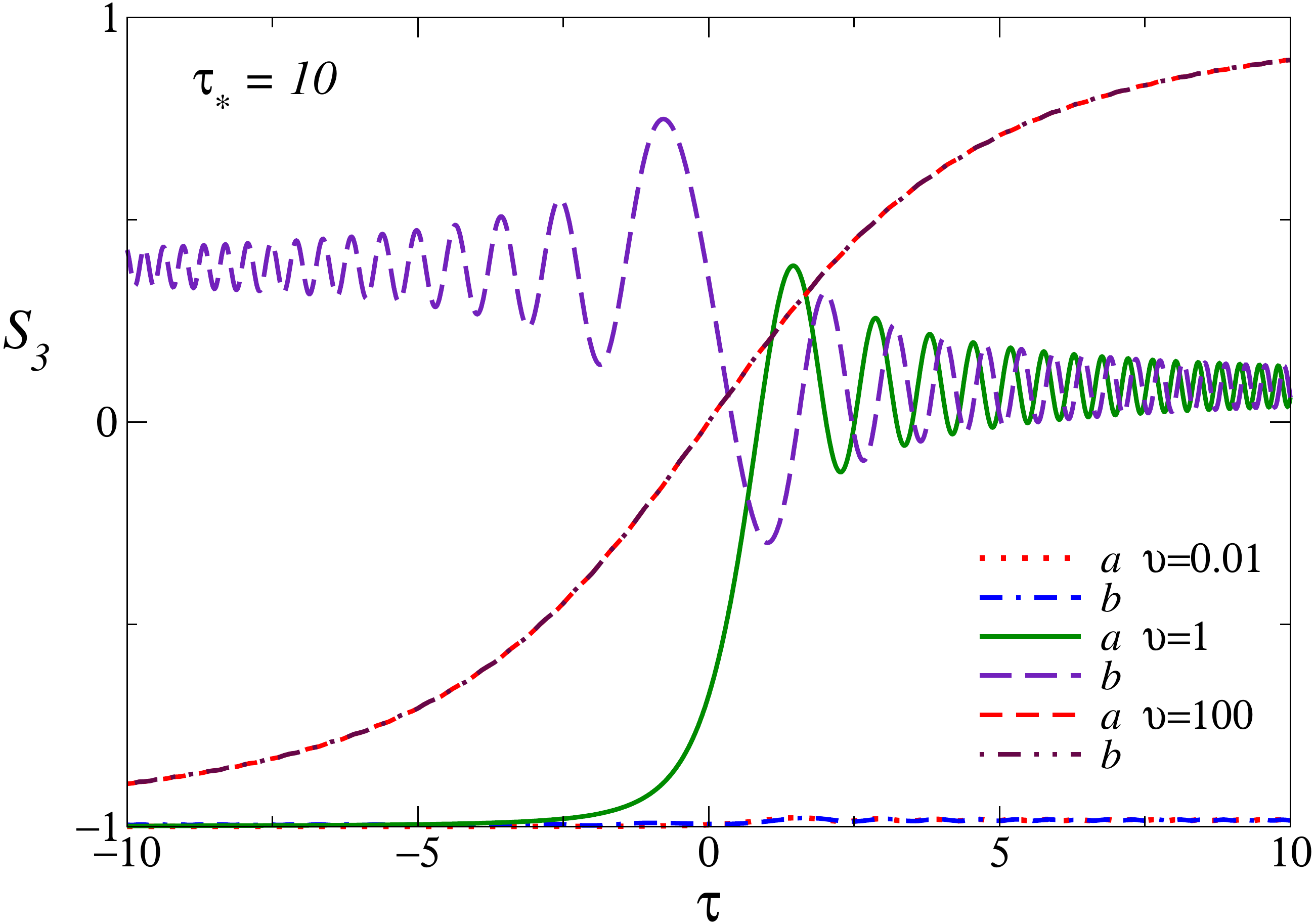}
  \caption{Evolution of $S_3$ during the round-trip protocol, for
    $\tau_\star=10$ (bottom) and $\tau_\star=20$ (top), and some
    values of $\upsilon$.}
  \label{lzfigs1}
\end{figure}

Since the scaling variable $\tau$ related to time takes the same
values in the intervals $-t_\star\le t\le t_\star$ and $t_\star\le
t\le 3 t_\star$, we separate the time dependence in two parts: ($a$)
for the first part where $\beta(t)$ and $\tau$ increases, and ($b$)
where $\beta(t)$ and $\tau$ decreases.  We monitor the dynamic
evolution along the protocol defined above by the expectation values
of the operators $\sigma^{(k)}$, i.e.
\begin{eqnarray}
  && S_{3}^{(a/b)}(\upsilon,\tau,\tau_\star) = \langle \Psi(t) |
  \sigma^{(3)} | \Psi(t) \rangle \,,\label{s3def}\\ &&
  S_1^{(a/b)}(\upsilon,\tau,\tau_\star) = \langle \Psi(t) | \sigma^{(1)} |
  \Psi(t) \rangle \,,\label{s1def}
\end{eqnarray}
and the adiabaticity function
\begin{equation}
  A^{(a/b)}(\upsilon,\tau,\tau_\star)
  = |\langle \, \Psi_0, \eta(t) \, | \,
    \Psi(t) \, \rangle|\,.
    \label{addef}
\end{equation}
\rev{Again, the superscripts $(a)$ and $(b)$ refer to the outward and
return trip, respectively.}  Note that the adiabatic limit of the
evolution is obtained by sending $\upsilon\to\infty$ keeping fixed
$\kappa$. Therefore,
\begin{equation}
  \lim_{\upsilon\to\infty} A^{(a/b)}(\upsilon, \kappa
  \sqrt{\upsilon}/2, \kappa_\star \sqrt{\upsilon}/2) = 1\,.
  \label{adiablimit}
\end{equation}

Some results for the {\em magnetization} $S_3$ are shown in
Fig.~\ref{lzfigs1} along the first and second branch of the protocol,
\rev{for various values of $\upsilon$, $\upsilon=0.1,\,1,\,10$,} and
$\tau_\star=10,\,20$. As expected, the case of large $\upsilon$ the
dynamic tends to be adiabatic, so that the values of $S_3$ along the
two ways tend to superimpose.  In the case of small $\upsilon$ the
dyamic tends to be frozen to the initial condition, moving only
slightly from the initial value. More complex behaviors are observed
for intermediate values of $\upsilon$.

\begin{figure}[!htb]
  \includegraphics[width=0.95\columnwidth]{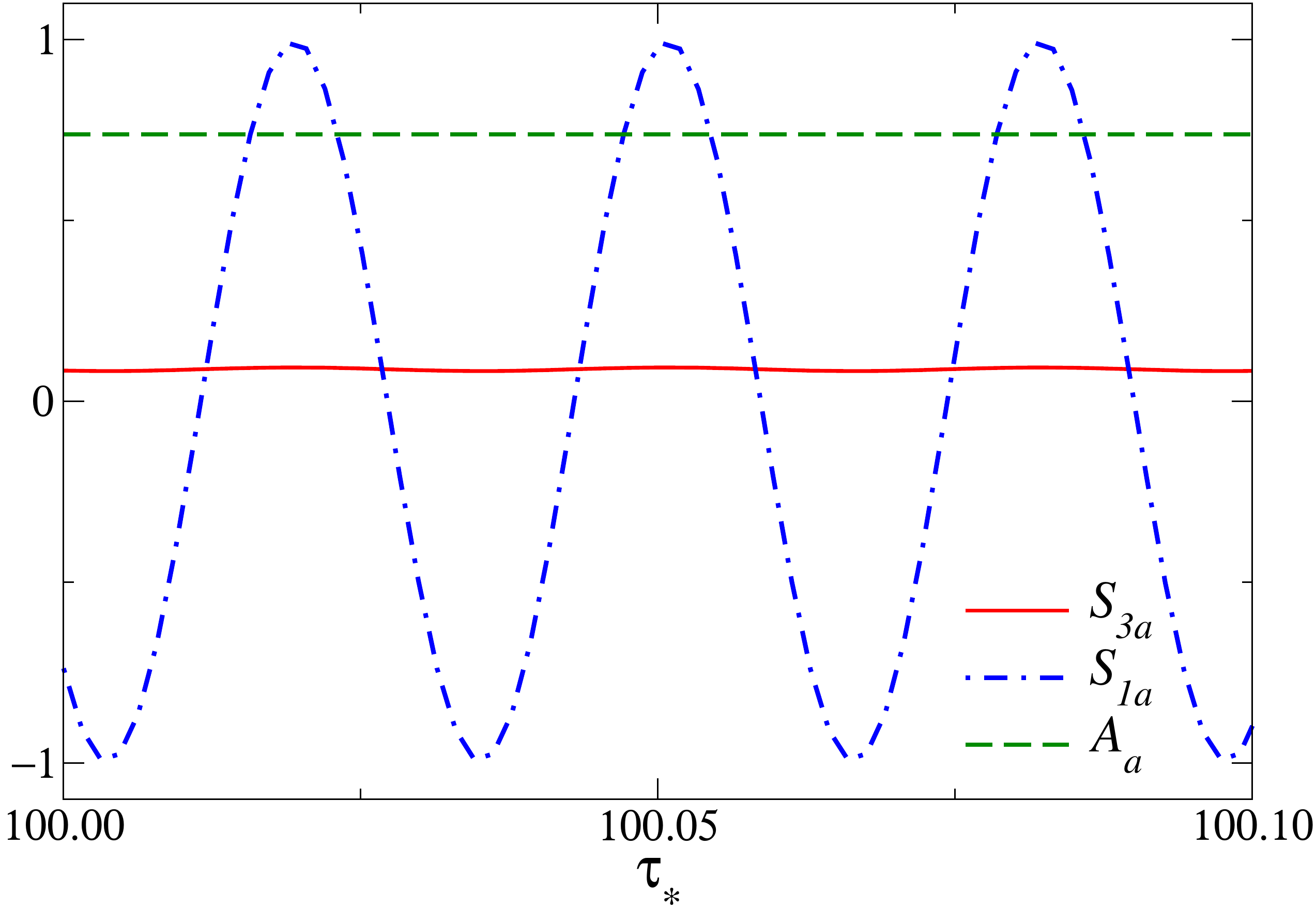}
  \includegraphics[width=0.95\columnwidth]{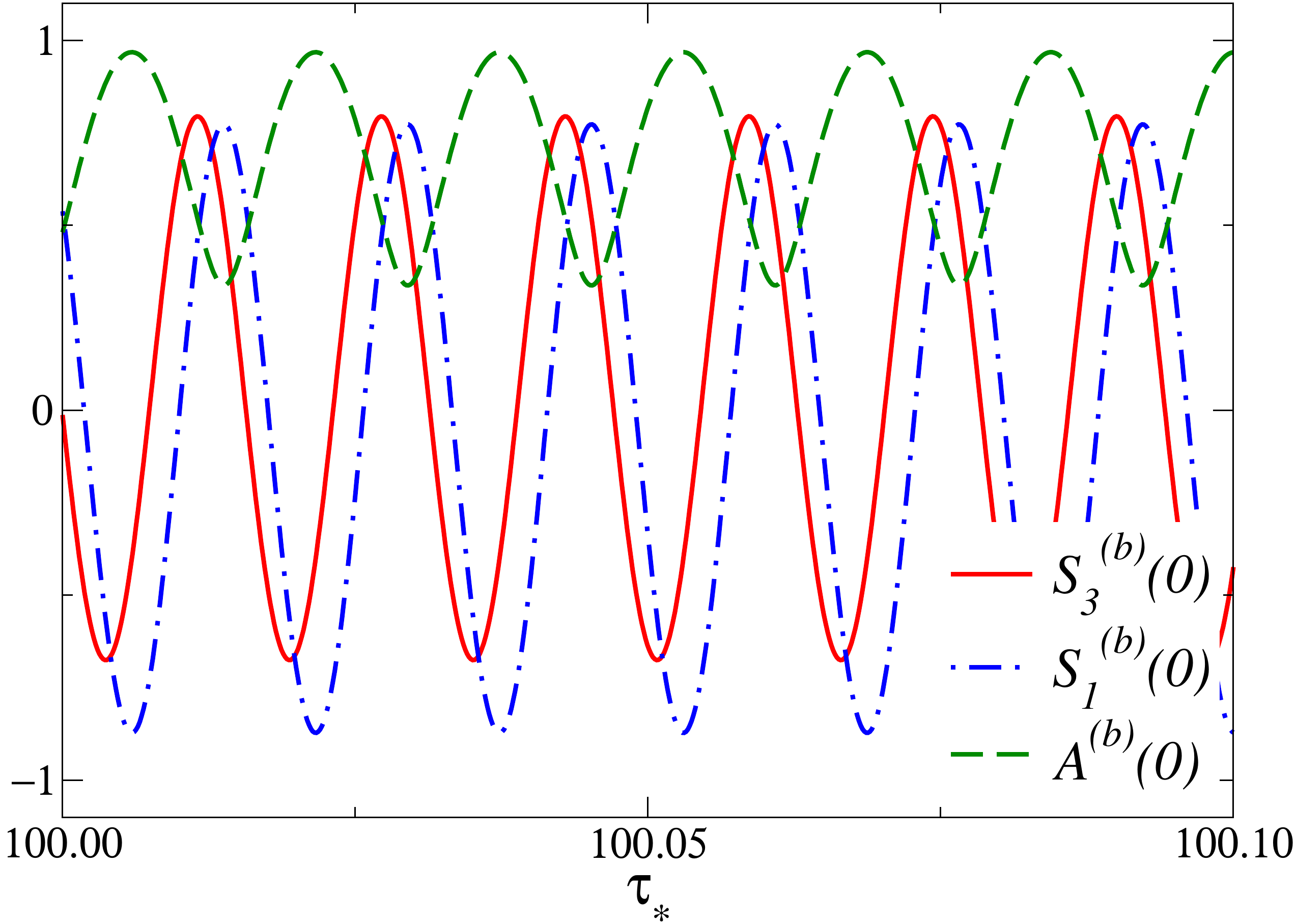}
  \includegraphics[width=0.95\columnwidth]{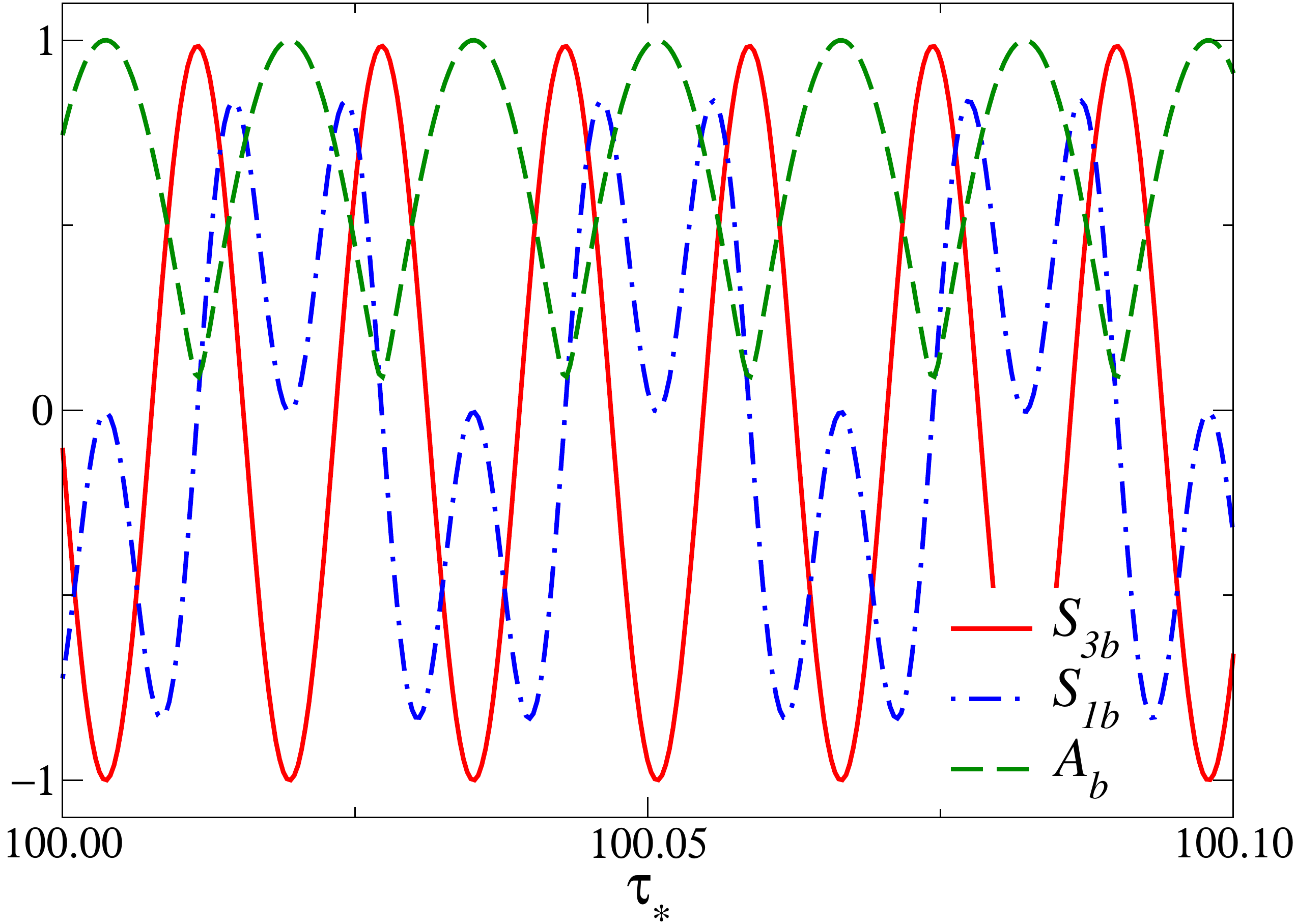}
  \caption{Dependence on $\tau_\star\equiv t_\star/\sqrt{t_s}$ of the
    {\em magnetizations} $S_{1/3}$ and the adiabaticity function $A$
    at the end of the first dynamic branch where $\beta(t)$
    is linearly increasing, and then back along the return way, when
    $\tau=0$ (intermediate) and at the end of the round-trip protocol
    (bottom), for $\upsilon=1$, and $\tau_\star\approx 100$. These
    results show clearly how the oscillations of $S_{1,a}$, and
    therefore of the relative phase of the two functions $\phi_i(t)$
    in Eq.~(\ref{psitbas}), at the end of the first branch are closely
    related to the oscillations of all observables along the return
    way of the round-trip protocol.}
  \label{lzfigs}
\end{figure}

We now analyze the dynamics of the round-trip protocol in the
large-$\tau_\star$ limit, showing that such limit is problematic for
this problem.  We consider the values of the above observables at the
end of the first and second part of the protocol:
\begin{eqnarray}
 S_{3/1a}(\upsilon,\tau_\star) &=&
  S_{3/1}^{(a)}(\upsilon,\tau_\star,\tau_\star)\,,\label{suddef}\\
  S_{3/1b}(\upsilon,\tau^\star) &=&
  S_{3/1}^{(b)}(\upsilon,-\tau_\star,\tau_\star)\,,\nonumber\\
  A_{a}(\upsilon,\tau_\star)
  &=& A^{(a)}(\upsilon,\tau_\star,\tau_\star)\,,\nonumber\\
 A_{b}(\upsilon,\tau_\star) &=&
  A^{(b)}(\upsilon,-\tau_\star,\tau_\star)\,.\nonumber
\end{eqnarray}

\begin{figure}[!htb]
  \includegraphics[width=0.95\columnwidth]{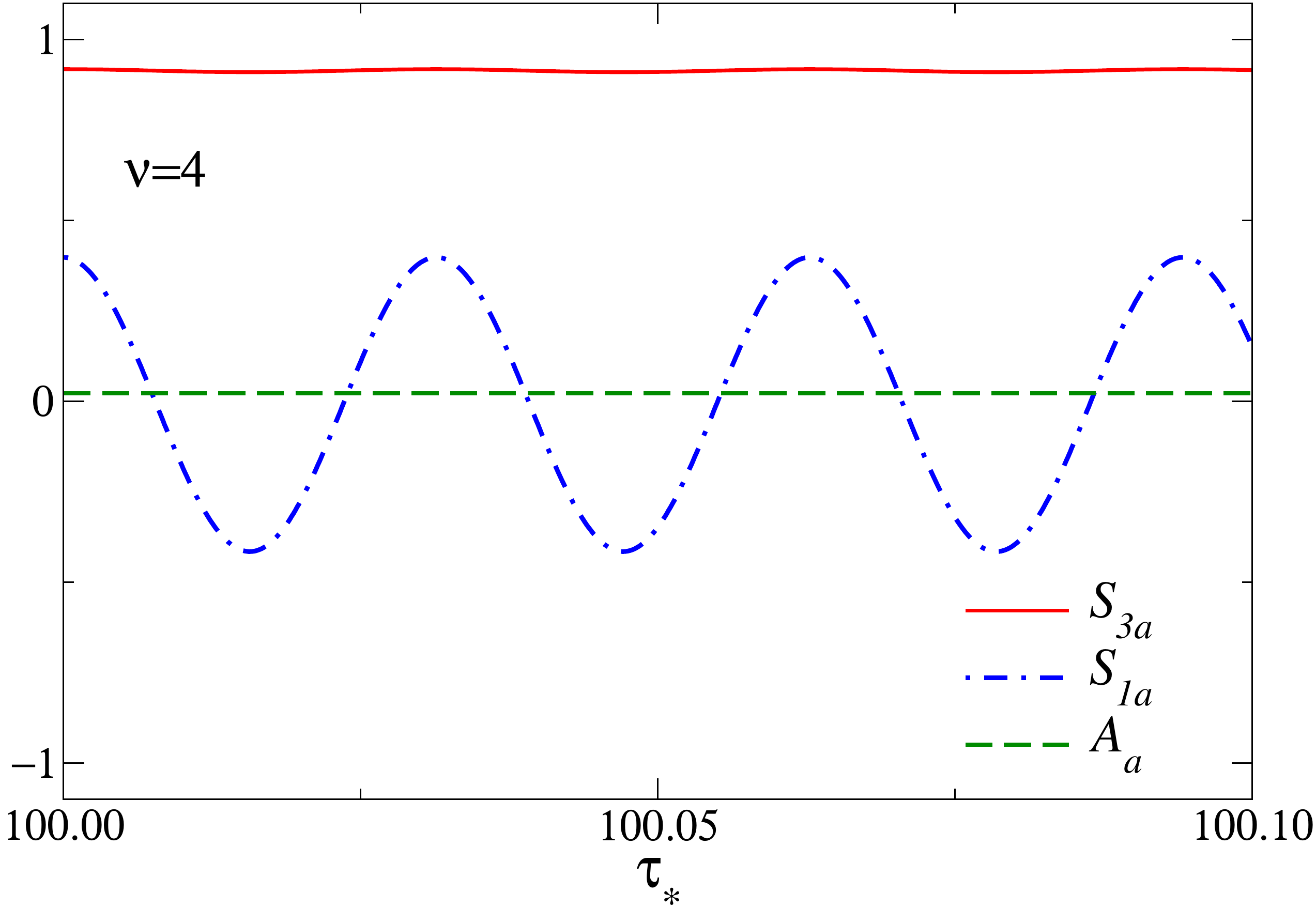}
  \includegraphics[width=0.95\columnwidth]{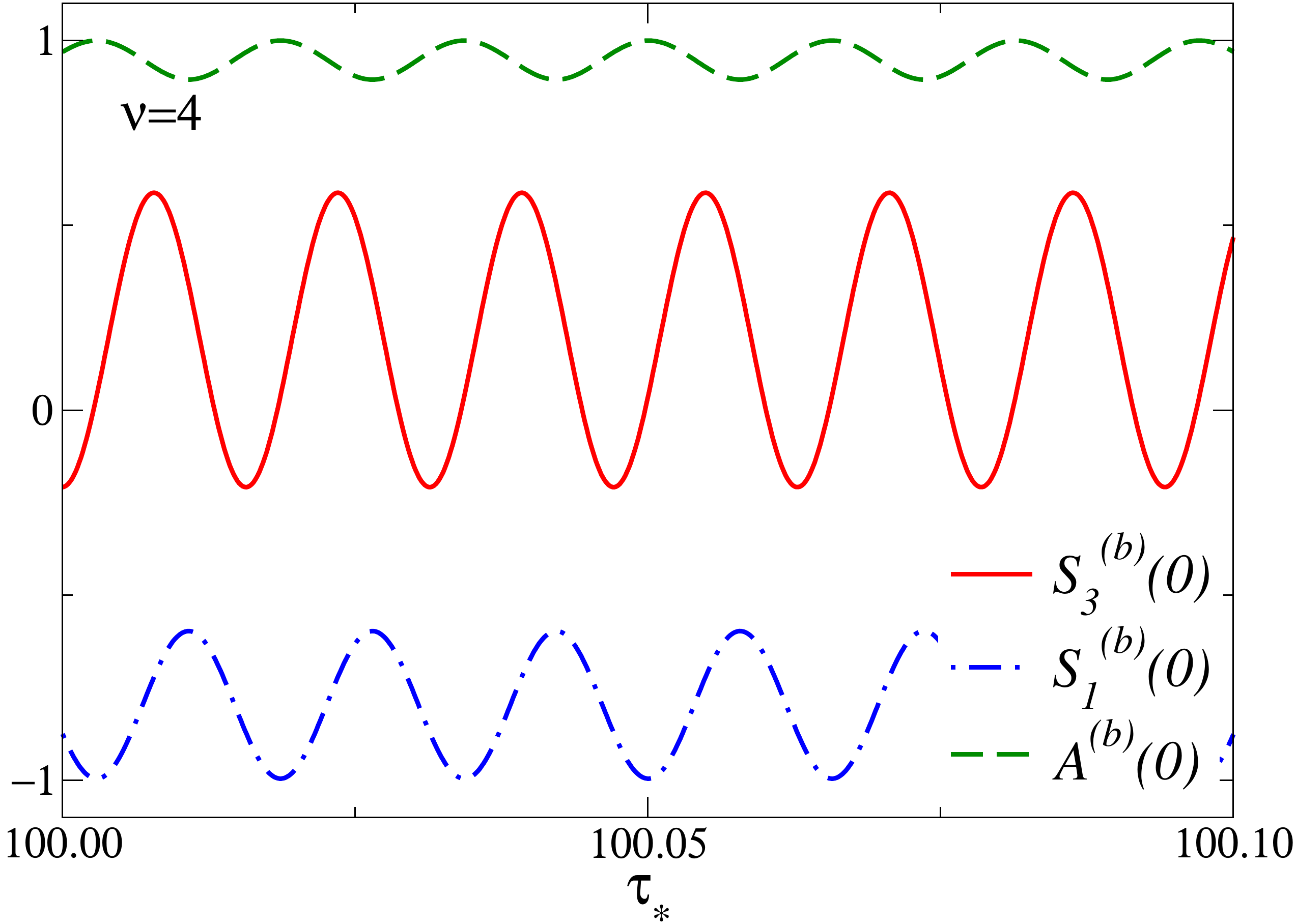}
  \caption{Dependence on $\tau_\star\equiv t_\star/\sqrt{t_s}$ of the
    {\em magnetizations} $S_{1/3}$ and the adiabaticity function $A$
    at the end of the first dynamic branch where $\beta(t)$
    is linearly increasing, and then along the return way, when
    $\tau=0$ (bottom), for $\upsilon=4$, and $\tau_\star\approx 100$.}
  \label{lzfigs2}
\end{figure}

\begin{figure}[!htb]
  \includegraphics[width=0.95\columnwidth]{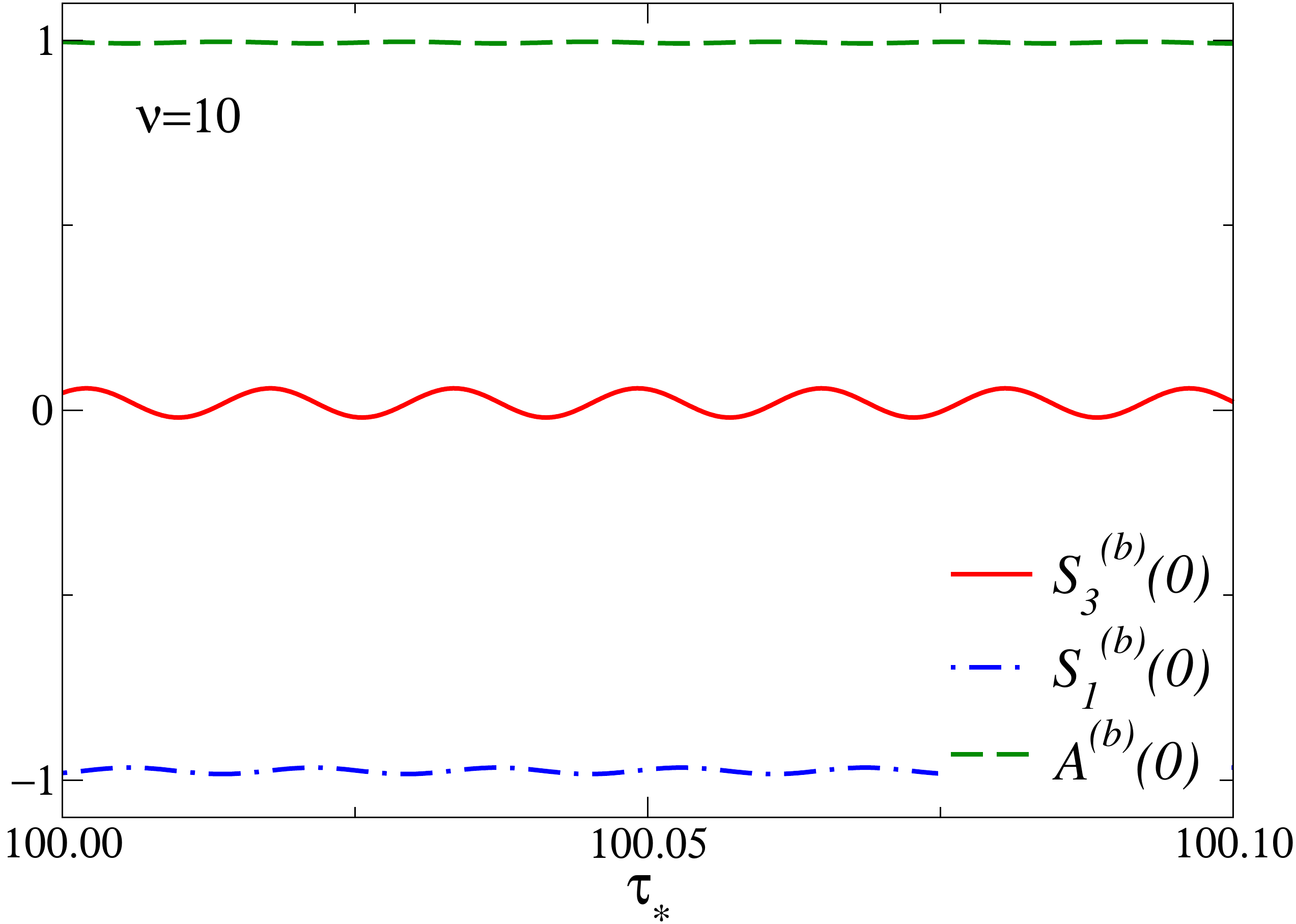}
  \caption{Dependence on $\tau_\star\equiv t_\star/\sqrt{t_s}$ of the
    {\em magnetizations} $S_{1/3}$ and the adiabaticity function $A$
    along the return way at $\tau=0$, for $\upsilon=10$, and
    $\tau_\star\approx 100$.}
  \label{lzfigs3}
\end{figure}

Some notable limits can be derived for the first branch of the
protocol using the asymptotic behaviors of the parabolic cylinder
functions $D_\nu(x)$~\cite{VG-96,ISN-22}, corresponding to the
standard Landau-Zener problem, see e.g. Refs.~\cite{VG-96,PRV-18-loc},
such as
\begin{eqnarray}
&&S_{3a}(\upsilon,\tau_\star\to\infty)
= 1 - 2 \,e^{-{\pi \upsilon/4}}\,,\qquad
\label{lzlimit}\\
&&A_a(\upsilon,\tau_\star\to\infty) =
\sqrt{1 - \,e^{-{\pi \upsilon/4}}} \,.\nonumber
\end{eqnarray}
Both $S_{3a}$ and $A_a$ approach their asymptotic behaviors with
oscillating corrections suppressed as $O(\tau_\star^{-1})$. For
example, in the case of the adiabaticity function we find
\begin{eqnarray}
  \Delta A_a &\equiv& 
  A_a(\upsilon,\tau_\star) - A_a(\upsilon,\infty)   \label{asyoscA}\\
&\approx&
  {f(\upsilon)\over \tau_\star} \cos[\tau_\star^2 - {u\over 4}\ln\tau_\star +
  g(\upsilon)]\,,
\nonumber
\end{eqnarray}
where $f$ and $g$ are time-independent functions of $\upsilon$ only.
Unlike $S_{3a}$ and $A_a$,
the quantity $S_{1a}$ does not show a regular large-$\tau_\star$ limit,
but rapid oscillations with diverging frequency in the large-$\tau^*$
limit.  Indeed, using again the asymptotic behaviors of the parabolic
cylinder functions $D_\nu(x)$~\cite{VG-96,ISN-22}, the asymptotic
large-$\tau_\star$ behavior of $S_{1,a}$ turns out to be
\begin{eqnarray}
  &&  S_{1a} \approx B(\upsilon)
  %  {\rm Re}\,e^{-i\varphi(\upsilon,\tau^\star)}\,,\qquad
  \cos\varphi(\upsilon,\tau_\star)\,,\qquad
\label{asys1}\\
&& B(\upsilon)=  2 \,e^{-\pi\upsilon/8}\, 
\sqrt{1 - e^{- \pi\upsilon/4}}\; \le 1 \,,\nonumber\\
&& \varphi(\upsilon,\tau_\star) =
\tau_\star^2  + {\upsilon\over 8} \ln(2\tau_\star^2)
- {\rm Arg}\left[\Gamma\left(i{\upsilon\over 8}\right)\right]
+ {3\pi\over 4}\,.
\nonumber
 \end{eqnarray}
In particular, $B(1) = 0.99611...$ and
\begin{equation}
  \varphi(1,\tau_\star)= \tau_\star^2 + {1\over 4} \ln\tau_\star
  + 4.08501...
  \label{var1ttau}
\end{equation}

Unlike $S_{3a}$ and $A_a$ that converge to a large-$\tau_\star$ limit,
the leading behavior of $S_{1a}$ is characterized by rapid
oscillations. Its oscillatory behavior is essentially related to the
relative phase $e^{-i\varphi(u,\tau)}$ of the functions
$\phi_1(u,\tau)$ and $\phi_2(u,\tau)$, cf. Eq.~(\ref{psitbas}).  Note
that oscillations become faster and faster in the large-$\tau^\star$
limit, with a time-dependent frequency $\omega(\tau_\star)$ diverging
as $\omega(\tau_\star)\approx \tau_\star$.  Therefore, unlike $S_{3a}$
and $A_a$ whose oscillations gets suppressed as $1/\tau_\star$
approximately, the quantity $S_{1a}$ does not possess a well defined
large-$\tau_\star$ limit, reflecting the fact that the relative phase
of the $\phi_i$ does not converge in the large-$\tau_\star$ limit.

This fact has dramatic implications for the behavior of the system
along the backward branch, making all quantities rapidly oscillating
at the return point, with a frequency related to that of the relative
phase at the end of the first branch.  This behavior is clearly shown
in Figs.~\ref{lzfigs}, where we report some results for the quantities
defined in Eqs.~(\ref{suddef}), at the end of the outward branch, and
along the return branch at $\tau=0$ and at the end of the round-trip
protocol, at fixed $\upsilon=1$ and as a function of the parameter
$\tau_\star$, for a relatively small interval around
$\tau_\star\approx 100$. As shown by the analogous curves reported in
Figs.~\ref{lzfigs2} and \ref{lzfigs3} for $\nu=4$ and $\nu=10$
respectively, the size of the oscillations depends on the value of
$\nu$, and, as expected, it tends to decreases in the adiabatic limit
when increasing $\nu$.

These results evidentiate the peculiar oscillations in the
large-$\tau_\star$ limit at finite values of $\nu$, which make
predictions on the return behavior practically impossible without an
extreme precision on the control of the parameters of the protocols.

\end{document}